\documentclass{emulateapj} 
\usepackage{apjfonts}        
\usepackage{graphicx}  
\usepackage{amsmath}   

\graphicspath{{./}}

\newcommand{\Msun}{\rm M_{\odot}}

\newcommand{\Lsun}{\rm L_{\odot}}

\newcommand{\LB}{\rm L_{\rm B}}
\newcommand{\Lb}{\rm L_{\rm B}}
\newcommand{\Lfir}{\rm L_{\rm {FIR}}}
\newcommand{\Lir}{\rm L_{\rm {IR}}}
\newcommand{\Lnir}{\rm L_{\rm {NIR}}}
\newcommand{\Lmir}{\rm L_{\rm {MIR}}}
\newcommand{\Lradio}{\rm L_{\rm {radio}}}
\newcommand{\Lx}{\rm L_{\rm {x-ray}}}

\newcommand{\MBH}{\rm M_{\rm{BH}}}

\newcommand{\Mstar}{\rm M_{\rm{star}}}
\newcommand{\msigma}{\MBH\rm{-}\sigma}

\newcommand{\zquasar}{J1148+5251}

\newcommand{\Gpc}{\rm {Gpc}}
\newcommand{\Mpc}{\rm {Mpc}}
\newcommand{\kpc}{\rm {kpc}}
\newcommand{\yr}{\rm {yr}}

\def\beq{\begin{equation}}
\def\eeq{\end{equation}}

\shorttitle{Modeling Dust in $z \sim 6$ Quasars}
\shortauthors{Li et al.}

\begin{document}   

\title{Modeling the Dust Properties of $z \sim 6$ Quasars with ART$^2$ ---
  All-wavelength Radiative Transfer with Adaptive Refinement Tree}

\author
{
Yuexing Li\altaffilmark{1},
Philip F. Hopkins\altaffilmark{1},
Lars Hernquist\altaffilmark{1}, 
Douglas P. Finkbeiner\altaffilmark{1},
Thomas J. Cox\altaffilmark{1},  \\
Volker Springel\altaffilmark{2}, 
Linhua Jiang\altaffilmark{3}, 
Xiaohui Fan\altaffilmark{3},
Naoki Yoshida\altaffilmark{4}
}

\affil{$^{1}$Harvard-Smithsonian Center for Astrophysics, Harvard
  University, 60 Garden Street, Cambridge, MA 02138, USA}
\affil{$^{2}$Max-Planck-Institute for Astrophysics,
  Karl-Schwarzschild-Str. 1, 85740 Garching, Germany} 
\affil{$^{3}$Department of Astronomy, University of Arizona, 933 N. Cherry
  Ave, Tucson, AZ 85721} 
\affil{$^{4}$Nagoya University, Dept. of Physics, Nagoya, Aichi
  464-8602, Japan} 
\email{yxli@cfa.harvard.edu}

\begin{abstract}

The detection of large quantities of dust in $z \sim 6$ quasars by
infrared and radio surveys presents puzzles for the formation and
evolution of dust in these early systems.  Previously \citep{Li2007},
we showed that luminous quasars at $z \gtrsim 6$ can form through
hierarchical mergers of gas-rich galaxies, and that these systems are
expected to evolve from starburst through quasar phases.  Here, we
calculate the dust properties of simulated quasars and their
progenitors using a three-dimensional Monte Carlo radiative transfer
code, ART$^2$ -- All-wavelength Radiative Transfer with Adaptive
Refinement Tree. ART$^2$ incorporates the radiative equilibrium
algorithm developed by \cite{Bjorkman2001} which treats dust emission
self-consistently, an adaptive grid method which can efficiently cover
a large dynamic range in spatial scales and can capture inhomogeneous
density distributions, a multiphase model of the interstellar medium
which accounts for the observed scaling relations of molecular clouds,
and a supernova-origin model for dust which can explain the existence
of dust in cosmologically young, starbursting quasars.  By applying
ART$^2$ to the hydrodynamic simulations of \cite{Li2007}, we reproduce
the observed spectral energy distribution (SED) and inferred dust
properties of SDSS J1148+5251, the most distant quasar detected in the Sloan
survey. We find that the dust and infrared emission are closely associated
with the formation and evolution of the quasar host.  As the system evolves
from a starburst to a quasar, the SED changes from being dominated by a
cold dust bump (peaking at $\sim 50\, \mu$m) to one that includes a
prominent hot dust component (peaking at $\sim 3\, \mu$m), and the
galaxy evolves from a cold to a warm ultraluminous infrared galaxy
(ULIRG) owing to heating and feedback from star formation and the
active galactic nucleus (AGN).  Furthermore, the AGN activity has
significant implications for the interpretation of observable aspects
of the hosts.  The hottest dust ($T \gtrsim 10^3$~K) is most
noticeable only during the peak quasar activity, and correlates with
the near-IR flux.  However, we find no correlation between the star
formation rate and far-IR luminosity during this phase owing to strong
AGN contamination.  Our results suggest that vigorous star formation in
merging progenitors is necessary to reproduce the observed dust
properties of $z \sim 6$ quasars, supporting a merger-driven origin
for luminous quasars at high redshifts and the starburst-to-quasar
evolutionary hypothesis.

\end{abstract}

\section{INTRODUCTION}

High-redshift quasars are important for understanding the formation
and evolution of galaxies and supermassive black holes (SMBHs) in the
early Universe.  In the past few years, nearly two dozen luminous
quasars have been discovered by the Sloan Digital Sky Survey (SDSS;
\citealt{York2000}) and the Canada-France High-z Quasar Survey 
(CFHQS; \citealt{Willott2007}) at $z \sim 6$, 
corresponding to a time when the Universe was less than a billion years old
(\citealt{Fan2003, Fan2004, Fan2006B, Fan2006A}).  As summarized by
\cite{Fan2006C}, these quasars are rare ($\sim 10^{-9}\, \Mpc^{-3}$ comoving);
believed to be powered by SMBHs with masses $\sim 10^9\, \Msun$ (e.g.,
\citealt{Willott2003, Barth2003}); reside near the end of the epoch of
reionization, as indicated by Gunn-Peterson absorption troughs
\citep{Gunn1965} in their spectra; and have similar spectral energy
distributions (SEDs) and comparable metallicity to lower-redshift counterparts    
(e.g., \citealt{Elvis1994, Glikman2006, Richards2006, Hopkins2007A}),
implying the early presence of ``mature'' quasars and the formation of
their hosts in rapid starbursts at even higher redshifts ($z \gtrsim
10$).

Follow-up, multi-wavelength observations have been carried out for
these $z \sim 6$ quasars, from X-ray (e.g., \citealt{Brandt2002,
Strateva2005, Vignali2005, Steffen2006, Shemmer2005, Shemmer2006}), to
optical /infrared (e.g., \citealt{Barth2003, Pentericci2003,
Freudling2003, White2005, Willott2005, Hines2006}), and radio
wavelengths (e.g., \citealt{Carilli2001, Bertoldi2003B, Walter2003,
Carilli2004, Wang2007}).  A noteworthy result of these studies is their
detection of dust in these high-redshift objects.  In particular, deep
infrared and radio surveys (e.g., \citealt{Robson2004, Bertoldi2003A,
Carilli2004, Charmandaris2004, Beelen2006, Hines2006}) have revealed a
large amount of cold dust in SDSS J1148+5251 (hereafter J1148+5251),
the most distant Sloan quasar detected at redshift $z = 6.42$
\citep{Fan2003}. The dust mass is estimated to be $\sim 1-7\times 10^8\,
\Msun$. The detection of dust is also reported in the first four CFHQS quasars 
at $z > 6$, including the new record holder CFHQS J2329-0301 at $z = 6.43$
\citep{Willott2007}. \cite{Maiolino2004} argue, from the observed dust
extinction curve of SDSS J1048+4637 at $z = 6.2$, that the dust in these
high-z systems may be produced by supernovae.

\cite{Jiang2006} have performed a comprehensive study of thirteen $z
\sim 6$ quasars by combining new {\em Spitzer} observations with
existing multi-band data.  It appears that nearly all 13 of these
quasars exhibit prominent infrared bumps around both $\lambda \sim 3
\mu$m and $\lambda \sim 50 \mu$m.  In a dusty galaxy, most of the
radiation from newly formed stars or an AGN is absorbed by dust and
re-emitted at infrared wavelengths.  The SEDs of star-forming galaxies
typically peak at 50 -- 80 $\mu$m (rest frame) and can be approximated
by a modified blackbody spectrum with dust temperatures below 100 K
(e.g., \citealt{Dunne2001}), while in quasars, some dust can be
heated directly by the AGN to temperatures up to $1200$ K
(e.g., \citealt{Glikman2006}) and dominate the near- to mid-IR
emission.  Therefore, the observations by \cite{Jiang2006} indicate
the presence of large amounts of both hot and cold dust in these
high-z systems.

However, the nature of the dust, and its formation and evolution in
the context of the quasar host, are unclear.  For example, the dust
distribution is unknown.  It has been suggested that the hot dust lies
in the central regions and is heated by an AGN to produce near-IR
emission \citep{Rieke1981, Polletta2000, Haas2003}, while the warm and
cold dust can extend to a few kpc and dominate at mid-IR and far-IR
wavelengths \citep{Polletta2000, Nenkova2002, Siebenmorgen2005}.
However, the relative importance of heating by stars and AGN activity
is uncertain.  This is essential to an accurate determination of the
star formation rate (SFR), but cannot be inferred simply from observed
SEDs.  Currently, the most common method used to derive a star
formation rate is to assume that most of the FIR luminosity comes from
young stars.  However, if the FIR luminosity is mainly contributed by
an AGN, then the SFR would be substantially reduced.  Finally, it is
not known how the SED and dust content of a quasar and its host evolve
with time.

In order to address these questions, we must combine a model for the
formation of a quasar with radiative transfer calculations that treat
dust emission self-consistently, to follow the physical properties,
environment, and evolution of quasar hosts and their dust content.

Earlier \citep{Li2007}, we developed a quasar formation model which
self-consistently accounts for black hole growth, star formation,
quasar activity, and host spheroid formation in the context of
hierarchical structure formation.  We employed a set of multi-scale
simulations that included large-scale cosmological simulations and
galaxy mergers on galactic scales, together with a self-regulated
model for black hole growth, to produce a luminous quasar at $z \sim
6.5$, which has a black hole mass ($\sim 2 \times 10^9\, \Msun$)
and a number of properties similar to \zquasar\ \citep{Fan2003}.  In our
scenario, luminous, high-redshift quasars form in massive halos that originate   
from rare density peaks in the standard $\Lambda$CDM cosmology, and
they grow through hierarchical mergers of gas-rich galaxies. Gravitational
torques excited in these mergers trigger large inflows of gas and produce
strong shocks that result in intense starbursts \citep{Hernquist1989B,
  Barnes1991, Barnes1992, Barnes1996, Hernquist1995, Mihos1996} and fuel rapid
black hole accretion \citep{DiMatteo2005, Springel2005B}.  Moreover, feedback 
from the accreting black hole disperses the obscuring material, briefly
yielding an optically visible quasar \citep{Hopkins2005A, Hopkins2005B,
Hopkins2005C, Hopkins2006A}, and regulating the $\msigma$ correlation
between the SMBHs and host galaxies \citep{DiMatteo2005, Robertson2006A}.  In
this picture, quasars are, therefore, descendents of starburst galaxies
\citep{Sanders1988, Norman1988, Scoville2003}. 

To make detailed contact with observations of $z \sim 6$ quasars, it
is necessary to theoretically predict the SEDs of these systems and
how they evolve with time.  Solving radiation transport in dusty,
starbursting quasars is, however, difficult owing to the non-locality
of the sources, the opacity and multiphase character of the
interstellar medium, and the complex morphology of these systems.
Over the last several decades, a variety of numerical techniques have
been developed to solve the radiative transfer problem with different
levels of approximation and in multi-dimensions.  Based on the
specific algorithms employed, the approaches can be classified into
two general categories: {\em finite-difference} and {\em Monte Carlo}
methods (e.g., \citealt{Jonsson2006}; see also \citealt{Pascucci2004}
who describe the codes as ``grid-based'' or ``particle-based'' by
analogy to hydrodynamic solvers).

{\em Finite-difference} codes solve the equations of radiative
transfer (RT) iteratively using finite convergence criteria.  They
either solve the moment equations for RT, originally formulated by
\cite{Hummer1971} for spherical geometry with a central point source
(e.g., \citealt{Scoville1976, Leung1976, Yorke1980, Wolfire1986,
Menshchikov1997, Dullemond2000}), or employ ray-tracing methods on a
discrete spatial grid for complex density configurations (e.g.,
\citealt{Rowan-Robinson1980, Efstathiou1990, Efstathiou1991,
Steinacker2003, Folini2003, Steinacker2006}). This technique provides
full error control but can be time-consuming.

{\em Monte Carlo} methods sample and propagate photons
probabilistically (e.g., \citealt{Witt1977, Lefevre1982, Lefevre1983,
Whitney1992, Witt1992, Code1995, Lopez1995, Lucy1999, Wolf1999,
Bianchi2000, Harries2000, Bjorkman2001, Whitney2003A, Jonsson2006,
Pinte2006}).  The Monte Carlo technique is more flexible than the
finite-difference one because it tracks the scattering, absorption and
re-emission of photons (or photon packets) in detail, and can handle
arbitrary geometries, but at the cost of computational expense to
reduce Poisson noise.  However, advances in computing technology and
algorithms have made high-accuracy Monte Carlo RT calculations feasible
and popular.

In particular, \cite{Bjorkman2001} have developed a Monte Carlo code
to handle radiative equilibrium and temperature corrections, which
calculates dust emission self-consistently.  It conserves the total
photon energy, corrects the frequency distribution of re-emitted
photons, and requires no iteration as the dust opacity is assumed to be 
independent of temperature.  This code has been used in a number of
applications, including protostars that have a disk and an envelope
with a single heating source in the center (e.g.,
\citealt{Whitney1992, Whitney1993, Whitney2003B, Whitney2003A,
Whitney2004}); circumstellar envelopes (e.g., \citealt{Wood1996B,
Wood1996A, Wood1998}); protoplanetary systems (e.g., \citealt{Wood2002,
Rice2003}); and galaxies that include a bulge and a disk with multiple
heating sources from these two populations (e.g., \citealt{Wood1997,
Wood2000}).  This code has also been able to generate optical--far-IR
SEDs that reproduce those of a sample of 21 X-ray selected AGNs
\citep{Kuraszkiewicz2003}, and the version of \cite{Whitney2003A} has been
applied to simulations of galaxy mergers with black hole feedback to study the
local ULIRGs \citep{Chakrabarti2007A} and
submillimeter galaxies \citep{Chakrabarti2007B}.

In order to produce accurate SEDs of quasars and their hosts formed by
galaxy mergers, as in \cite{Li2007}, the RT code must satisfy the
following requirements: (1) to be able to handle arbitrary geometries
and distributed heating sources; (2) resolve the large dynamic ranges
in spatial scales and densities in the merger simulations; (3)
incorporate a multiphase description of the interstellar medium
(ISM); and (4) employ a dust model appropriate for a young starburst
system.

In many earlier applications, radial logarithmic or uniformly spaced
meshes were used to generate density grids. However, in merging
galaxies, the ISM is clumpy and irregular owing to shocks and tidal
features produced during the interactions, and has multiple density
centers, making logarithmic algorithms inefficient.  In such a
situation, an adaptive grid approach appears ideal for resolving
localized high-density regions, while still covering the large volumes
of merging systems \citep{Wolf1999, Kurosawa2001, Harries2004,
Jonsson2006}.  

In addition, dust models commonly used in previous works are based on
observed extinction curves for the Milky Way (e.g.,
\citealt{Weingartner2001, Calzetti1994, Kim1994, Mathis1977,
Calzetti2000}), in which the dust is assumed to be produced mainly by
old, low-mass stars with ages $> 1$ Gyr \citep{Mathis1990,
Whittet2003, Dwek2005}.  However, a large amount of dust ($\sim 1-7
\times 10^8\, \Msun$) is detected in the $z \simeq 6.42$ quasar host
\citep{Bertoldi2003A} at a time when the Universe was only $\sim 850$
Myr old.  It has been suggested by observations (e.g.,
\citealt{Maiolino2004, Maiolino2006, Moseley1989, Dunne2003,
Morgan2003B, Sugerman2006}) and theoretical studies (e.g.,
\citealt{Todini2001, Nozawa2003, Schneider2004, Nozawa2007, Bianchi2007}) that  
supernovae can provide fast and efficient dust formation in the early
Universe, motivating other choices for the dust model in the
RT calculations.

Finally, in a multiphase description of the ISM \citep{McKee1977}, as
adopted in our simulations \citep{Springel2003A, Springel2003B, Hopkins2006A},  
the ``hot-phase'' (diffuse) and ``cold-phase'' (dense molecular and
${\rm HI}$ core) components co-exist under pressure equilibrium but
have different mass fractions and volume filling factors (i.e., the
hot-phase gas is $\le 10\%$ in mass but $\gtrsim 99\%$ in volume), 
so both phases contribute to the dust extinction and should therefore be
included in the RT calculations. 

We have refined the Monte Carlo RT code developed by \cite{Bjorkman2001} and
\cite{Whitney2003A} by implementing: an adaptive grid algorithm similar to
that of \cite{Jonsson2006} for the density field and the arbitrarily
distributed sources; a multiphase ISM model \citep{Springel2003A} which
accounts for observed scaling relations of molecular clouds for the dust
distribution; and a supernova-origin dust model for the opacity using the
grain size distribution of \cite{Todini2001}. We refer to our new
code as ART$^2$ (All-wavelength Radiative Transfer with Adaptive
Refinement Tree). ART$^2$ is capable of producing SEDs and images in a wide
range of wavelengths from X-ray to millimeter. In the present paper we focus
on the dust properties from optical to submillimeter bands. As we show in what
follows, ART$^2$ reproduces the spectrum of a single galaxy with bulge and
disk calculated with the original code. More important, it captures the
inhomogeneous density distribution in galaxy mergers and reproduces the
observed SED and dust properties of \zquasar\ based on the simulations of
\cite{Li2007}.  Therefore, ART$^2$ can be used to predict multi-wavelength
properties of quasar systems and their galaxy progenitors.

This paper is organized as follows. In \S~2, we describe our
computational methods and models, including the multi-scale
simulations of \cite{Li2007} for $z \sim 6$ quasar formation, and our
implementation of ART$^2$ which incorporates radiative equilibrium as
in \cite{Bjorkman2001}, an adaptive grid, a multiphase ISM, and a
supernova-origin dust model. In \S~3, we present the multi-wavelength
SED from optical to submillimeter of a simulated quasar at $z \sim
6.5$.  The dust distribution and properties of the quasar are
discussed in \S~4 and we describe their evolution in \S~5.  We consider
the robustness of our results in \S~6 and summarize in \S~7.

\section{Methodology}

In order to capture the physical processes underlying the formation
and evolution of quasars in the early Universe, and to compare their
multi-wavelength properties to observations, we combine quasar
formation and radiative transfer calculations.  We first perform a
set of novel multi-scale simulations that yield a luminous quasar at
$z \sim 6.5$ \citep{Li2007}. We then apply the 3-D, Monte Carlo
radiative transfer code, ART$^2$, to the outputs of the hydrodynamic
galaxy mergers simulations to calculate the SED of the system.  The
models and simulations of quasar formation are described in detail in
\cite{Li2007}.  Here, we briefly summarize the modeling of quasar
formation, and the specifications of ART$^2$.

\subsection{Formation Model of $z \sim 6$ Quasars}

\subsubsection{Multi-scale Simulations}

The $z \sim 6$ quasars are rare (space density $\sim 1\, \Gpc^{-3}$
comoving), and appear to be powered by supermassive black holes of
mass $\sim 10^9\, \Msun$. Therefore, simulations of high-redshift
quasar formation must consider a large cosmological volume to
accommodate the low abundance of this population, have a large dynamic
range to follow the hierarchical build-up of the quasar hosts, and
include realistic prescriptions for star formation, black hole growth,
and associated feedback mechanisms. The multi-scale simulations in
\cite{Li2007} include both N-body cosmological calculations in a
volume of $3\, \Gpc^3$ to account for the low number density of
quasars at $z \sim 6$, and hydrodynamical simulations of individual
galaxy mergers on galactic scales to resolve gas-dynamics, star
formation, and black hole growth.

First, we perform a coarse dark matter-only simulation in a volume of
$3\, {\rm Gpc^{3}}$. The largest halo at $z=0$, within which early,
luminous quasars are thought to reside \citep{Springel2005A}, is then
selected for resimulation at higher resolution.  The evolution of this
halo and its environment is re-simulated using a multi-grid zoom-in
technique \citep{Gao2005} that provides much higher mass and spatial
resolution for the halo of interest, while following the surrounding
large-scale structure at lower resolution. The merging history of the
largest halo at $z \sim 6$, which has then reached a mass of $\sim 7.7
\times10^{12}\, \Msun$ through seven major (mass ratio $<$ 5:1) mergers
between redshifts 14.4 and 6.5, is extracted. These major mergers are
again re-simulated hydrodynamically using galaxy models 
which include a \cite{Hernquist1990} halo and an exponential disk 
scaled appropriately for redshift \citep{Robertson2006A}, and adjusted to
account for mass accretion through minor mergers. Each of these eight
galaxy progenitors has a black hole seed assumed to originate from the
remnants of the first stars (\citealt{Abel2002, Bromm2004, Tan2004,
Yoshida2003, Yoshida2006, Gao2007}).

The simulations were performed using the parallel, N-body/Smoothed
Particle Hydrodynamics (SPH) code GADGET2 \citep{Springel2005D}, which
conserves energy and entropy using the variational principle
formulation of SPH \citep{Springel2002}, and which incorporates a
sub-resolution model of a multiphase interstellar medium to describe
star formation and supernova feedback \citep{Springel2003A}. Star
formation is modeled following the Schmidt-Kennicutt Law
(\citealt{Schmidt1959, Kennicutt1998}). Feedback from supernovae is
captured by an effective equation of state for star-forming gas
\citep{Springel2003A}.  A prescription for supermassive black hole
growth and feedback is also included, where black holes are
represented by collisionless ``sink'' particles that interact
gravitationally with other components and accrete gas from their
surroundings.  The accretion rate is estimated from the local gas
density and sound speed using a spherical Bondi-Hoyle
\citep{BondiHoyle1944, Bondi1952} model that is limited by the
Eddington rate. Feedback from black hole accretion is modeled as
thermal energy injected into the surrounding gas \citep{Springel2005B,
DiMatteo2005}.  We note that implementations of our model for black
hole growth and feedback that do not explicitly account for 
Eddington-limited accretion achieve similar results to the
method employed by us (e.g. compare the works of \citealt{DiMatteo2007}
and \citealt{Sijacki2007}).

These hydrodynamic simulations adopted the $\Lambda$CDM model with
cosmological parameters from the first year Wilkinson Microwave Anisotropy
Probe data (WMAP1, \citealt{Spergel2003}), ($\Omega_{\rm{m}}$,
$\Omega_{\rm{b}}$, $\Omega_{\Lambda}$, $h$, $n_s$, $\sigma_8$)= (0.3, 0.04,
0.7, 0.7, 1, 0.9). In this paper, we use the same parameters.

\subsubsection{Hierarchical Assembly of the Quasar System}

\begin{figure}
\begin{center}
\vspace{0.5cm}
\includegraphics[width=3.4in]{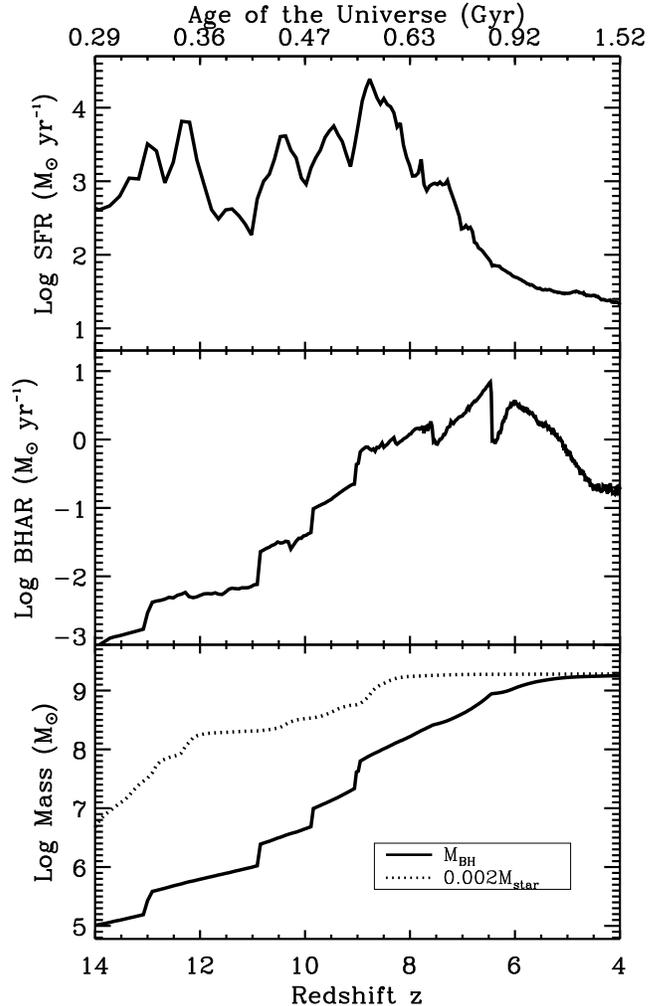}
\vspace{0.5cm}
\caption{Evolution of the star formation rate, black hole accretion rate, and
  masses of the black holes and stars, respectively, of the simulated quasar
  system at $z \sim 6.5$, adopted from \cite{Li2007}. }   
\label{Fig_quasar} 
\end{center}
\end{figure}

In the simulation analyzed here, the quasar host galaxy builds up
hierarchically through seven major mergers of gas-rich progenitors
between $z = 14.4 - 6.5$.  Gravitational interactions between the
merging galaxies form tidal tails, strong shocks and efficient gas
inflows that trigger intense starbursts. The highly concentrated gas
fuels rapid black hole accretion. Figure~\ref{Fig_quasar} shows the
evolution of some aspects of the system. Between $z \sim$~14--9, the
merging galaxies are physically small and the interactions occur on
scales of tens of kiloparsecs. By $z \sim$~9-7.5, when the last major
mergers take place, the interactions have increased dramatically in
strength.  Galaxies are largely disrupted in close encounters, tidal
tails of gas and stars extend over hundreds of kiloparsecs, and
powerful bursts of star formation are triggered, resulting in an
average star formation rate of $\sim 10^3\, \Msun\, \yr^{-1}$ that
peaks at $z\sim 8.5$. During this phase, the black holes are heavily
obscured by circumnuclear gas. The luminosity from the starbursts
outshines that from the accreting black holes. So, we refer to this
period ($z \sim 14-7.5$) as the ``starburst phase.''

Once the progenitors have coalesced, the multiple SMBHs from the
galaxies merge and grow exponentially in mass and feedback energy via
gas accretion.  During this period ($z \sim 7.5-6$, hereafter referred
to as ``quasar phase''), the black hole luminosity outshines that of
the stars.  At redshift $z \approx 6.5$, when the galaxies coalesce,
the induced high central gas densities bring the SMBH accretion and
feedback to a climax. The black hole reaches a mass of
$\sim 2\times 10^{9}\, \Msun$, and has a peak bolometric luminosity
close to that of \zquasar.  Black hole feedback then drives a powerful
galactic wind that clears the obscuring material from the center of
the system.  The SMBH becomes visible briefly as an optically-luminous
quasar similar to \zquasar. Once the system relaxes, the SMBH and the host
satisfy the relation, $\MBH \approx 0.002\, \Mstar$, similar to that 
measured in nearby galaxies (\citealt{Magorrian1998, Marconi2003}), as a
result of co-eval evolution of both components \citep{Li2007,
Robertson2006A, Hopkins2007D}.  

After $z < 6$ (the ``post-quasar phase''), feedback from star
formation and the quasar quenches star formation and self-regulates
SMBH accretion.  Consequently, both star formation and quasar activity
decay, leaving behind a remnant which rapidly reddens. The object will
eventually evolve into a cD-like galaxy by the present day.  (For an
overview of this scenario, see e.g., \citealt{Hopkins2007C,
Hopkins2007B}.)

The photometric calculations by \cite{Robertson2007} show that this quasar
system satisfies a variety of photometric selection criteria based on
Lyman-break techniques. The massive stellar spheroid descended from these $z
\sim 6$ quasars could be detected at $z \sim 4$ by existing surveys,
while the galaxy progenitors at higher redshifts will likely require future
surveys of large portions of the sky ($ \gtrsim$ 0.5\%) at wavelengths
$\lambda \gtrsim 1\, \mu$m owing to their low number densities. 

\subsection{ART$^2$: All-wavelength Radiative Transfer with Adaptive
  Refinement Tree}  

ART$^2$ is based on a unification of the 3-D Monte Carlo radiative
equilibrium code developed by \cite{Bjorkman2001} and
\cite{Whitney2003A}, an adaptive grid, a multiphase ISM model, and a
supernova-origin dust model.  Below, we describe our implementation of
ART$^2$.

\subsubsection{Monte Carlo Radiative Transfer for Dust in Radiative Equilibrium}

\begin{figure}
\begin{center}
\includegraphics[width=3.5in]{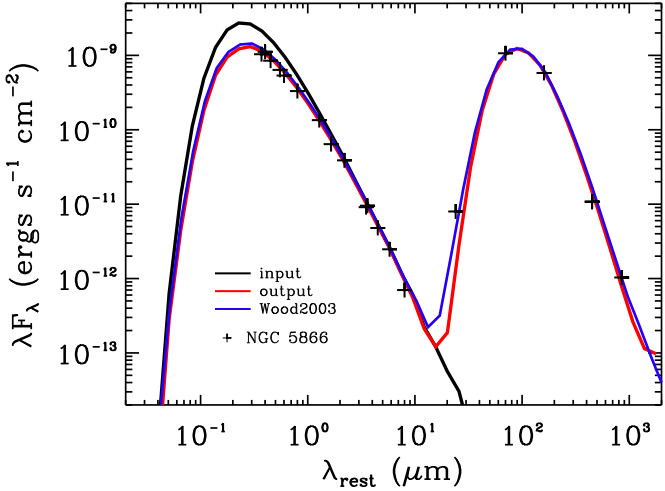}
\vspace{0.5cm}
\caption{Test run of the radiative equilibrium algorithm adopted from
  \cite{Bjorkman2001} on 3-D Cartesian coordinates for a simple galaxy with
  a bulge and a disk. The input spectrum of the stars is a blackbody
  with a temperature of 15000 K (black curve). The red curve is our output,
  the blue curve is simulation data from Kenny Wood, while the crosses are 
  observations of a lenticular galaxy NGC 5866, which has an unusual
  extended dust disk seen exactly edge-on. Both Wood's data and the
  observations come with the code release package from Kenny
  Wood (http://www-star.st-and.ac.uk/$\sim$kw25/research/montecarlo/gals/gals.html).} 
\label{Fig_sed_jaffe} 
\end{center}
\end{figure}

The Monte Carlo RT method follows the propagation, scattering,
absorption, and reemission of "photon packets" (groups of
equal-energy, monochromatic photons that travel in the same
direction), by randomly sampling the various probability distribution
functions that determine the optical depth, scattering angle, and
absorption rates.  For the dusty environments we are concerned with,
the re-emitted spectrum depends on the temperature of the dust, which
is assumed to be in thermal equilibrium with the radiation field.  In
the traditional scheme, the frequencies of the re-emitted photons are
sampled from local reemission spectra with fixed temperatures.  The
dust radiative equilibrium is ensured by performing the Monte Carlo
transfer iteratively until the dust temperature distribution
converges.  Such methods often require a large number of iterations
and are therefore computationally expensive \citep{Bjorkman2001,
Pascucci2004}.

\cite{Bjorkman2001} proposed a solution to this problem by formulating
an ``immediate reemission'' algorithm, in which the dust temperature
is immediately updated upon absorption of a photon packet, and the
frequencies of re-emitted photons are sampled from a spectrum that
takes into account the modified temperature.  The advantage of this
approach is that dust radiative equilibrium and the radiative transfer
solutions are obtained simultaneously without iteration. This
algorithm is described in detail in \cite{Bjorkman2001}; here we
briefly outline the steps.
 
Assuming that each photon packet carries an energy $E_\gamma$, and after
absorption of $N_i$ packets in the $i-th$ grid cell, the total energy absorbed
in the cell is
\begin{equation}
  E_i^{\rm abs} = N_i E_\gamma \; .
\label{eq:Eabs} 
\end{equation} 

In radiative equilibrium, this energy must be re-radiated, with a thermal
emissivity of $j_\nu = \kappa_\nu \rho B_\nu(T)$, where $\kappa_\nu$ is the
absorptive opacity, $\rho$ is the dust density, and $B_\nu(T)$ is the Planck
function at temperature $T$,  

\begin{equation}
 B_\nu(T)=\frac{2h_{\rm P}}{c^2}\frac{\nu^3}{e^{h_{\rm p}\nu/(kT)}-1}  \; ,
\label{eq:planck} 
\end{equation} 
where $h_{\rm P}$ is Planck's constant, $c$ the speed of light, and $k$ is
Boltzmann's constant. The emitted energy in the time interval $\Delta t$ is
\begin{eqnarray}
\label{eq:Eem} 
 E_i^{\rm em} &=& 4\pi\Delta t \int dV_i \int \rho \kappa_\nu B_\nu(T) \,d\nu \cr
              &=& 4\pi\Delta t \kappa_{\rm P}(T_i) B(T_i) m_i \;,
\end{eqnarray} 
where $\kappa_{\rm P}=\int\kappa_\nu B_\nu\,d\nu / B$ is the Planck mean
opacity, $B = \sigma T^4 / \pi$ is the frequency integrated Planck
function, and $m_i$ is the dust mass in the cell.

Equating the absorbed (\ref{eq:Eabs}) and emitted (\ref{eq:Eem})
energies, we obtain the dust temperature as follows after absorbing $N_i$ packets:
\begin{equation}
    \sigma T_i^4= { {N_i L }\over {4 N_\gamma \kappa_{\rm P}(T_i) m_i} } \;,
\label{eq:REtemp} 
\end{equation}
where $N_\gamma$ is the total number of photon packets in the simulation, and $L$
is the total source luminosity. Note that 
because the dust opacity is temperature-independent, the product 
$\kappa_{\rm P}(T_i) \sigma T_i^4$ increases monotonically with temperature.  
Consequently, $T_i$ always increases when the cell absorbs an additional 
packet.

The added energy to be radiated owing to the temperature increase $\Delta T$
is determined by a temperature-corrected emissivity $\Delta j_\nu$ in the
following approximation when the temperature increase, $\Delta T$, is small:
\begin{equation}
\Delta j_\nu \approx \kappa_\nu \rho \Delta T {{dB_\nu(T)} \over {dT}} \;.
\end{equation}
The re-emitted packets, which comprise the diffuse radiation field, then
continue to be scattered, absorbed, and re-emitted until they finally escape
from the system. This method conserves the total energy exactly, and does not
require any iteration as the emergent SED, $\nu L_\nu = \kappa_\nu B_\nu(T)$,
corresponds to the equilibrium temperature distribution \citep{Bjorkman2001}.  

This Monte Carlo radiative equilibrium code works as follows: first, the
photon packets are followed to random interaction locations, determined by the
optical depth. Then they are either scattered or absorbed with a probability
given by the albedo ($a = n_s \sigma_s/(n_s \sigma_s + n_a
\sigma_a)$, where $n$ and $\sigma$ are number density and cross section for
either scattering or absorption, respectively). If the packet is scattered, a
random scattering angle is obtained from the scattering phase function. If
instead the packet is absorbed, it is reemitted immediately at a new frequency
determined by the envelope temperature, using the algorithm described
above. After either scattering, or absorption plus reemission, the photon
packet continues to a new interaction location. This process is repeated until
all the packets escape the dusty environment. Upon completion of the Monte
Carlo transfer, the code produces emergent SEDs and images at any given
inclination for a wide range of broadband filters, including those of {\em
  Hubble Space Telescope} (NICMOS bands), {\em Spitzer} (IRAC and MIPS bands),
and {\em SCUBA} submillimeter bands.

\cite{Bjorkman2001} tested this algorithm extensively by comparing a
set of benchmark calculations to those of \cite{Ivezic1997} for
spherical geometry. \cite{Baes2005} critically study the frequency
distribution adjustment used by \cite{Bjorkman2001}, and give a firm
theoretical basis for their method, although it may fail for small dust
grains.    
 
One drawback of this code, however, is its fixed geometry and limited
dynamic range.  It can handle only 2-D or 3-D spherical-polar grids,
which are not suitable for capturing the arbitrary geometry and large
dynamic ranges characteristic of merging galaxies, which
have multiple density centers, and where gas and stars extend hundreds
of kpc and shocks produce highly condensed gas on scales of
pcs.  We have developed an improved version of this code by
implementing an adaptive Cartesian grid on top of the code released by
Wood~\footnote{http://www-star.st-and.ac.uk/$\sim$kw25/research/montecarlo/gals/gals.html}.

To ensure that our implementation of the algorithm is correct, we
rerun the test problem that comes with the code release package of
Wood using a uniform Cartesian grid.  The test problem consists of a
simple galaxy with bulge and disk components.  As is shown in
Figure~\ref{Fig_sed_jaffe}, our code reproduces the result of Wood
very well.

\subsubsection{Adaptive-mesh Refinement Grid}
\label{subsec_grid}

\begin{figure*}
\begin{center}
\includegraphics[width=3.0in]{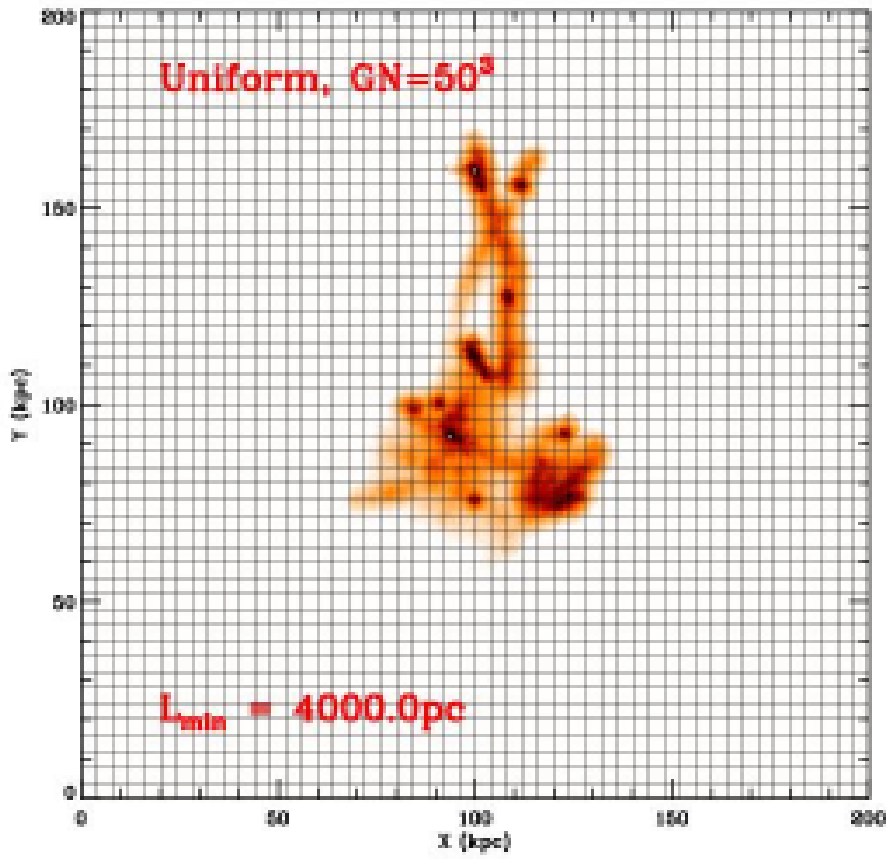}
\hspace{0.5cm}
\includegraphics[width=3.0in]{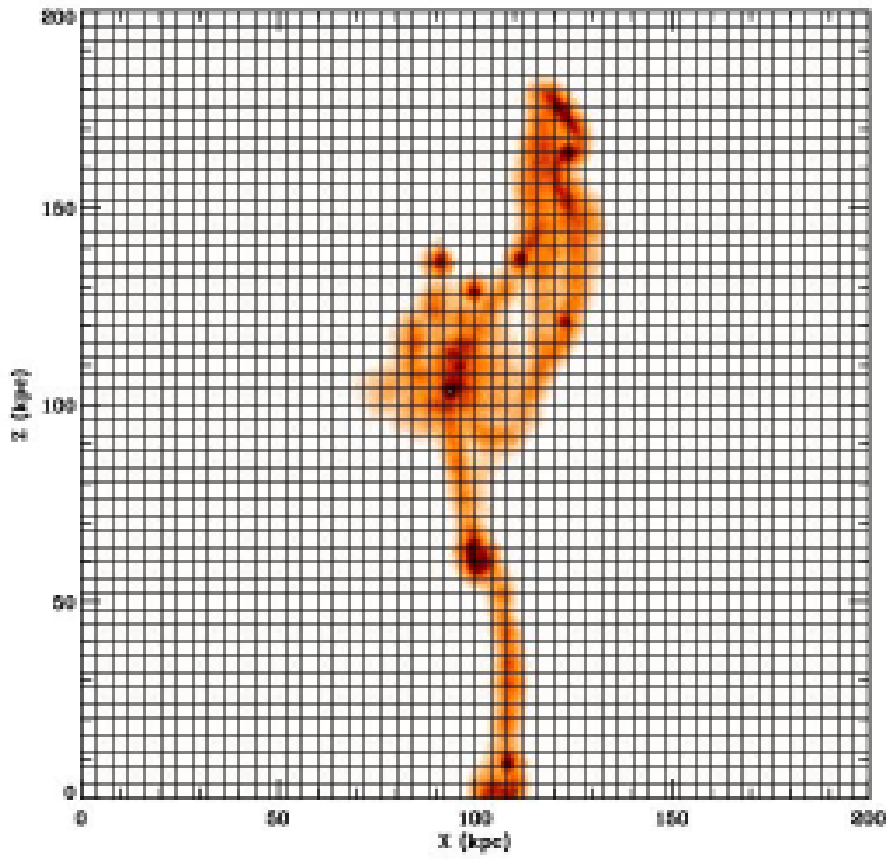} \\
\includegraphics[width=3.0in]{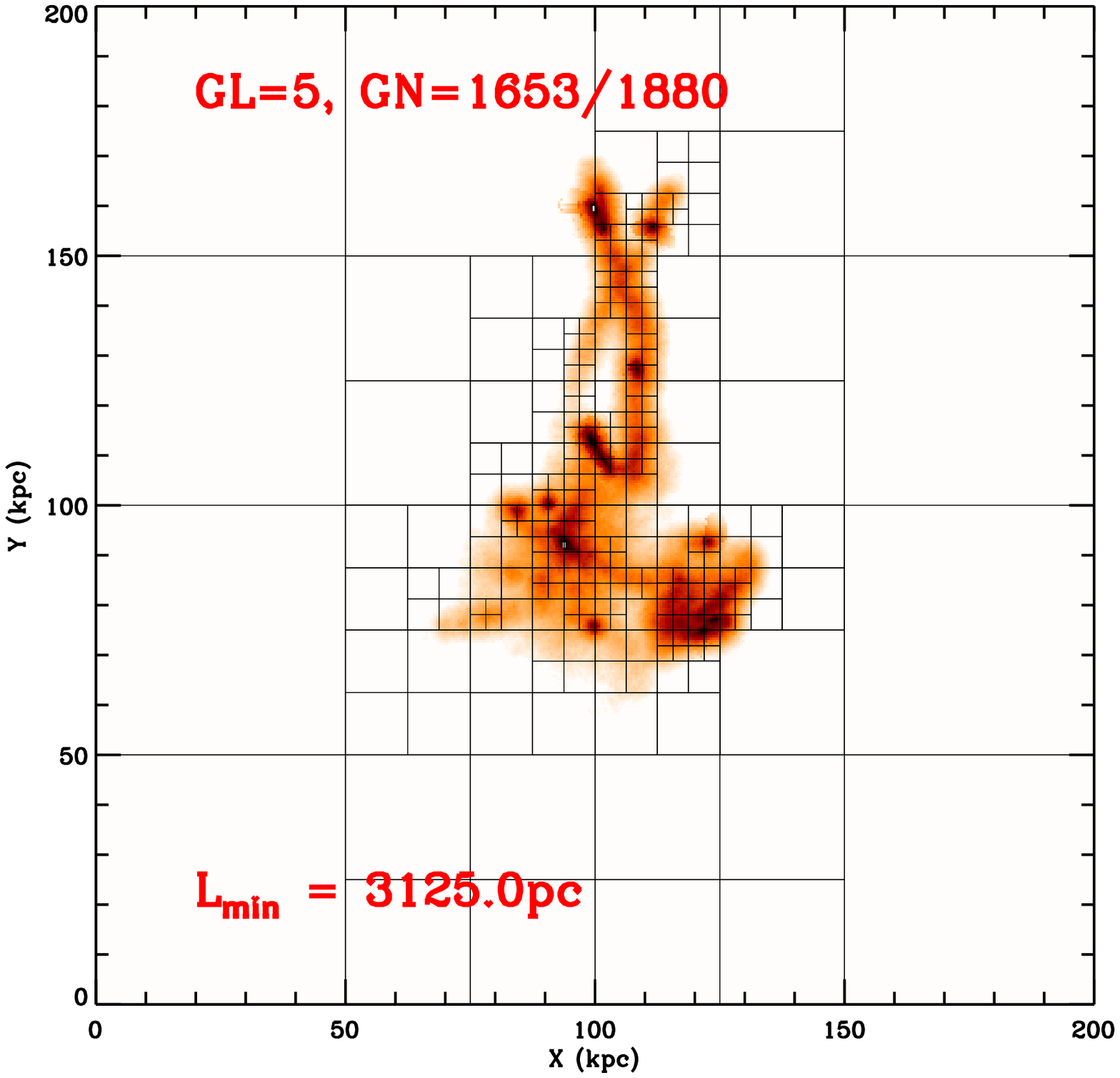} 
\hspace{0.5cm}
\includegraphics[width=3.0in]{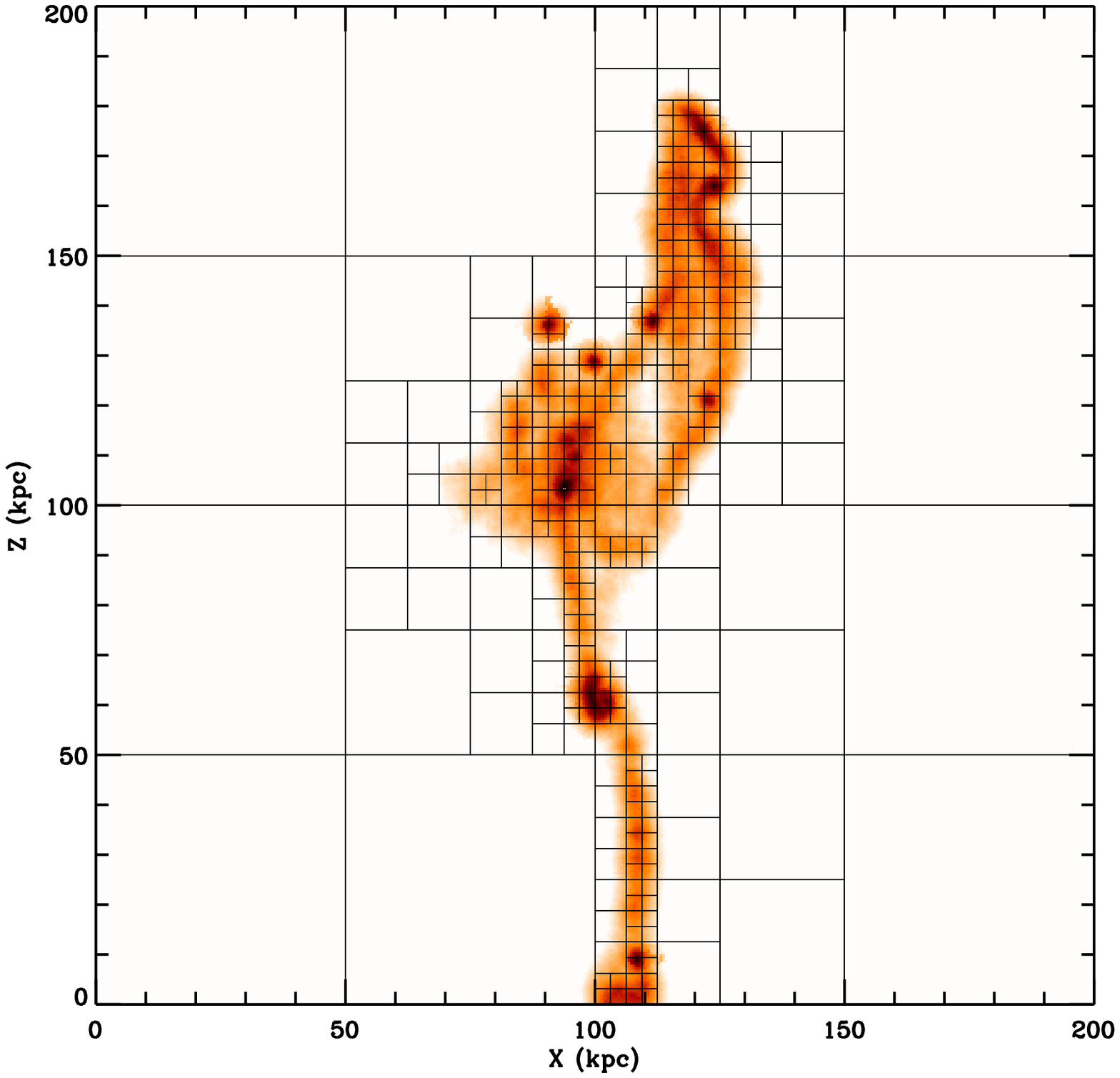} \\
\includegraphics[width=3.0in]{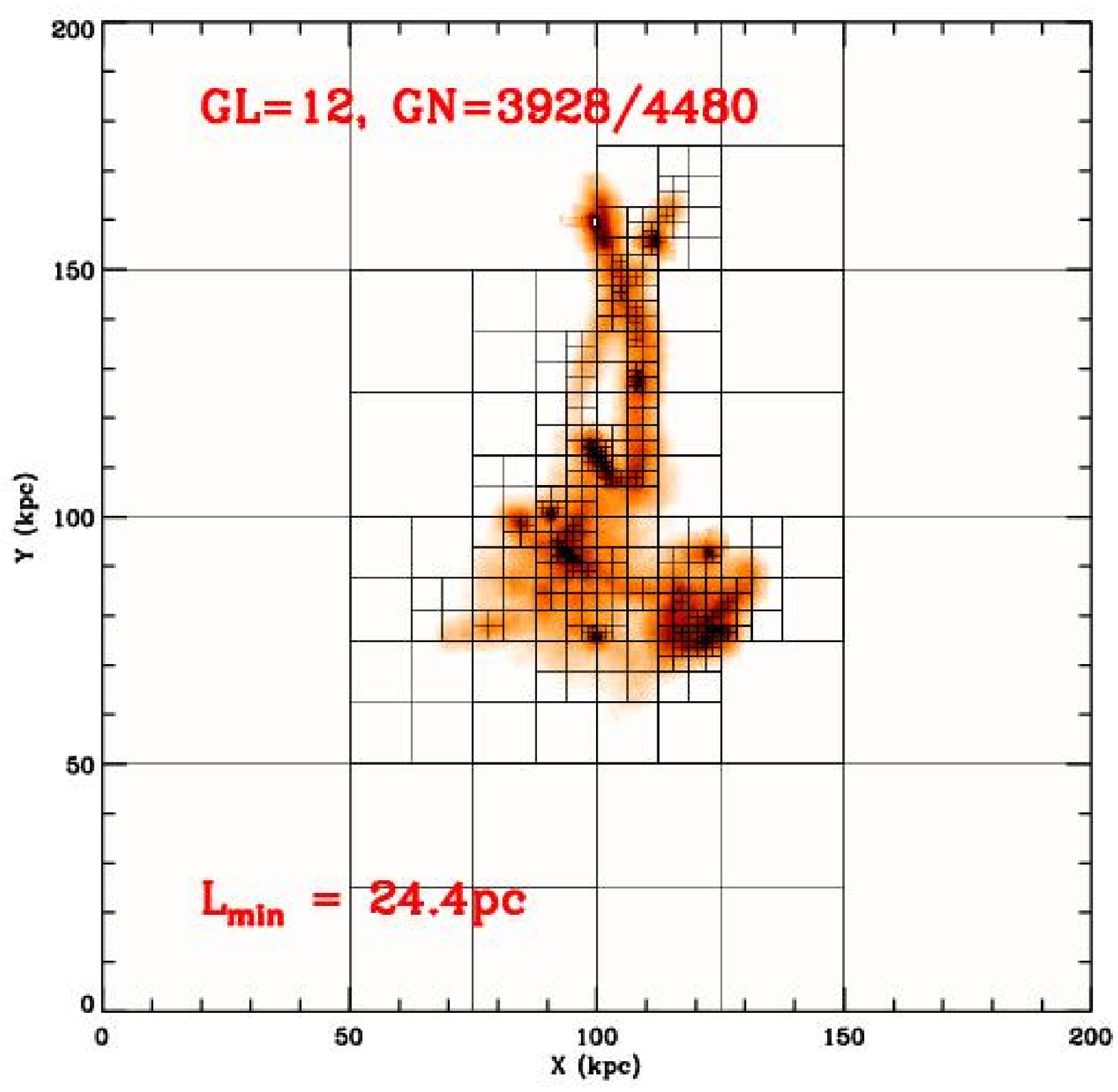} 
\hspace{0.5cm}
\includegraphics[width=3.0in]{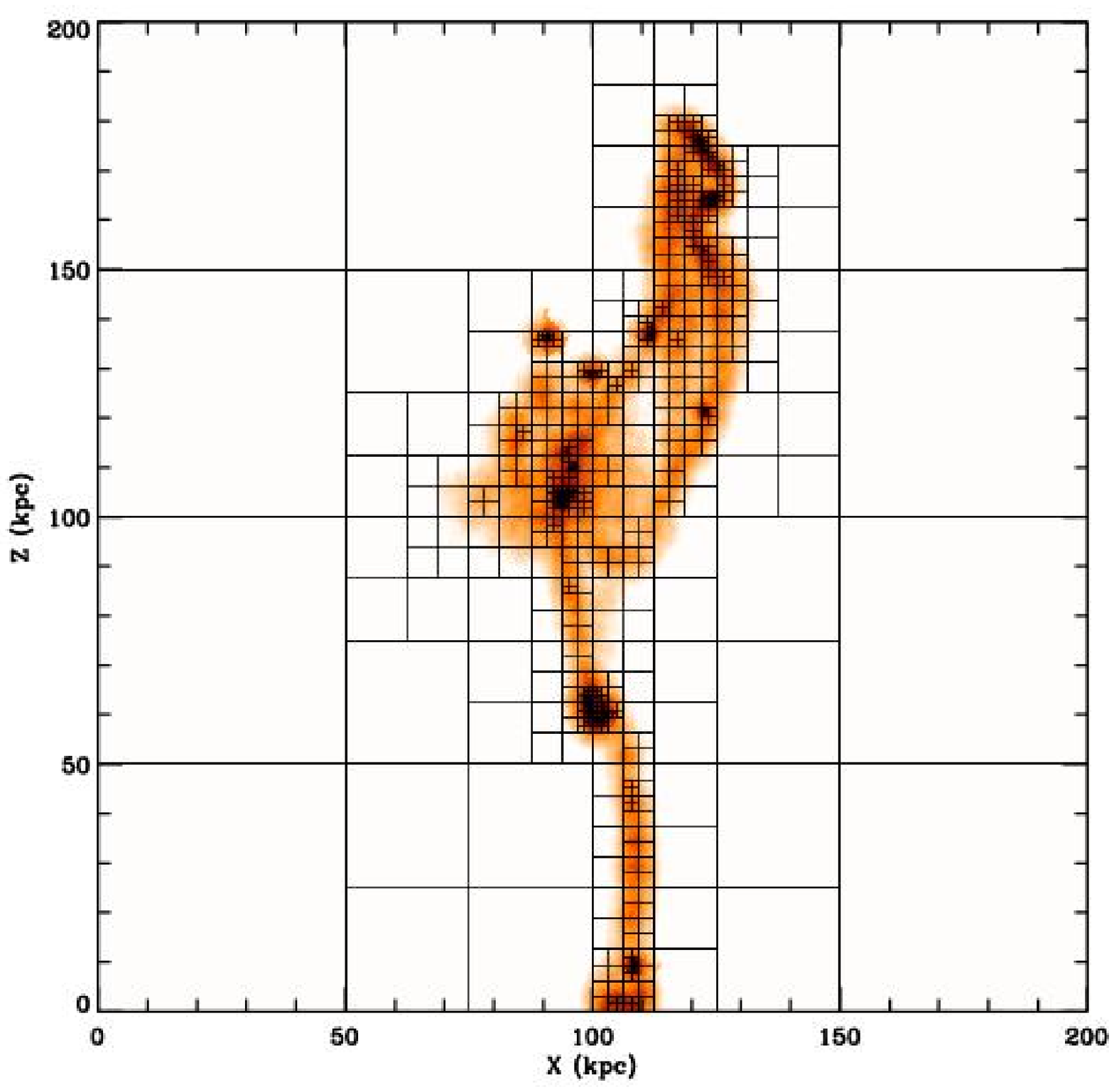} 
\vspace{0.5cm}
\caption{Example of the adaptive grid applied to a snapshot of the galaxy
  merger from \cite{Li2007}. From top to bottom is: uniform grid, and adaptive
  grids with maximum refinement levels of RL=5, and RL=12, respectively. The
  label GN indicates the total grid number covering the entire volume (in case
  of adaptive grids, it provides the number of the finest grid and that of the
  total grid). This plot demonstrates that adaptive grids with
  sufficient refinement capture the density distribution well, while the
  uniform grid fails to do so unless an unreasonably large number of cells is
  employed. Throughout this paper, we use a maximum refinement level
  of 12, which has a grid resolution comparable to that of the hydrodynamic
  simulations in \cite{Li2007}.}
\label{Fig_grid} 
\end{center}
\end{figure*}

The SPH simulations using GADGET2 \citep{Springel2005D} output
hydrodynamic information as particle data. However, the ray tracing in
the RT calculation is done on a grid. Therefore, it is necessary to
interpolate the particle-based density field onto a
grid. \cite{Jonsson2006} performed radiative transfer calculations on
SPH simulations of galaxy mergers using an adaptive grid as
implemented in his Monte Carlo RT code, SUNRISE.  Unfortunately,
self-consistent calculations of dust radiative equilibrium and emission
are not yet included in SUNRISE.

Our algorithm for constructing adaptive grids is similar to that of
\cite{Jonsson2006}. We typically start with a base grid of a $4^3$ box
covering the entire simulation volume. Each cell is then adaptively
refined by dividing it into $2^3$ sub-cells. The refinement is stopped
if a predefined maximum refinement level, RL, is reached, or if the
total number of particles in the cell becomes less than a certain
threshold, whichever criterion is satisfied first. The maximum refinement
level used in the present work is 12, and the maximum particle number allowed
in the cell is 32, half the number of the SPH smoothing kernel neighbors used
in the GADGET2 simulations. The resolution of the finest level is therefore
$L_{\rm min}=L_{\rm box}/2^{(\rm RL+1)}$, where $L_{\rm box}$ is the box length,
and ${\rm RL}$ is again the maximum refinement level. For example, for the
parameters used in this simulation, $L_{\rm box}=200$ kpc, ${\rm RL}=12$, we
have $L_{\rm min}= 24.4$ pc. 

The adaptive-mesh refinement grid serves as an efficient tree for mass
assignment owing to fast neighbor finding within the grids for SPH
smoothing. After the grid is constructed, the gas properties at the
center of each grid cell, such as density, temperature, and
metallicity, are calculated using the SPH smoothing kernel
\citep{Hernquist1989A} of the original simulation. All physical quantities  
are assumed to be uniform across a single cell. 

Figure~\ref{Fig_grid} gives an example of the adaptive grid applied to a
snapshot from the galaxy merger simulations in \cite{Li2007}, and compared
with a uniform grid.  This particular snapshot represents the time when the
system reaches the peak quasar phase, when the galaxies are in the final
stages of coalescence. The system is highly dynamical as much gas is falling
into the center, while feedback from the central massive black hole drives an
outflow. The gas distribution is thus inhomogeneous.

As is apparent in Figure~\ref{Fig_grid}, a uniform grid of $50^3$ (top
panel) barely captures the density distribution with a spatial
resolution of $4\, \kpc$. The resolution is worse than an adaptive grid with a 
moderate maximum refinement level (i.e., RL=5, middle panel), which has a
finest cell length of $\sim 3.1\, \kpc$. The grid with a maximum refinement
level of 12 shown in the bottom panel fully 
captures the large dynamic range of the gas density distribution in
three dimensions. It has a minimum cell length of $\sim 24.4$ pc for
the finest cells, which is comparable to the spatial resolution of the
original hydrodynamic simulation of \cite{Li2007}. This resolution is
equivalent to a ${\sim 10000 }^3$ uniform grid, which is impractical
with existing computational facilities.

\subsubsection{A Multiphase Model for the Interstellar Medium}
\label{subsec_ism}

In determining the dust distribution, we adopt the multiphase model of
\cite{Springel2003A} for the ISM.  The ISM is then comprised of
condensed clouds in pressure equilibrium with an ambient hot gas, as
in the picture of \cite{McKee1977}.  In the hydrodynamic simulations,
individual SPH particles represent regions of gas that contain cold,
dense cores embedded in a hot, diffuse medium. The hot and cold phases
of the ISM co-exist in pressure equilibrium but have different mass
fractions and volume filling factors (i.e., the hot-phase gas is $\le
10\%$ in mass but $\gtrsim 99\%$ in volume).  In our previous studies,
\cite{Li2007} and \cite{Hopkins2006A} used only the hot-phase density
to determine e.g. the obscuration of the central AGN, as the majority
of sight lines will pass through only this component owing to its
large volume filling factor.  However, this method gives only an
effective lower limit on the column density.

Here, we consider two components of the dust distribution, having
different dust-to-gas ratios and being associated with the two phases
of the ISM.  Within each grid cell in the RT calculation, the hot gas
is uniformly distributed, while the cold, dense cores are randomly
embedded.  Because of the much higher density and smaller volume of
the cold phase, it is impractical to resolve these clouds either in
hydrodynamic simulations or in our radiative transfer calculations.
Consequently, we implement an observationally-motivated,
sub-resolution prescription to treat the cold clouds, constrained by
the observed mass spectrum and size distribution of molecular clouds
in galaxies.

Observations of giant molecular clouds (GMCs) in galaxies show that
the GMCs follow simple scaling relations \citep{Larson1981},
namely a power-law mass distribution, ${\rm d}N/{\rm d}M \propto
M^{-2}$ (e.g., \citealt{Fuller1992, Ward-Thompson1994, Andre1996,
Blitz2006}), as well as a power-law mass-radius relation $M \propto
R^2$ (.e.g, \citealt{Sanders1985, Dame1986, Scoville1987, Solomon1987,
Rosolowsky2005, Rosolowsky2007}). It has been suggested by theoretical
modeling that the mass function is produced by turbulence in
self-gravitating clouds (e.g., \citealt{Elmegreen1996, Elmegreen2002,
Ballesteros-Paredes2002, Li2003}), while the mass-radius relation is
attributed to virial equilibrium \citep{Larson1981}.

Here we incorporate these two empirical relations into our ISM model
for the RT calculations. For a given cell, the two-phase break-down in
\cite{Springel2003A} determines the hot and cold phase gas density
according to pressure equilibrium. The cold clouds are assumed to
follow the Larson scaling relations:

\begin{eqnarray}
\label{eq:ms}
\frac{dn}{dM} &=& A M^{-\alpha}, \\
M &=& BR^{\beta},
\end{eqnarray}
where $dn/dM$ is the number density of the clouds differential in cloud
mass $M$, and $R$ is the cloud radius. From these equations, one
obtains the cloud size distribution
\begin{eqnarray}
\frac{dn}{dR} &=& \beta A B^{1-\alpha} R^{-(\alpha\beta+1-\beta)} \nonumber\\
&=& C R^{-\gamma},
\end{eqnarray}
where $C = \beta A B^{1-\alpha}$, and $\gamma = \alpha\beta+1-\beta$.

Assume that the minimum and maximum values of the cloud mass are $M_0$ and $M_1$,
then the normalization constant of the mass spectrum for each cell can be
determined using
\begin{equation}
\int \frac{dn}{dM}MdM = x_c\rho_c,
\end{equation}
or
\begin{equation}
A = x_c\rho_c\frac{2-\alpha}{M_1^{2-\alpha} - M_0^{2-\alpha}} \, ,
\end{equation}
where $\rho_c$ and $x_c$ are the cold gas density and volume filling factor,
respectively. The normalization constant of the $M$-$R$ relation may be
determined using 
\begin{equation}
x_c = \int \frac{4\pi}{3}R^3\frac{dn}{dM}dM,
\end{equation}
or 
\begin{equation}
B = \left[\frac{4\pi A}{3\eta x_c}\left(M_1^\eta -
  M_0^\eta\right)\right]^{\beta/3},
\end{equation}
where 
\begin{equation}
\eta = 1 + \frac{3}{\beta} - \alpha.
\end{equation}
The minimum and maximum cloud radii in the cell are therefore
\begin{eqnarray}
R_0 &=& \left(\frac{M_0}{B}\right)^{1/\beta} \nonumber\\
R_1 &=& \left(\frac{M_1}{B}\right)^{1/\beta}.
\end{eqnarray}

For a photon traveling a distance $L$ in the cell, the average number of
cold clouds of radius $R$ the photon will intersect is given by
\begin{eqnarray}
\label{eq:NR}
\frac{dN}{dR} &=& \pi R^2 L \frac{dn}{dR} \nonumber\\
&=& \pi C R^{2-\gamma}.
\end{eqnarray}
Integrating over the cloud radius, one obtains
\begin{equation}
N = \pi LC \frac{R_1^{3-\gamma} - R_0^{3-\gamma}}{3-\gamma}.
\end{equation}
Therefore, the average distance the photon must travel to hit a cold cloud
(the mean free path) is
given by
\begin{equation}
L_{m} = \frac{3-\gamma}{\pi
  C\left(R_1^{3-\gamma}-R_0^{3-\gamma}\right)}.
\end{equation}

In the dust RT calculation, we assume that dust is associated
with both the cold and hot phase gases through certain dust-to-gas ratios. When a
photon enters a cell, we first determine the distance $L_h$ it travels in the
hot phase gas before hitting a cold cloud, which is an exponential
distribution function 
\begin{equation}
p(L) = \frac{1}{L_{m}}\exp{\left(-\frac{L}{L_{m}}\right)}.
\end{equation}
Therefore, $L_h=-L_{m}\ln \xi$, where $\xi$ is a random number
uniformly distributed between 0 and 1. The radius of the cloud the photon just
hits is also determined randomly assuming the distribution function of
Equation~(\ref{eq:NR}); i.e.,
\begin{equation}
R = \left[R_0^{3-\gamma} +
  (R_1^{3-\gamma}-R_0^{3-\gamma})\xi\right]^{1/(3-\gamma)}. 
\end{equation}

The distance $L_c$ the photon travels in this cold cloud is again a random
variable given by $L_c=2R\sqrt{\xi}$, because clouds are assumed to be
uniformly distributed. These equations completely define the statistical
procedure for determining the dust column densities associated 
with $L_h$ and $L_c$ as:
\begin{eqnarray}
\label{eq:NhNc}
N_h &=& \rho_h L_h \nonumber\\
N_c &=& \frac{3BR^{\beta-3}L_c}{4\pi}.
\end{eqnarray}

With a given dust opacity curve, these equations allow one to calculate the
optical depths, $\tau_h$ and $\tau_c$ for the hot and cold dust,
respectively. They relate the photon path lengths in the multiphase ISM to the
total optical depth $\tau_{\rm tot}$, which is then compared with a randomly
drawn number $\tau_i$, to determine whether the photon should be stopped for
scattering or absorption. In detail, the Monte Carlo ray-tracing
procedure for the radiative transfer therefore involves the following steps:

\begin{enumerate}

\item
A photon packet is emitted from either a blackhole or a stellar source with
random frequencies consistent with the source spectra. The photon is emitted
with a uniformly distributed random direction. The probability of a photon
being emitted by any given source is determined by its luminosity relative to
the total. 

\item
A random optical depth over which the photon must travel before an interaction
with the dust occurs, $\tau_i=-\ln\xi$, is drawn to determine the interaction
location. The interaction includes scattering and absorption. In our method,
the photon energies are not weighted, only one event is allowed. That
is, at any given interaction site, the photon is either scattered or absorbed,
but not both.   

\item
Starting from the location of the photon emission, the cumulative optical
depth of the photon, $\tau_{\rm tot}$, is calculated stochastically using
Equation \ref{eq:NhNc} for both hot and cold dusts. If the photon is stopped
for interaction within a single cell, then $\tau_{\rm tot}$ is the sum of
contributions from possibly multiple segments of both hot and cold dusts
within this cell. If the photon passes through multiple cells before an
interaction occurs, then $\tau_{\rm tot}$ is the sum of all contributions from 
relevant segments in these cells.  

\item
At each boundary between the hot and cold phase gas clouds, or at the boundary
of the grid cell, the next interaction point is determined by the comparison
between $\tau_i$ and $\tau_{\rm tot}$. If $\tau_i \le \tau_{\rm tot}$, then
the photon is either scattered or absorbed, with a probability given by the
scattering albedo. The exact interaction location is then determined inside
either hot or cold phase gas, such that $\tau_{\rm tot}$ becomes exactly
$\tau_i$. If the photon is scattered, its direction and polarization state are
altered using the Henyey-Greenstein phase function, and the ray-tracing of the
new photon is repeated from step 2. If the photon is absorbed, depending on
whether the absorption site occurs in the hot or cold phase dust, the
temperature of the appropriate dust cell is raised, and a new photon is
reemitted according to the scheme of \citet{Bjorkman2001}. The ray-tracing of
the newly emitted photon again restarts from step 2. 

\item 
If the photon escapes from the system without reaching the optical depth
$\tau_i$, it is then collected in the output spectrum and image. The next
photon will be picked up from the source, and the whole Monte Carlo procedure
from step 1 will be restarted. 
\end{enumerate}

After all $N_{\gamma}$ photons have been traced, one obtains the
dust temperature distribution for both hot and cold dusts in radiative
equilibrium, as well as the output spectra and images. Note that such a 
stochastic procedure outlined here may not be as accurate as physically
tracking the locations and sizes of the cold phase gas clouds, which can be
rather difficult, if not impossible. However, this method works
efficiently for a large number of cold clouds uniformly distributed in the
cells, and it gives correct average extinction properties for a multiphase
ISM model we are interested here.

Hereafter, we refer to ``HPG-dust'' as the dust that originates from
the hot-phase gas, and ``CPG-dust'' as that originating from the 
cold-phase gas. Note that the gas only determines the distribution and mass
of the dust through a given dust-to-gas ratio. The dust temperature is
not associated with the gas temperature, and is calculated
self-consistently according to radiative equilibrium.  We further
refer to ``cold dust'', ``warm dust'' and ``hot dust'' as dust with
temperatures $T \lesssim 100$~K, $100 \lesssim T \lesssim 1000$~K, and $ 1000
\lesssim T \lesssim 1200$~K, respectively.
 
In the RT calculations, we adopt a mass spectrum with $\alpha=1.8$, as
suggested by the observations of \cite{Blitz2006}, and an observed
mass-radius relation with $\beta=2.0$, which is also a result of the
virial theorem. The resulting mass-radius relation in our simulations
has a normalization in the range $\sim 10 - 10^4\, \Msun/{\rm
pc^2}$. For a cloud size of $\sim 1\, {\rm pc}$, the normalization is
$\sim 300\, \Msun/{\rm pc^2}$, similar to observations of the Milky Way
\citep{Scoville1987, Rosolowsky2007}. It has been shown that the
normalization of the mass-radius relationship depends on galactic
environment (e.g., \citealt{Elmegreen1989, Rosolowsky2005,
Blitz2007}).  For example, it is about $\sim 50\, \Msun/{\rm pc^2}$ in the
Large Magellanic Cloud but could be two orders of magnitude higher in
ULIRGs. In extreme starburst galaxies, the normalization could go up
to $10^4\, \Msun/{\rm pc^2}$, as we find in our simulations. 

We assume the cold clouds to be in the mass range of $M_0=10^3\,
M_\odot$ and $M_1=10^7\, M_\odot$, which is similar to that of
protoclusters clouds in star forming galaxies, as shown in
\cite{Li2004, Li2005A, Li2005B, Li2006}. For the cold clouds, we have
enhanced the pressure by a factor of 10 over the thermal pressure to
account for the effects of turbulence (e.g., \citealt{Blitz2006,
Chakrabarti2007A}). This enhancement factor is referred to as ``CP''
hereafter. The dust-to-gas ratio of the cold clouds is chosen
to be the same as the Milky Way value (1:124; \citealt{Weingartner2001}), as
found in a large sample of ULIRGs \citep{Dunne2001, Klaas2001}, while that of
the hot, diffuse gas is chosen to be 1\% of the Milky Way value, consistent
with the dust survival rate of sputtering in a hot, diffuse ISM
\citep{Burke1974, Reynolds1997}. Observations of some obscurred or red AGNs
also suggest a ratio significantly lower than that of Milky Way (e.g.,
\citealt{Maiolino2001, Kuraszkiewicz2003, Hall2006}). We will perform a
systematic parameter study of these choices in \S~\ref{sec_param}. However, we
note that these values are found to best reproduce the observed quasar SED at
$z\sim 6.5$. Because the dust opacity is proportional to the gas metallicity,
therefore the dust opacity is weighted by the metallicity of the gas for both
hot and cold gas phases.

\subsubsection{Supernova-origin Dust Model}
\label{subsec_sndust}

\begin{figure}
\begin{center}
\includegraphics[width=3.5in]{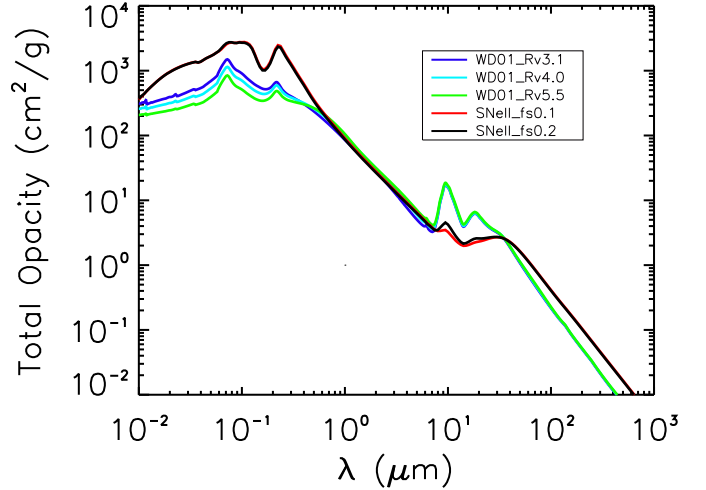} 
\vspace{0.5cm}
\caption{
Comparison of the dust absorption opacity curves from a supernova-origin dust
model (SN model, with silicate fraction $f_{\rm s}=10\%$, 20\% in red and
black, respectively) and those of \cite{Weingartner2001} (WD01 model,
with $R_{\rm V}=3.1$, 4.0, 5.5, in blue, cyan, and green, respectively). Note
there are two differences between these two models: the silicate feature at $\sim
9.7\, \mu$m, and a higher opacity in the optical band.}   
\label{Fig_opac}
\end{center}
\end{figure}

\begin{figure}
\begin{center}
\includegraphics[width=3.5in]{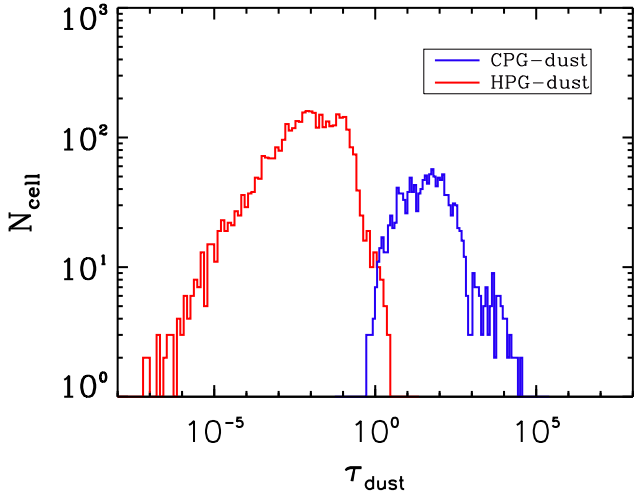}
\vspace{0.5cm}
\caption{Distribution of the optical depths of both HPG- and CPG-dust,
  respectively, at $\lambda=0.1\, \mu$m where the dust opacity peaks as shown
  in Figure~\ref{Fig_opac}. The density grid corresponds to the simulation
  snapshot at $z=6.5$ shown in Figure~\ref{Fig_grid} (bottom panel). The
  ``HPG-dust'' and ``CPG-dust'' refer to the dust associated with hot-phase and
  cold-phase gas, respectively, as described in \S~\ref{subsec_ism}. }    
\label{Fig_tau} 
\end{center}
\end{figure}

Our understanding of the formation and distribution of dust has benefited
from dust maps of the  Milky Way (e.g., \citealt{Gehrz1989, Schlegel1998}). In
particular, \cite{Gehrz1989} concludes from a detailed Galactic 
survey of dust-producing stars that dust in the Milky Way originates in three
principle ways: (1) by condensation in winds of evolved,
post-main-sequence objects, which accounts of $\sim 90\%$ of the stellar dust;  
(2) by condensation in ejecta from massive novae, supernovae and Wolf-Rayet
stars, which amounts to $< 10\%$; and (3) by slow accretion in molecular 
clouds (see also \citealt{Marchenko2006} for a review). 

Over the past several decades, various dust models have been developed
based on the observed extinction curves of the Large Magellanic Cloud
(LMC), the Small Magellanic Cloud (SMC) and the Milky Way (e.g., see
reviews by \citealt{Savage1979, Mathis1990, Calzetti1994,
Dorschner1995, Calzetti2000, Whittet2003, Draine2003}). In these
classical models, dust is assumed to form in the envelopes of
old, low-mass stars such as asymptotic giant branch (AGB) stars
with ages $\gtrsim 1$ Gyrs \citep{Mathis1990, Morgan2003A, Dwek2005,
Marchenko2006}.

However, this picture may be different for young, high-redshift
objects. Recent deep millimeter and submillimeter observations of
several $z \sim 6$ quasars, which trace the far-infrared thermal dust
emission from these systems, show large masses of dust in these quasar
hosts when the Universe was less than 1 Gyr old (e.g.,
\citealt{Bertoldi2003A, Charmandaris2004, Robson2004, Carilli2004,
Maiolino2004, Beelen2006, Hines2006, Jiang2006, Willott2007}).
In particular, 
\cite{Maiolino2004} find that the extinction curve of the reddened
quasar SDSS J1048+46 at $z = 6.2$ is different from those observed at
$z < 4$ (similar to that of the Small Magellanic Cloud,
\citealt{Hopkins2004}), but matches the extinction curve expected for
dust produced by supernovae. 

It has been suggested that core-collapse supernovae (Type-II SNe) may
provide a fast and efficient mechanism for dust production. The
observational evidence for dust formation in SNe comes from
observations of SN1987A (e.g., \citealt{Gehrz1987, Moseley1989,
Roche1993, Spyromilio1993, Colgan1994}), Cassiopeia A
(\citealt{Dunne2003, Dwek2004}; see however \citealt{Krause2004} who
argue that the dust emission is not associated with the remnant), Kepler's
supernova remnant \citep{Morgan2003B}, and SNe 2003gd
\citep{Sugerman2006}. Theoretically, several groups have calculated
dust formation in the ejecta of Type-II SNe \citep{Todini2001,
Nozawa2003, Schneider2004, Hirashita2005, Dwek2007}. In particular,
\cite{Todini2001} have developed a dust model based on standard nucleation
theory and tested it on the well-studied case of SN1987A. They find that SNe
with masses in the range of $12 - 35\, \Msun $ produce about 1\% of the mass
in dust per supernova for primordial metallicity, and $\sim 3\%$ for
solar metallicity.

In the present work, we adopt the dust size distribution of
\cite{Todini2001} for solar metallicity and a $M=22\, \Msun$ SN model, as in
Figure~5 in their paper. This size distribution is then combined with the dust
absorption and scattering cross sections of \cite{Weingartner2001}, to
calculate dust absorption opacity curves \citep{Finkbeiner1999A, Finkbeiner1999B}. 
We note that \cite{Bianchi2007} re-analysize the model of \cite{Todini2001} and 
follow the evolution of newly condensed grains from the time of formation to
their survival. This new feature shows better agreement with
observations and further supports our motivation to use a SN dust model. 
Figure~\ref{Fig_opac} shows dust absorption
opacity curves, in the range $10^{-2} - 10^4\, \mu$m, from both
the supernova-origin dust model (hereafter SN model) and
\cite{Weingartner2001} (hereafter WD model), respectively. The SN
models include silicate fractions $f_{\rm s}=10\%$ (in red) and $f_{\rm
s}=20\%$ (in black), which show slight differences in the silicate
feature at $\sim 9.7\, \mu$m ($f_{\rm s}=20\%$ model has a slightly
higher opacity). The WD models include all the curves commonly used
with extinction $R_{\rm V}=\frac{A(V)}{A(B)-A(V)}=3.1$, 4.0, 5.5,
which differ in the opacity in the UV-NIR ($0.01\, \mu \rm{m} - 1\,
\mu \rm{m}$) bands ($R_{\rm V}=3.1$ model has the highest
opacity).

Between the SN and WD models, there are two noticeable differences; one is 
the silicate feature at $\sim 9.7\, \mu$m, and the second is the UV-NIR
band. On one hand, the WD model has strong peaks around $\sim 9.7\, \mu$m
(silicate feature), while the SN model produces an opacity lower by
nearly one order of magnitude, because the ejecta of SNe with mass
above $15\, \Msun$ predominantly form amorphous carbon grains, which
decreases the silicate fraction in the dust grains. On the other hand,
the SN model increases the opacity in the optical band by a factor of
a few owing to the smaller grain size. We note that the 
small grain sizes may cause
quantum fluctuations 
in the temperature and high-temperature transients (e.g.,
\citealt{Draine2001}). However, the dust produced by the SN model is dominated  
by graphite grains, which have a size distribution from 
$\sim 100 - 2000$~\AA, comparable to the size range of the dust grains in the
WD model, so the temperature fluctuation is expected to be insignificant.   
Note we do not include polycyclic aromatic hydrocarbon (PAH) features at $\sim
7.7\, \mu$m, which was modeled in detail by \cite{Li2002} and
\cite{Draine2007}. The PAH feature is a good diagnostic of starbursts 
at low redshifts \citep{Houck2005}, however it is very difficult to
detect in $z \gtrsim 6$ systems. Also dust spin is not included in our
modeling, which could be significant around $10$~mm
\citep{Draine1998, Finkbeiner2002}.
 
The differences between these two models, in particular the silicate
feature, may be diminished by choosing different grain compositions
and size distributions \citep{Laor1993}. For example, the WD model
would converge to the SN model if the silicate fraction is reduced to
$\sim 5\%$.  We note that \cite{Elvis2002} propose an alternative
mechanism for dust production in quasars. They suggest that the
physical conditions (i.e., low temperature and high density) in the
quasar outflow are similar to those in the envelopes of AGB stars, and
hence may produce dust similar to that made by AGB stars. This
scenario does not have the same timescale issue as does the AGB model, but
only requires heavy metals such as carbon and silicate, which should
be available even in high-redshift quasars, as noted in the
introduction. Therefore, quasar winds may also serve as efficient
factories for producing dust with extinction curves similar to those
of the WD model. Currently there are no observations available to
distinguish between these models at high redshifts. We will return to
this in \S~\ref{subsec_dustcomp}.
 
Figure~\ref{Fig_tau} shows an example of the resulting optical depths
of both HPG- and CPG-dust at wavelength $\lambda=0.1\, \mu$m where the
dust opacity reaches its maximum, as shown in Figure~\ref{Fig_opac}. This plot
demonstrates that the optical depth is resolved very well in the hot, diffuse
gas which occupies $\sim 99\%$ of the volume, while the optical depth of the
cold, dense cores is usually much larger than unity, in particular in optical
bands with high frequencies.  The energy absorbed by the cold dust in these
wavelengths is then re-emitted at infrared or longer wavelengths.

\section{The Spectral Energy Distribution of A Quasar System at $z \simeq 6.5$}
\label{sec_sed}

\begin{figure*}
\begin{center}
\includegraphics[width=5.0in]{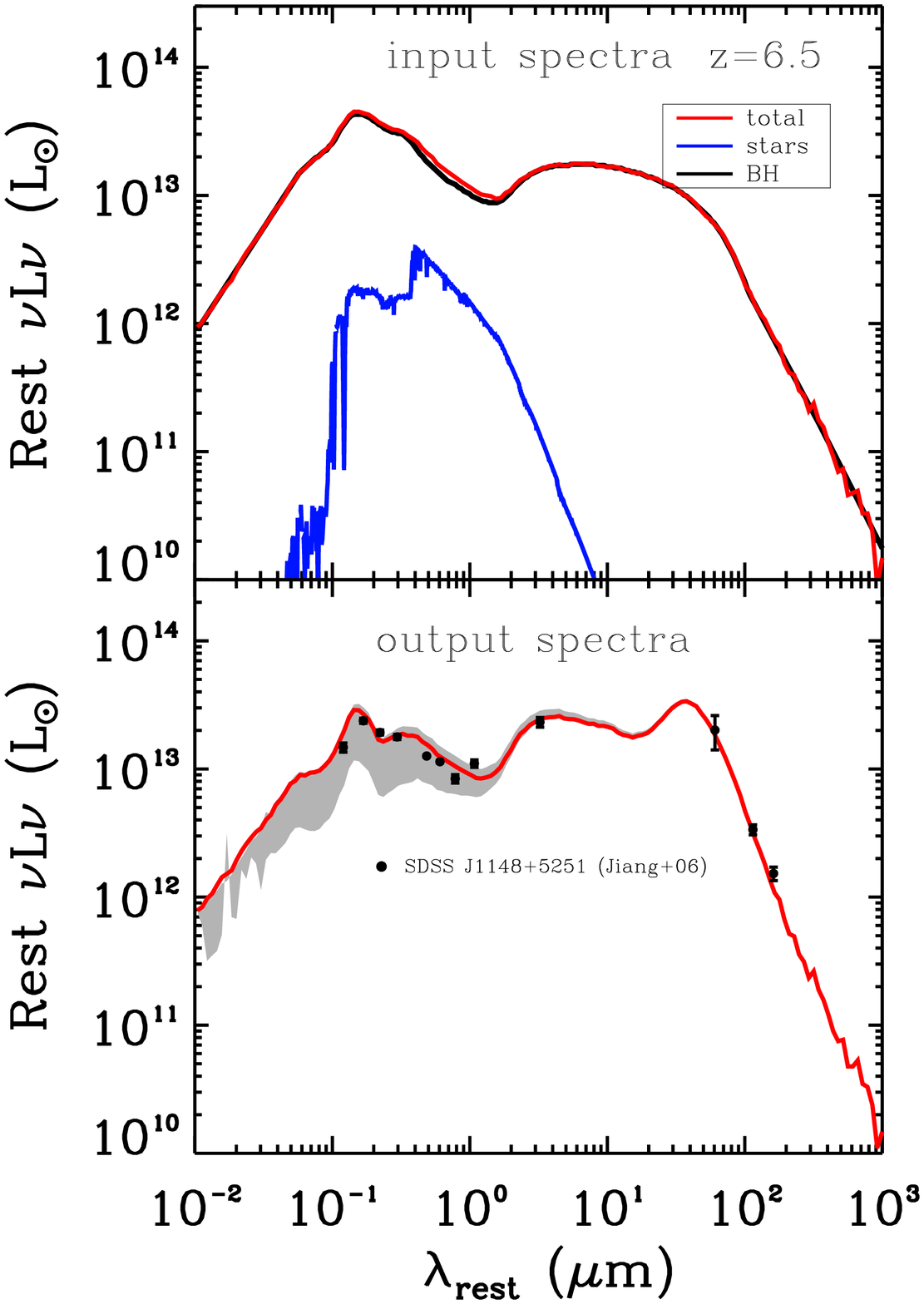}
\vspace{1.5cm}
\caption{
{\small SEDs of a quasar system at $z \simeq 6.5$ determined by applying ART$^2$ to
the SPH simulations of \cite{Li2007}. The input spectra (top
panel) include the stellar spectrum (in blue) calculated with STARBURST99
\citep{Leitherer1999, Vazquez2005}, and the composite spectrum of the AGN (in
black, \citealt{Hopkins2007A}).  Note this snapshot corresponds to the peak
quasar phase of the system, the stellar radiation is insignificant compared to
that from the AGN at this evolution stage.   
The red curve is the total input spectrum. The
output spectra are shown in the bottom panel. The red curve is
the total SED assuming an isotropic distribution of photon energies in all
directions, while the grey region indicates the range from two orthogonal
angles corresponding to the z-axis and xy-plane of the Cartesian grid,
respectively.  The filled black circles with error bars are observations from
\cite{Jiang2006}. } }        
\label{Fig_sed_z6}
\end{center}
\end{figure*}

We now calculate the spectral energy distribution from UV/optical to
submillimeter of the modeled quasar in \cite{Li2007} by applying
ART$^2$ to our SPH simulations, which provide the gas density field
and heating sources for the RT calculations. The input spectrum
includes that from stars and black holes, as shown in
Figure~\ref{Fig_sed_z6} (top panel). The stellar spectrum is
calculated using the stellar population synthesis code STARBURST99
\citep{Leitherer1999, Vazquez2005}. The age, mass and metallicity of
the stars are taken from the SPH simulations of quasar formation in
\cite{Li2007}, while the stellar initial mass function (IMF) is
assumed to be a top-heavy Kroupa IMF \citep{Kroupa2002}, which
characterizes a starburst better than the classical Salpeter IMF
\citep{Salpeter1955}.

The input black hole spectrum is a composite template from
\cite{Hopkins2007A}, which consists of a broken power-law (e.g.,
\citealt{Laor1993, Marconi2004}) and an IR component thought
to come from the hottest dust around the AGN, which cannot 
be fully resolved in our hydrodynamic simulations. The normalization
of this spectrum is the total bolometric luminosity of the black
holes. Compared to the luminosity calculation in \cite{Li2007}, the
black hole spectrum here is normalized to 
the bolometric luminosity of \zquasar. Heating by cosmic microwave
background radiation at different redshifts $z$ is also taken into
account by including a uniform radiation field with temperature of
$T_{\rm cmb}=2.73\times(1+z)$ K.

The emergent spectrum has a wavelength range of $10^2 - 2\times 10^7
\AA\ $ in the rest frame, and 50 viewing angles (10 in polar angle evenly
divided in ${\rm cos}\theta$ and 5 in azimuthal $\phi$). 
We use $10^7$ photon packets isotropically emitted from sources, and the
maximum refinement level of the adaptive grid is RL=12, which is
above the requirement for convergence (see \S~\ref{subsec_res} for
resolution studies).

The calculated rest-frame SEDs of the system during the peak quasar phase at
$z=6.5$ are shown in Figure~\ref{Fig_sed_z6} (bottom panel). The output SED
depends on the viewing angle. The red curve is the isotropically averaged SED
(e.g., it can be understood as an average SED multiplied by the solid angle
$4\pi$). This averaged SED agrees very well with observations of \zquasar\
\citep{Jiang2006}. A prominent feature of this SED is the two infrared bumps
peaking around $3\, \mu$m and $50\, \mu$m. The $50\, \mu$m bump is produced by
cold dust ($T \sim 50$ K) heated by strong star formation, as commonly seen in
the SEDs of starburst galaxies \citep{Sanders1996, Siebenmorgen2007}. The $3\,
\mu$m bump is produced by the hot dust ($T \sim 1000$ K) heated by the central
AGN. This unique feature appears to be ubiquitous in most of the quasar SEDs
over a wide range of redshifts (e.g., \citealt{Elvis1994, VandenBerk2001,
Telfer2002, Vignali2003, Richards2006, Jiang2006}).

To demonstrate the line-of-sight dependence, Figure~\ref{Fig_sed_z6} also
shows the range of SEDs viewed from two orthogonal angles of the Cartesian
grid: along the z-axis (upper range) and the xy-plane (lower range),
respectively. This range indicates difference in column density along
the sight lines, and shows that dust extinction in the UV/optical
bands differs by a factor of $\sim 3$, but no difference at wavelengths
longward of $1\, \mu$m. This suggests that the dust is close to optically thin
along these two viewing angles during this time, and that infrared or
submillimeter observations of luminous quasars may not be diagnostics for the
orientation of the host. We have also checked the three major
components of the output SED, namely the scatter, escape and reemission, and
found that the photon energy is conserved. This confirms the photon
conservation algorithm used in the radiative transfer calculation. Moreover,
we find that the emergent SED in the UV/optical bands ($0.01 - 1\, \mu$m) is
dominated by scattering. 

It is interesting to comment on the spectral feature at wavelength
$\lambda_{\rm {rest}} = 9.7\mu$m where the silicate cross section
peaks \citep{Draine2003}. This feature can produce either emission or
absorption in the  spectrum depending on the optical depth of the medium at
this wavelength \citep{Bjorkman2001}. In order to have absorption at this
wavelength, the medium has to be optically thick (i.e., $\tau_{\rm
{9.7\mu m}} >> 1$). Generically, this would also result in deep
absorption in the optical/NIR bands ($\sim 0.01 - 1\, \mu$m) owing to
their much higher dust opacity, which is almost two orders of
magnitude higher than that at $9.7\mu$m according to the dust opacity
curve in Figure~\ref{Fig_opac}.  Our spectra exhibit absorption when
the system is in the starburst phase (e.g., $z \gtrsim 10$) when the object
is highly obscured. As the system proceeds to the quasar phase (e.g.,
$z < 8$), the emission feature becomes increasingly prominent as dust
becomes more and more transparent to the radiation.

In observations, both absorption and emission features at
$\lambda_{\rm {rest}} = 9.7\mu$m have been detected for a wide range
of objects up to $z \sim 3$ \citep{Armus2004, Papovich2006}. An
absorption feature is frequently seen in dusty star-forming
galaxies such as M82 (e.g., \citealt{Sturm2000}), Arp 220 (e.g.,
\citealt{Spoon2004}), and ULIRG IRAS08572+3915 which exhibits the most
extreme absorption \citep{Spoon2006}. A comprehensive collection of
the SEDs of starburst nuclei and ULIRGs is reviewed by
\cite{Siebenmorgen2007}, who also provide a library of 7000 SEDs for
dusty galaxies. 

On the other hand, an emission feature is also reported in many
observations \citep{Hao2005, Siebenmorgen2005, Sturm2005}. In
particular, \cite{Hao2007} analyze a sample of 196 local AGNs and
ULIRGs observed by the Infrared Spectrograph (IRS; \citealt{Houck2004}) on
board the Spitzer Space Telescope \citep{Werner2004} to study the distribution
of strengths of the $9.7\mu$m silicate feature. These authors find a wide
range of silicate strengths: quasars are characterized by silicate emission
and Seyfert 1s equally by emission or weak absorption. Seyfert 2s are
dominated by weak silicate absorption, and ULIRGs are characterized by
strong silicate absorption (mean apparent optical depth of about
1.5). \cite{Spoon2007} find that the same sample of galaxies is
systematically distributed along two distinct branches: one with
AGN-dominated spectra and one with deeply obscured nuclei and
starburst-dominated spectra. These authors suggest that the separation
may reflect a fundamental difference between the dust geometries in
the sources: clumpy for AGNs versus non-clumpy obscuration for
starbursts. For example, \cite{Levenson2007} suggest that the
extremely deep absorption in IRAS08572+3915 requires a source to be
embedded in a smooth distribution of material that is both
geometrically and optically thick.  Our simulations show that the dust
distribution around the quasar is quite clumpy, and that the SEDs
changes from starburst to AGN-dominated as the system evolves from the
starburst to quasar phases (see Figure~\ref{Fig_sed_z}). These are
consistent with the observations.

\section{Parameter Studies}
\label{sec_param}

In this Section, we explore the large parameter space involved in this
radiative transfer calculation, and systematically study the resolution
convergence, parameters in the two-ISM model, input spectra for
the BHs, the dust models, and the dust-to-gas ratios for both the cold- and
hot-phase gas. 

\subsection{Resolution Studies}
\label{subsec_res}

\begin{figure}
\begin{center}
\includegraphics[width=3.2in]{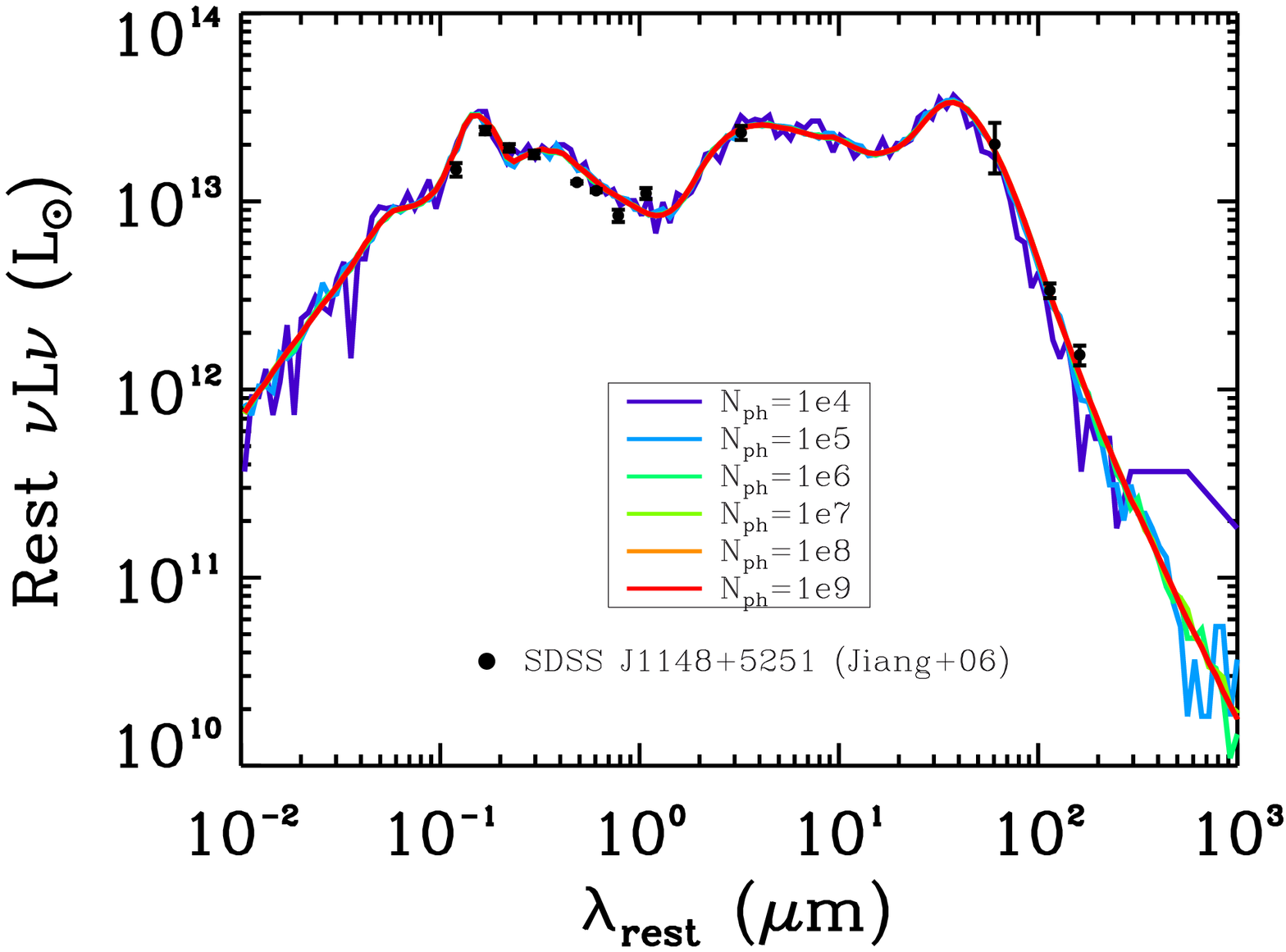}\\
\vspace{0.5cm}
\includegraphics[width=3.2in]{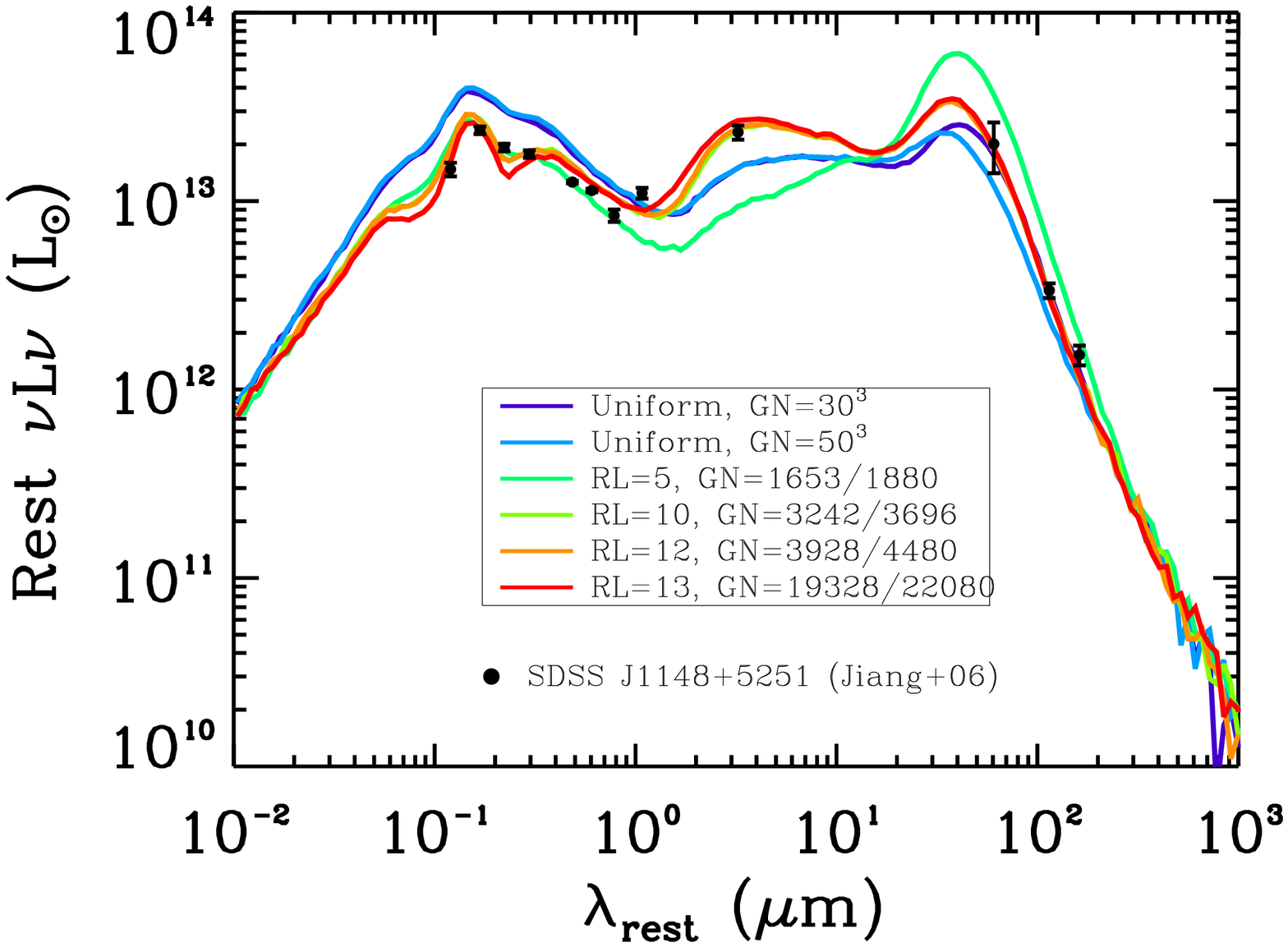}
\vspace{1cm}
\caption{
Resolution studies of the number of photons (top panel), and the grid
refinement level (bottom panel), respectively. The top panel shows the SEDs
produced with photon number $N_{\rm ph}=10^4 - 10^9$. These SEDs have similar
shapes, and they converge when $N_{\rm ph} \gtrsim 10^6$. A smaller photon
number results in larger fluctuations in the SED owing to greater Poisson
error. The bottom panel compares the SEDs produced with a uniform grid and those
with adaptive grids of different refinement level. Compared to an adaptive grid
method, SEDs using uniform grids with grid number ${\rm GN}=30^3$ and $50^3$
do not have sufficient dynamic range to resolve both the cold and hot dust, which
result in an underestimate of the dust emission longward of $10\, \mu$m. As
the grid refinement level increases, hot dust is
better resolved, 
contributing to the hot dust bump at $1 - 10\, \mu$m. The SEDs converge when the
refinement level goes above 10, which has a minimum cell size approaching 
that of the spatial resolution of the original hydrodynamic simulations. Level
12 is the standard refinement level used in this paper.}     
\label{Fig_res}
\end{center}
\end{figure}

Figure~\ref{Fig_res} shows resolution studies for photon number (top panel)
and grid size (bottom panel). The SEDs converge when the photon number is
larger than $10^6$. If the photon number is low, then Poisson noise is 
significant, which results in large fluctuations in the SEDs. Therefore,
throughout the paper, we use a photon number of $10^7$ for SED production, and
$10^8$ for images in order to have higher signal-to-noise.

For the adaptive grids, as the refinement level increases, the hot
dust in the central region around the AGN is better resolved, 
contributing to the hot dust bump in $1 - 10\, \mu$m. The SEDs converge when
the refinement level goes above 10, which has a minimum cell size close to
the spatial resolution ($\sim$30 pc) of the original hydrodynamic
simulations. Compared to the adaptive grid
method, SEDs using uniform grids with reasonable computing expense
have poor resolution. As demonstrated in Figure~\ref{Fig_grid}, a
$50^3$ uniform grid has a resolution of $4$ kpc, which can only resolve the
dust in the diffuse gas in the outskirts of the system, but not the clumpy,
dense regions around the central AGN.  Consequently, the resulting SEDs do not
entirely resolve dust emission in $1 - 10\, \mu$m which comes 
partly from hot dust near the AGN, and the cold dust bumps are lower by up to
one order of magnitude.

\subsection{ISM Model Parameters}

\begin{figure}
\begin{center}
\includegraphics[width=3.5in]{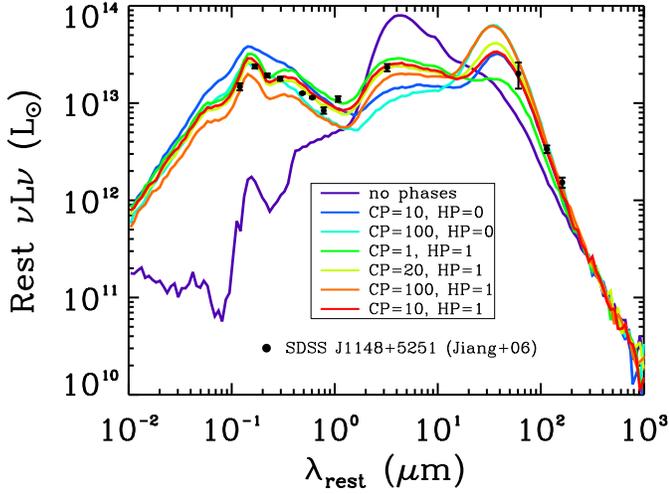}
\vspace{0.5cm}
\caption{Parameter study of the two-phase break down of the ISM with
  hot and cold phases. In the legend, HP=1 indicates existence of
  hot-phase gas, while CP=10 indicates that the cold pressure is enhanced by a
  factor of 10 (see \S~\ref{subsec_ism} for more details). The purple curve
  represents the ``no phase  break-down'' case in which the cold gas is not
  considered; there is only hot-phase gas whose density is the same as that
  given directly by the SPH simulations. The blue and cyan curves represent cases
  in which no hot-phase gas is present (only cold-phase gas is considered), but the
  pressure enhancement factor for cold-phase gas varies from 10 to 100,
  respectively. The rest of the colored curves represent cases in which both
  hot and cold phases co-exist, with cold gas pressure varying from 1 to
  100. In the RT calculations, we use standard values CP=10 and HP=1.}     
\label{Fig_phase}
\end{center}
\end{figure}

\begin{figure}
\begin{center}
\includegraphics[width=3.5in]{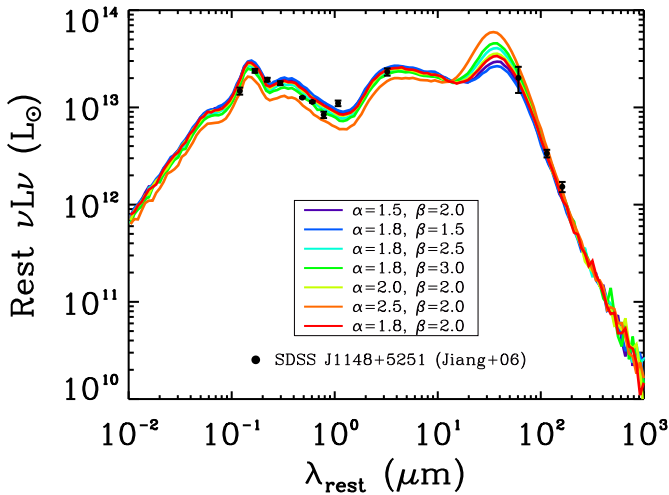}
\vspace{0.5cm}
\caption{Parameter study of the 
  power-law indices of the mass spectrum $\alpha$ and
  mass-radius relation $\beta$ of the cold clouds (see \S~\ref{subsec_ism} for
  more details). The output SED is not sensitive to the ranges of
  $\alpha$ and $\beta$ considered here. In the RT simulations, we adopt
  $\alpha=1.8$ and $\beta=2.0$ as suggested by observations. }     
\label{Fig_MR}
\end{center}
\end{figure}

Figure~\ref{Fig_phase} shows the SEDs with various parameters for the
two-phase breakdown of the ISM. When there is no phase break-down
(purple curve), only hot-phase gas is considered (no cold gas). In
this case, the hot gas has both large volume filling factor and
mass fraction, so the extinction is extremely high, which leads
to significant emission in the NIR but little emission in the $20 -
1000\, \mu$m range (the cold dust is sparse). In cases with cold gas
only (no hot-phase gas), the SEDs show big cold dust bumps but no hot
dust emission, as indicated by the blue and cyan curves. A comparison
of these two curves shows that a larger cold phase pressure leads to a
higher volume filling factor for the cold dust, resulting in stronger
cold dust emission.

In the cases where both hot and cold phases coexist, the amount of
cold dust emission depends on the cold gas pressure, as indicated by
the other colored curves. We find that cold pressure in the range of
10 -- 100 is able to produce cold dust bumps that fit the
submillimeter observations of \zquasar.  However, because the SED with
CP=10 has both the hot and cold dust bumps that agree better with the
mean SEDs of luminous quasars in the Sloan samples of
\cite{Richards2006}, we choose CP=10 as a standard value in our
calculations. 

Figure~\ref{Fig_MR} shows a study of the output SEDs obtained by
varying the power-law indices of both the mass spectrum and the
mass-radius relation of the cold clouds. Increasing $\alpha$ would
steepen the cloud mass function, leading to more smaller cold clouds
that boost the cold dust bump.  A similar effect is seen by increasing
$\beta$. However, the SEDs are not very sensitive to the change of
either $\alpha$ or $\beta$ in the ranges of 1.5 -- 2.5 and 1.5 -- 3.0,
respectively. Changing $\alpha$ from 1.5 to 2.5, or $\beta$ from 1.5
to 3.0 only results in a change in the SED by a factor of about 2.  In
the RT calculations, we adopt the standard values of $\alpha=1.8$ and
$\beta=2.0$, as suggested by observations \citep{Blitz2006,
Rosolowsky2007}.

\subsection{Input Spectrum}

\begin{figure}
\begin{center}
\includegraphics[width=3.5in]{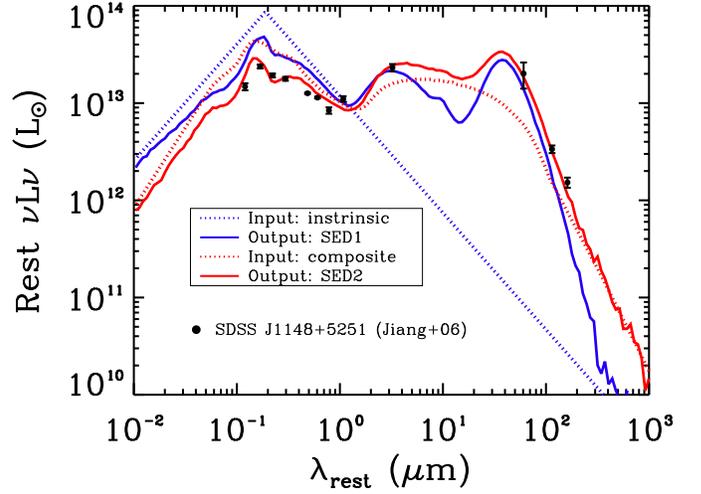}
\vspace{0.5cm}
\caption{A parameter study of the input spectrum for the black hole. The solid
  curves represent the output SEDs, while the dotted lines represent the input
  black hole spectra. The blue and red curves indicate the use of an
  intrinsic, broken power-law as in \cite{Marconi2004}, and a composite
  spectrum from \cite{Hopkins2007A}, respectively. Both input spectra have the
  same bolometric luminosity. A power-law input spectrum would require
  pc-scale resolution to resolve the dust near the AGN in order to produce the
  near-IR emission, which is below the resolution of our hydrodynamic
  simulations. A composite spectrum therefore serves as a sub-resolution
  recipe to resolve the dust emission within parsecs of the AGN.}     
\label{Fig_inspec}
\end{center}
\end{figure}

Figure~\ref{Fig_inspec} shows a comparison of the emergent SEDs using  
different input black hole spectrum, namely a broken power-law as in
\cite{Marconi2004} (blue curve), and a composite spectrum from
\cite{Hopkins2007A} (red curve). All other parameters being equal, the SED
using the power-law spectrum shows more emission in the optical bands and
lower emission in the IR than the SED with the composite input spectrum. These
differences owe their origin to differences in the input spectrum.  The
composite spectrum includes emission from hot dust residing in the
vicinity of the AGN that is below the resolution limit of our hydrodynamic
simulations. This figure emphasizes the care that must be taken when including
an AGN and performing radiative transfer on scales that are not well-resolved,
and demonstrates that a composite black hole spectrum as in
\cite{Hopkins2007A} provides a viable sub-resolution prescription for the dust 
emission contributed within parsecs of the AGN. 

\subsection{Dust Models}
\label{subsec_dustcomp}

\begin{figure}
\begin{center}
\includegraphics[width=3.5in]{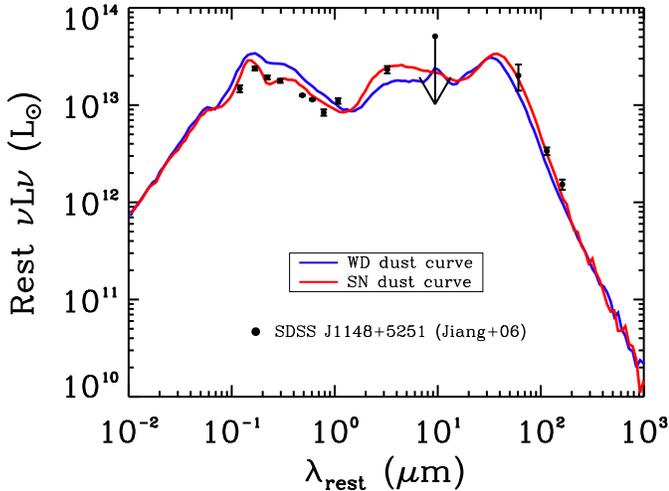}
\vspace{0.5cm}
\caption{
Comparison of the SEDs from using the dust extinction curves of
\cite{Weingartner2001} with $R_{\rm V}=3.1$ (WD model) and that from
Type-II supernovae (SN model). The arrow at $\sim$ 9.43 $\mu$m indicates a
2$\sigma$ upper limit. \zquasar\ was observed with MIPS at 70 $\mu$m but with
no detection. The arrow indicates a $2\sigma$ upper limit. The current
observations cannot distinguish between these two dust models.}     
\label{Fig_dustcomp}
\end{center}
\end{figure}

Figure~\ref{Fig_dustcomp} shows the emergent SEDs using dust extinction curves 
from the WD model \citep{Weingartner2001} with $R_{\rm V}=3.1$ and our
SN model, respectively.  The SEDs agree well at wavelengths $\lambda \gtrsim
10\, \mu$m, but differ by a factor of a couple in the optical /NIR bands $0.1
- 10\, \mu \rm{m}$. It appears that WD model would require a higher dust-to-gas
ratio than that of the Milky Way by a factor of a few in order to produce the
observed SED of \zquasar. In such case, the WD model would produce a stronger
peak at the $9.7\, \mu$m silicate feature by a similar factor relative to the
SN model. Observation of \zquasar\ at $\sim 9.7\, \mu$m would be helpful in
distinguishing between these two models. However, the current data-point at
that wavelength (observed by MIPS at 70 $\mu$m) is only a $2\sigma$ upper
limit, insufficient to constrain the model. 

As discussed in \S~\ref{subsec_sndust}, \cite{Elvis2002} suggest that
quasars may also be copious producers of dust, as condensation in
quasar outflows is similar to dust formation in the envelopes of AGB
stars. In this case, the dust extinction curve would be similar to the
WD model. However, the contribution of dust from quasar winds in the
early Universe remains unknown. The recent report by
\cite{Stratta2007} of dust extinction in the host galaxy of GRB 050904
at $z = 6.3$ independently supports a supernova-origin dust model. 
If both supernovae from starbursts and quasar outflows play important
roles in dust production in these young quasar systems at $z \sim 6$,
then the resulting dust extinction curve would be intermediate to the
WD and SN models shown in Figure~\ref{Fig_dustcomp}. More observations
and deeper surveys for dust in high-z galaxies and quasars will be
necessary to constrain dust formation mechanisms at early 
cosmic times.

\subsection{Dust-to-gas Ratios}

\begin{figure}
\begin{center}
\includegraphics[width=3.5in]{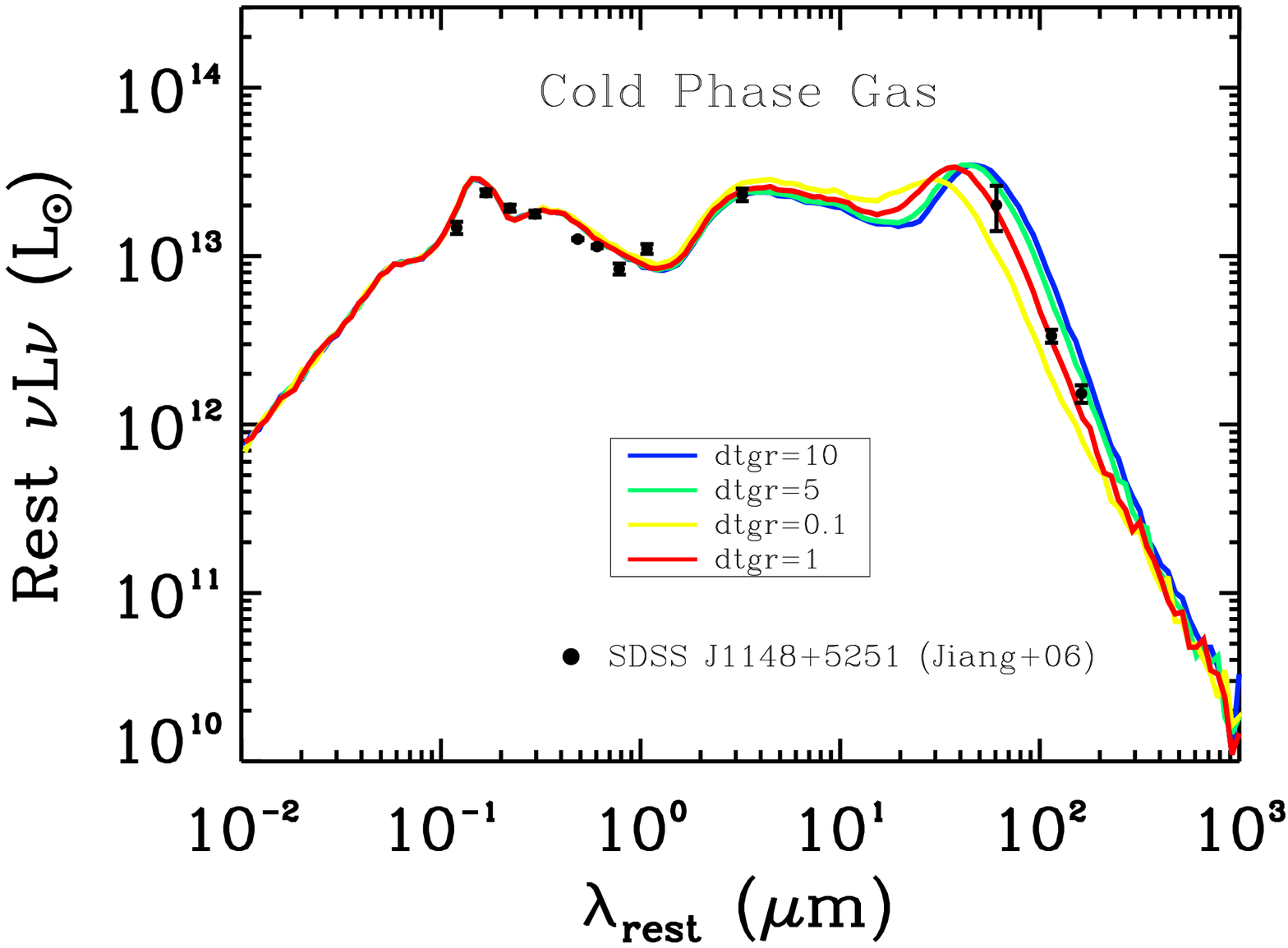}\\
\vspace{0.5cm}
\includegraphics[width=3.5in]{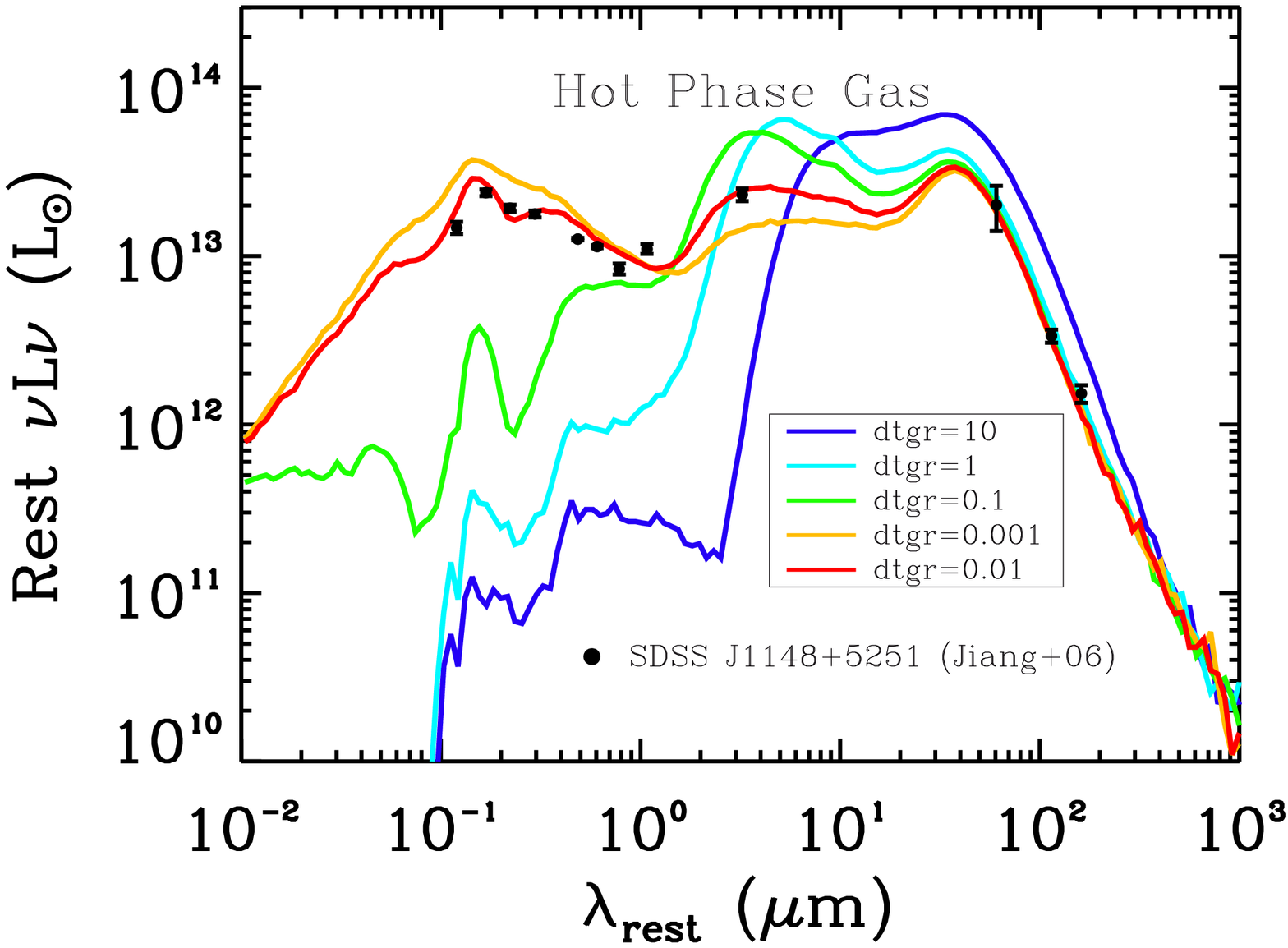}
\vspace{0.5cm}
\caption{A parameter study of the dust-to-gas ratio relative to Milky Way
  value, for the Cold Phase Gas (CPG, top panel) and Hot Phase Gas (HPG,
  bottom panel), respectively. This figure shows that the emergent SED is more
  sensitive to the dust-to-gas ratio of the HPG than to the CPG owing to the
  large covering factor of the HPG. In the regular RT calculations, the
  default values are 1 and 1\% Milky Way value for CPG and HPG, respectively.} 
\label{Fig_dtg}
\end{center}
\end{figure}

It has been suggested that ULIRGs have dust-to-gas ratio (mostly cold gas as
in our modeling) close to that of the Milky Way (e.g., \citealt{Dunne2001,
Klaas2001}). While some observations of obscurred, X-ray selected AGNs seem to
suggest that the dust-to-gas ratio in these objects has a wide range, from $\sim
10^{-3}$ to a few of MW value (e.g., \citealt{Maiolino2001, Kuraszkiewicz2003,
Hall2006}). Although this range may apply mainly to the hot, fully or
partially ionized gas in the circumnuclear regions of the AGNs, it would be
interesting to see how such a range affects the output SED in our calculations. 

Figure~\ref{Fig_dtg} shows the comparsion of the emergent SEDs with different
dust-to-gas ratios in a wide range, for both cold (top panel) and
hot-phase gases (bottom panel), respectively. The values in the plot are
relative to the MW value. A change of two orders of magnitude in the
dust-to-gas ratio of the cold phase gas results in a difference in the cold
dust bump only by a factor of a few. However, a similar change in the
dust-to-gas ratio in the hot phase gas would result in substantial difference
in the output SED. This study shows the extinction is dominated by the hot
dust, which has a much larger covering factor that the cold dust. We find that
using 1 and 1\% of MW value for the cold and hot phase gas, respectively,
reproduce the observation of SDSS J1148 reasonably well. Therefore, we elect
to use them as default values for our standard RT calculations. Note this
choice of 1\% of MW value for the hot, diffuse gas has little effect for the
starburst or ULIRG phase, because a substantial fraction of gas during that
stage is in the cold clouds. However, it may affect the SEDs of major quasar
phase, as the gas is heated by the AGN. We should point out that in our ISM
model, we do not have the warm phase as in the picture of
\cite{McKee1977}. Such a warm phase medium may have a dust-to-gas ratio 
similar to that of MW. We will explore such a treatment in future work.

\section{The Dust Distribution}
\label{sec_dust}

\begin{figure}
\begin{center}
\includegraphics[width=3.1in]{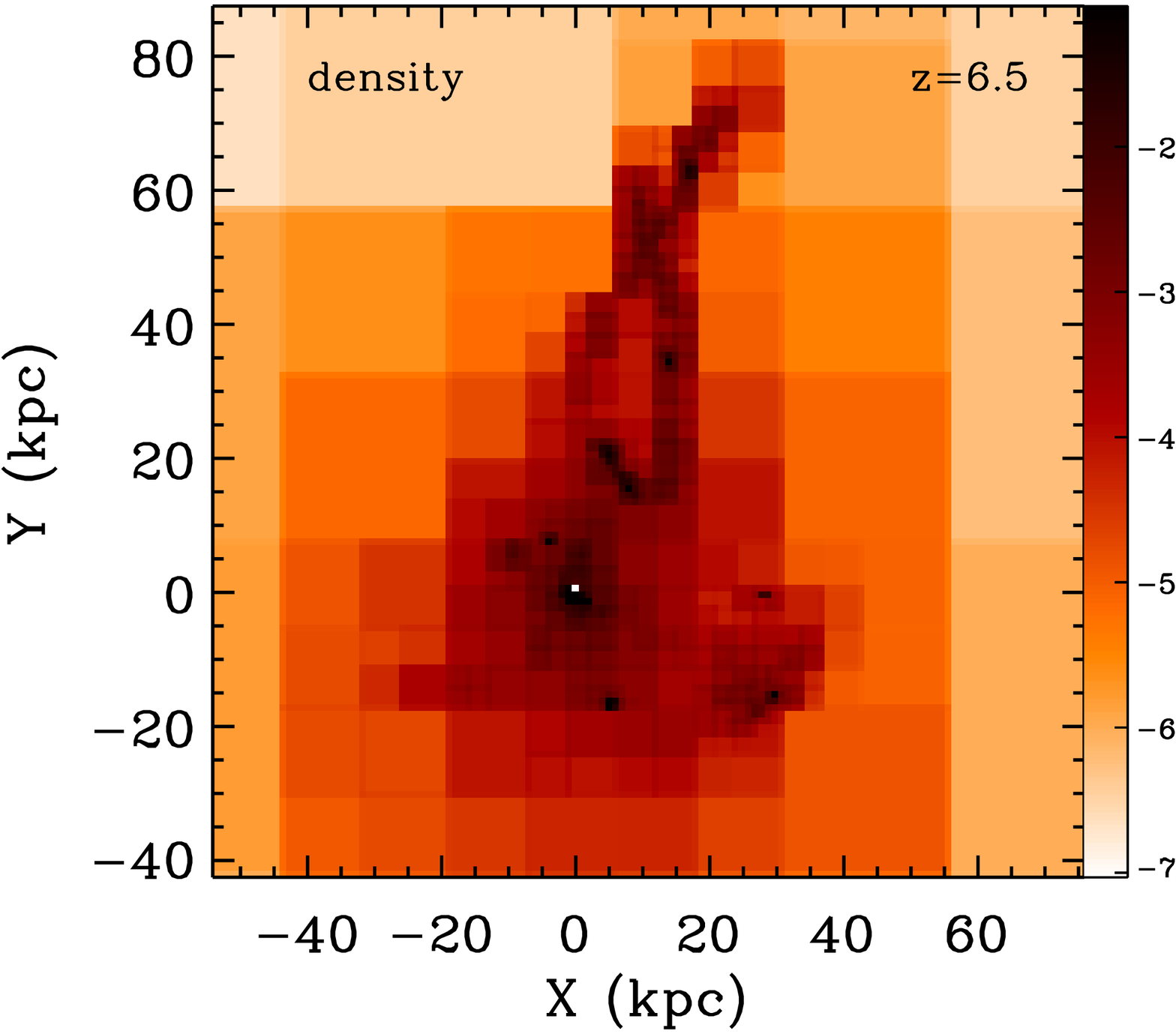} \\
\includegraphics[width=3.1in]{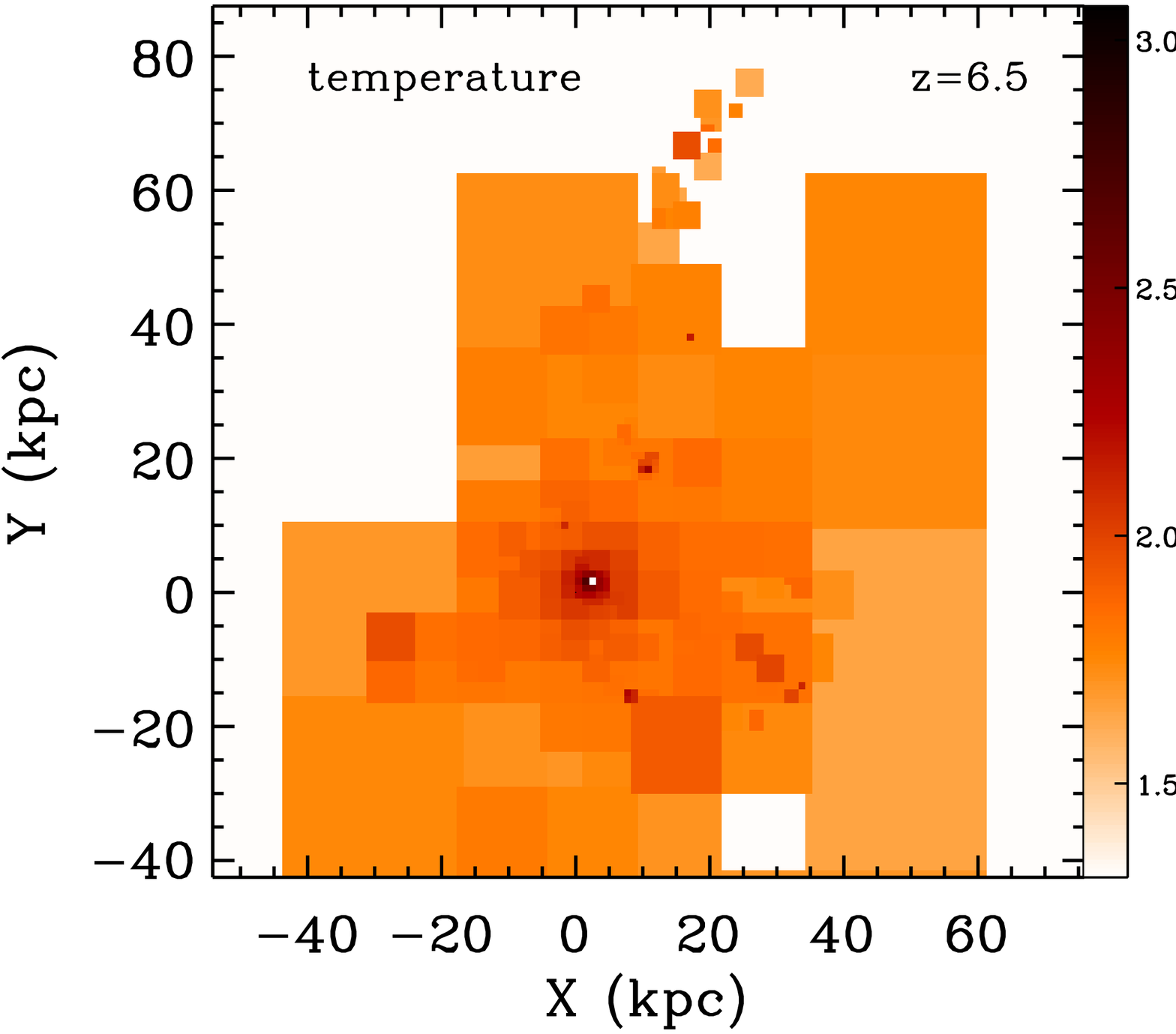} 
\vspace{1cm}
\caption{
Maps of the projected density (top panel) and temperature (bottom
panel) of the HPG-dust (dust associated with hot phase gas). The
origin of the map is the location of the central quasar, and the coordinates
are comoving. Note that the gas only determines the distribution and mass of
the dust; the dust temperature is not associated with the gas temperature but
is calculated self-consistently from the radiation field (see
\S~\ref{subsec_ism} for more details).}    
\label{Fig_dustmap}
\end{center}
\end{figure}

The distribution of dust is essential to investigating dust formation
mechanisms and heating sources. Figure~\ref{Fig_dustmap} shows the projected
spatial distribution of the dust density and temperature in the system at
$z=6.5$. The projected quantity is calculated with $\int f(l){\rm d}l/L$, where
$f(l)$ is density or temperature distribution along the line of sight, while
$L$ is the total length of the sight line. Gravitational torques in the
interaction produce strong shocks and tidal features, making the dust
distribution highly inhomogeneous and clumpy. The dynamic range in density can
be up to six orders of magnitude. Both the dust density and temperature peak
in the central regions near the AGN. 

\begin{figure}
\begin{center}
\vspace{1cm}
\includegraphics[width=3.2in]{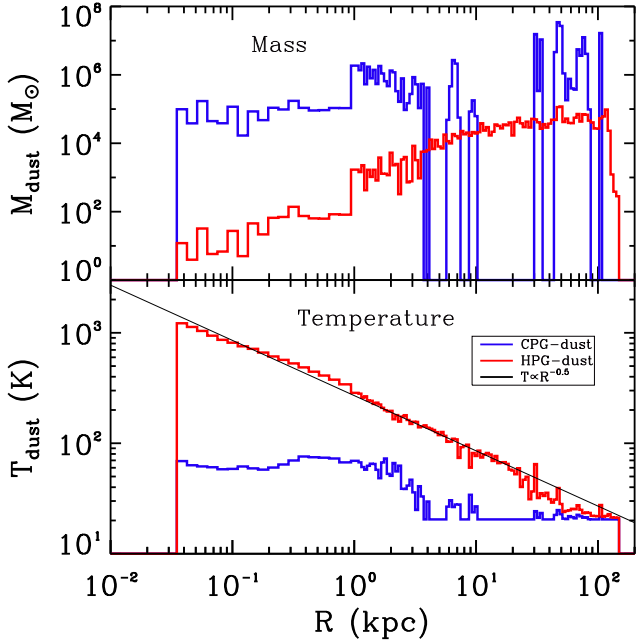}
\vspace{1cm}
\caption{
Radial distribution of dust mass (top panel) and temperature (bottom
panel) as a function of the distance from the central AGN. The blue and red
curves represent the CPG- and HPG-dust, respectively, while the black curve in
the bottom panel is a power-law temperature profile, $T \propto
R^{-1/2}$ as expected from heating by the central AGN.}    
\label{Fig_rdust}
\end{center}
\end{figure}

The detailed radial distributions of the dust mass and temperature are
quantified in Figure~\ref{Fig_rdust}. The blue and red curves in this figure 
represent CPG- and HPG-dust, dust associated with cold- and hot-phase gas,
respectively, as defined in \S~\ref{subsec_ism}. The CPG-dust has high
density but a small volume filling factor; it is generally optically thick to
radiation. On the other hand, the HPG-dust has a lower density but fills $\gtrsim
99\%$ of the volume, and is generally optically thin. Both the dust mass and
temperature vary with distance from the central AGN. The cold dust is
typically surrounded by hot dust. In the inner hundred parsecs, the cold cores
are highly condensed and are optically thick even to the hard radiation from
the black hole. However, the HPG-dust at the edges surrounding these cold
cores is heated directly by the central AGN. This is indicated by the
power-law temperature profile in the lower panel of Figure~\ref{Fig_rdust}, $T
\propto R^{-1/2}$ as expected from Equation (\ref{eq:REtemp}): $T^4(R) \propto
E(R) \propto R^{-2}$. The hottest dust is heated up to $\sim 1200$~K, which is
below the dust sublimation temperature $\sim 1600$~K
\citep{Hoenig2006}. Cooler dust ($T \sim$ tens to hundreds K) is distributed
in an extended region from $\sim$ 100 pc to several kiloparsecs. The 
heating sources come both from stars and the central AGN. Beyond 10 kpc, the
dust temperature drops to below 100 K. Note that the minimum temperature in
Figure~\ref{Fig_rdust} is $\sim 20$~K, which is the CMB temperature at z=6.5.  

\begin{figure}
\begin{center}
\includegraphics[width=3.2in]{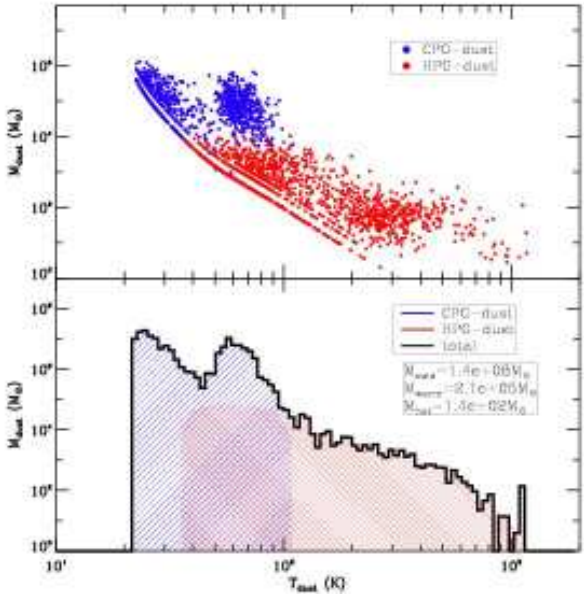}
\vspace{1cm}
\caption{
Mass-temperature distribution of both the CPG- (in blue) and HPG-dust (in
red). In the bottom panel, $M_{\rm cold}$, $M_{\rm warm}$, and $M_{\rm hot}$
represent the total mass of cold-, warm- and hot-dust with $T \lesssim 100$~K,
$100 < T < 1000$~K, and $1000 \lesssim T \lesssim 1200$~K, respectively.}   
\label{Fig_mdust}
\end{center}
\end{figure}

Figure~\ref{Fig_mdust} shows histograms of the dust mass and
temperature. The dust temperature ranges from $\sim 10 - 10^3$ K. The
amount of the hottest dust ($\sim 10^3$ K) is $\sim 1.4\times 10^2\, \Msun$,
while that of the cold dust ($T \lesssim 10^2$ K) is $\sim 1.4\times 10^8\,
\Msun$. This is in agreement with the estimates of the dust detected in the
host of \zquasar\ \citep{Bertoldi2003A, Beelen2006, Jiang2006}. 

The large amount of cold dust ($\sim 1.4\times 10^8\, \Msun$) located
within 3 kpc from the AGN provides efficient cooling for the formation of
molecular gas. In \cite{Narayanan2007}, we calculate carbon monoxide emission
using a non-local thermodynamic equilibrium radiative transfer code
\citep{Narayanan2006b, Narayanan2006c}, and find that CO gas forms in this
region, with a total mass of $\sim 10^{10}\, \Msun$, similar to observations 
by \cite{Walter2004}. This suggests that cold dust and molecular gas are
closely associated, and both depend on the star formation history. Our results
show that significant metal enrichment takes place early in the quasar
host, as a result of strong star formation in the progenitors, and that   
intense starbursts ($\rm {SFR} \gtrsim 10^3\, \Msun, \yr^{-1}$)
within $\lesssim 10^8$ yr are necessary to produce the observed
properties of dust and molecular gas in the host of \zquasar, a conclusion
which is also supported by the analytical models of \cite{Dwek2007}. 

\section{Evolution of the Quasar System}

\subsection{Transition from cold to warm ULIRG}

\begin{figure}
\begin{center}
\includegraphics[width=3.5in]{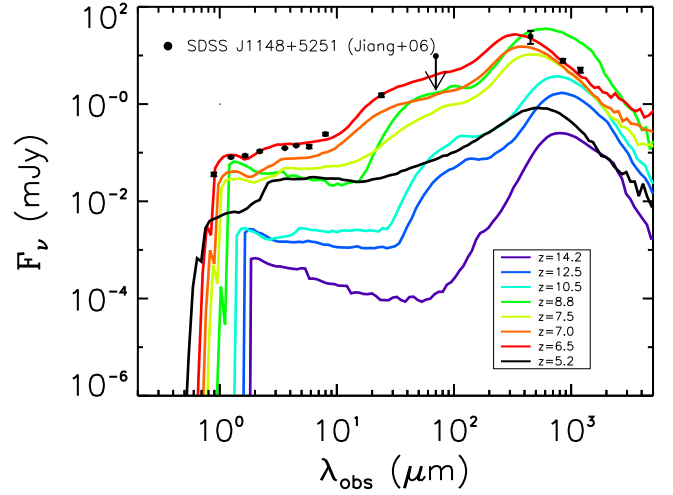}
\vspace{1cm}
\caption{Evolution of the SEDs of the quasar system and its
  galaxy progenitors in the observed frame. The colored curves represent SEDs
  from $z \sim 14$ to  $z \sim 5.2$, while the black dots are again
  observations from \cite{Jiang2006}, as described in the legend. Absorption
  of the Lyman line series and continuum by the intergalactic medium
  \citep{Madau1995} is taken into account, which results in a sharp drop at
  short wavelengths.}  
\label{Fig_sed_z}
\end{center}
\end{figure}

\begin{figure}
\begin{center}
\includegraphics[width=3.5in]{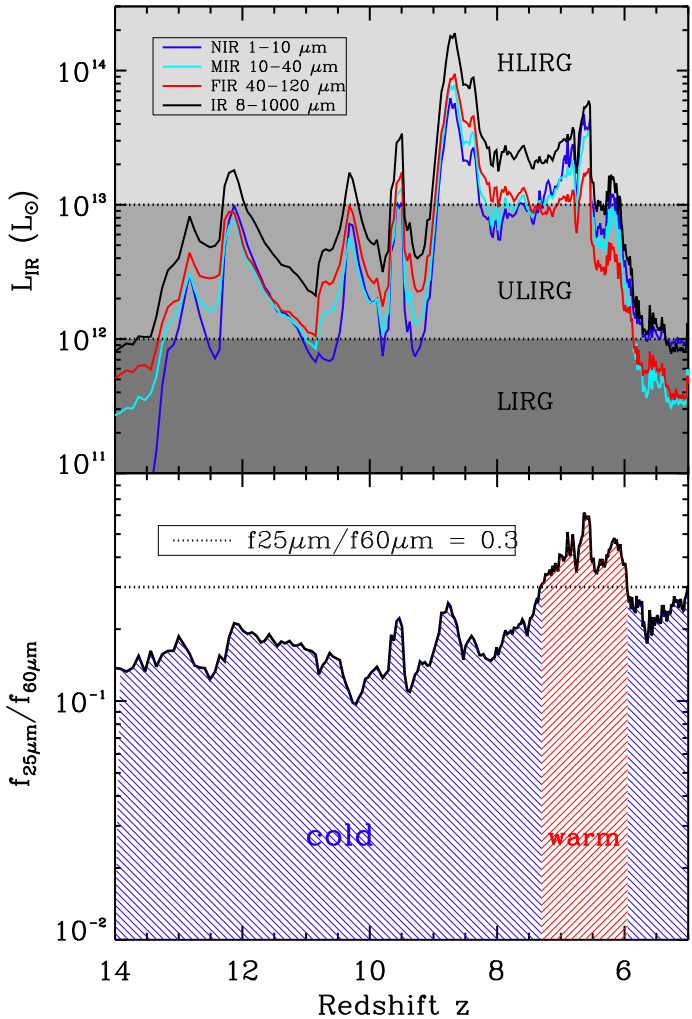} 
\vspace{0.8cm}
\caption{Evolution of rest-frame infrared luminosities of the quasar
  system (top panel) and the rest-frame color $f_{25\mu\rm m}/f_{60 \mu\rm m}$
  (bottom panel). In the top panel, the solid curves represent the luminosity in
  near-IR ($1 - 10\, \mu$m), mid-IR ($10 - 40\, \mu$m), far-IR ($40 - 120\,
  \mu$m), and IR ($8 - 1000\, \mu$m), respectively. The grey shades indicate
  regimes for LIRG ($\Lir > 10^{11}\, \Lsun$), ULIRG ($\Lir > 10^{12}\,
  \Lsun$), and HLIRG ($\Lir > 10^{13}\, \Lsun$),
  respectively, as classified by \cite{Sanders1996}. The ratio $f_{25\mu\rm
  m}/f_{60 \mu\rm m}$ in the bottom panel is an indicator of the coldness of
  the SED with a critical value of 0.3 \citep{Sanders1996}. The quasar system
  in our model evolves from ``cold'' ULIRG ($f_{25\mu\rm m}/f_{60 \mu\rm
  m}<0.3$) to ``warm'' ULIRG (including HLIRG, $f_{25\mu\rm m}/f_{60 \mu\rm m}
  \ge 0.3$) as it transforms from starburst to quasar phases.} 
\label{Fig_LIR_color}
\end{center}
\end{figure}

In \cite{Li2007}, we show that the host of the $z\sim 6$ quasar
undergoes hierarchical mergers starting from $z\sim 14$, and the
quasar descends from starburst galaxies. The evolution of the SEDs of
the system in the observed frame is shown in
Figure~\ref{Fig_sed_z}. Note that absorption of the Lyman lines and
continuum by the intergalactic medium \citep{Madau1995} is taken into
account. During early stages at $z \sim 14$, the SED of the quasar
progenitor shows only a cold dust bump that peaks around $\sim
60\mu$m, which is characteristic of starburst galaxies
\citep{Sanders1996}. As the system evolves from starburst to quasar
phases, dust extinction in the UV-optical bands increases, boosting
the emission reprocessed by dust at wavelengths longward of $1\,
\mu$m. The gradually increasing radiation from the accreting black
holes heats the nearby dust to high temperatures, contributing to
the hot dust bump which peaks around $\sim 3\mu$m (rest frame). At the maximum  
quasar phase at $z \simeq 6.5$ indicated by the red curve, the hot
dust bump SED reaches its peak with a temperature of $\sim 1200$~K, as
we have seen in the previous section. Such an SED represents 
luminous, blue quasars in the samples of \cite{Jiang2006} and
\cite{Richards2006}. As the system ages and reddens in the post-quasar
phase (indicated by the black curve), the total luminosity of the
system drops. However, there is still some residual hot dust,
and the infrared luminosity is dominated by the NIR and MIR.  This
resembles the class of infrared-bright, optically-red quasars found in
recent surveys (e.g., \citealt{Brand2006}). Overall, the evolution of
the SED from a starburst to a quasar can be characterized by the slope of
the infrared SED ($3 - 50\, \mu$m), as it decreases from the starburst
to quasar phases owing to the increase of NIR emission from the hot
dust heated by the AGN.

Infrared luminosities are powerful tools to study starburst galaxies and
quasars. From the results in the previous sections, we see that the emission
in near-IR ($1 - 10\, \mu$m), mid-IR ($10 - 40\, \mu$m) and far-IR ($40 -
120\, \mu$m) comes from re-emission by hot, warm, and cold dust, 
respectively. Note that the meaning of FIR varies in the literature. Here we
use the definition of FIR given by \cite{Condon1992} in order to compare our
results with the observations by \cite{Carilli2004} who used the same FIR
range. 

Using $\Lir$ ($8 - 1000\, \mu$m), \cite{Sanders1996}  
classified infrared luminous galaxies into three categories, namely luminous
infrared galaxy (LIRG, $\Lir > 10^{11}\, \Lsun$), ultra-LIRG (ULIRG, 
$\Lir > 10^{12}\, \Lsun$), and hyper-LIRG (HLIRG, $\Lir > 10^{13}\,
\Lsun$). Figure~\ref{Fig_LIR_color} (top) shows the evolution of the infrared
luminosities of the quasar system in our simulations. The luminosities
increase with the star formation rate and black hole accretion rate. The
system is ultraluminous most of the time, with periods in HLIRG phases
associated with bursts of star formation or quasar activity. When the last
major mergers take place between $z \sim 9 - 7.5$, the strong shocks and highly
concentrated gas fuel rapid star formation and black hole growth. This
also produces a large amount of dust heated by the central AGN and
stars. As a result, the infrared luminosities increase
dramatically, pushing the system to HLIRG class. 

The infrared luminosities strongly peak at $z\sim 8.7$ when the
star formation rate reaches $\gtrsim 10^4\, \Msun\, \yr^{-1}$,
suggesting a significant contribution from stars in heating the dust
to emit in the range $1 - 1000\, \mu$m. During the major quasar phase
($z \sim 7.5 - 6$) the star formation declines to $\sim 10^2\,
\Msun\, \yr^{-1}$, and the emission in NIR and MIR outshines that in FIR,
demonstrating that AGN can play a dominant role in heating the dust and
producing NIR and MIR emission.  Furthermore, the AGN can also contribute
to FIR emission.  For example, while the star formation rate drops by a
factor of $\sim 500$ from $z\sim 8.7$ (when star formation rate peaks)
to $z\sim 6.5$ (when black hole accretion rate peaks, see
Figure~\ref{Fig_quasar}), the $\Lfir$ declines by only a factor of $\sim
50$, indicating substantial contribution to the $\Lfir$ by the AGN. 
This significant AGN contribution has important implications for estimating
the star formation rate. We will discuss this issue at length in the next
section. During the ``post-quasar'' phase at $z<6$, as a result of
feedback which suppresses both star formation and black hole
accretion, and gas depletion which reduces the amount of dust, the
infrared luminosities drop rapidly.

The flux ratio $f_{25\mu\rm m}/f_{60 \mu\rm m}$ is a color indicator
for the coldness of the SED, and can be used as a diagnostic for AGN
activity, as suggested by \cite{Sanders1996}. For example, starburst
galaxies usually have cold colors $f_{25\mu\rm m}/f_{60 \mu\rm
m}<0.3$, while quasars are warm with $f_{25\mu\rm m}/f_{60 \mu\rm
m}\gtrsim0.3$.  The evolution of the color $f_{25\mu\rm m}/f_{60
\mu\rm m}$ of the simulated quasar system is shown in
Figure~\ref{Fig_LIR_color} (bottom panel). The color index is below
0.3 most of the time, i.e. during both starburst- and post-quasar
phases. However, it rises above 0.3 during the short quasar
phase. This figure clearly demonstrates that the system evolves from a
cold to warm ULIRG (including HLIRG) as it transforms from starburst
to quasar phases. Similar trend is also reported by
\cite{Chakrabarti2007A}. Our results provide further theoretical support for 
the starburst-to-quasar conjecture, as suggested by observations
\citep{Sanders1996, Scoville2003}.

\subsection{AGN Contamination and the {\rm SFR} -- $\Lfir$ Relation}

\begin{figure}
\begin{center}
\includegraphics[width=3in]{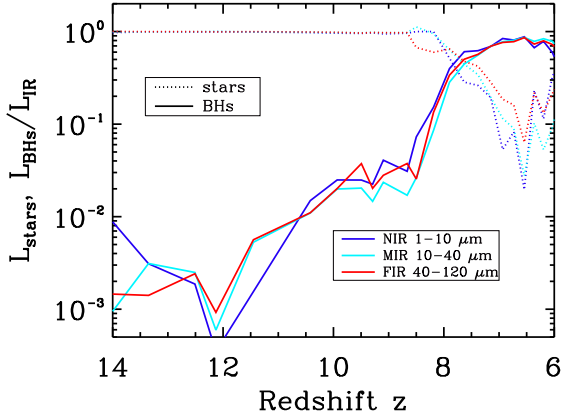} 
\vspace{1cm}
\caption{Contribution to the infrared luminosities, $\Lnir$, $\Lmir$, and
    $\Lfir$, from both AGN and stars, respectively. During the 
    starburst phase, stars are the main heating source for the infrared
    emission. However, during the peak quasar phase, AGN heating dominates. 
     } 
\label{Fig_lir_stars_bhs}
\end{center}
\end{figure}

\begin{figure}
\begin{center}
\includegraphics[width=3.2in]{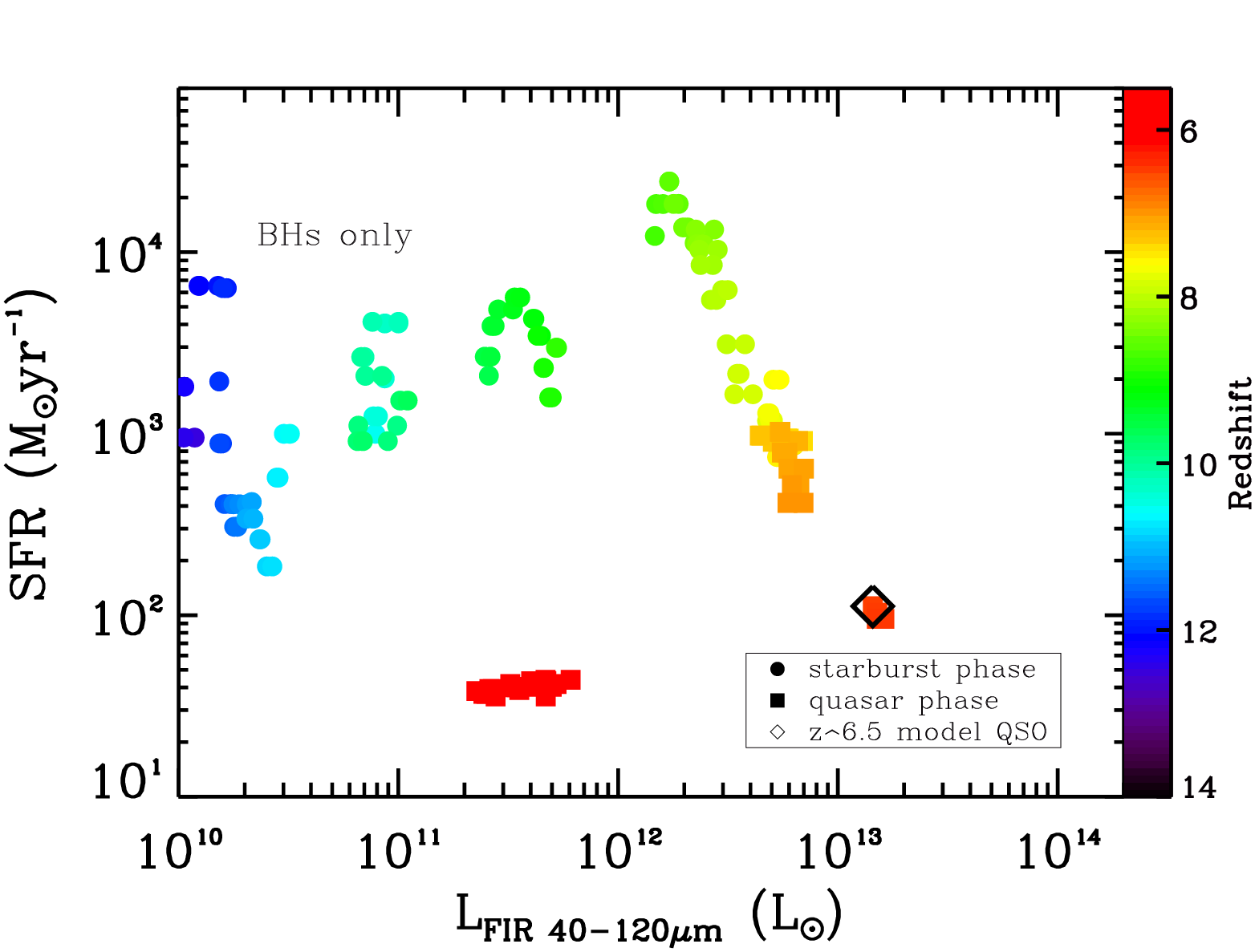} \\
\vspace{0.5cm}
\includegraphics[width=3.2in]{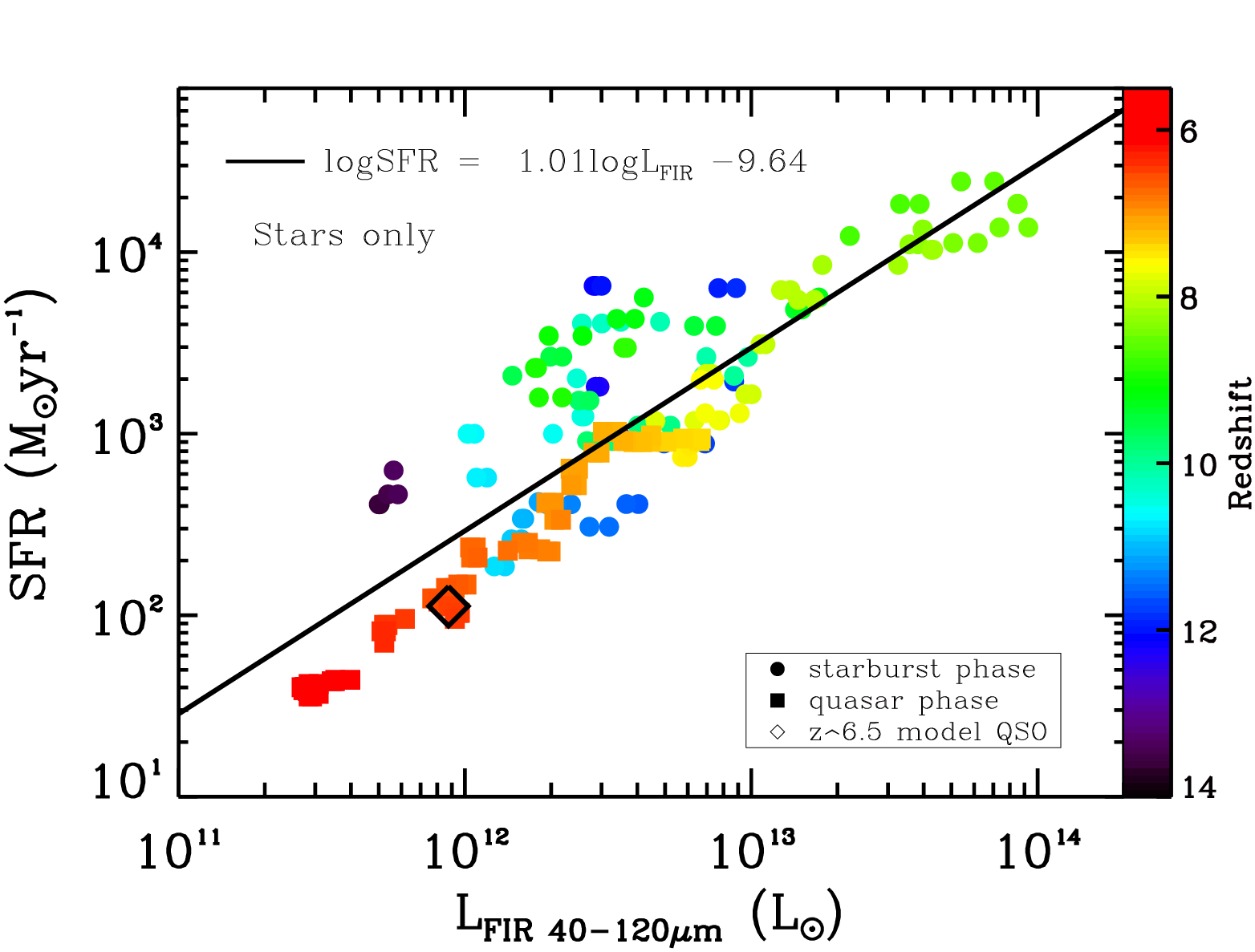} \\
\vspace{0.5cm}
\includegraphics[width=3.2in]{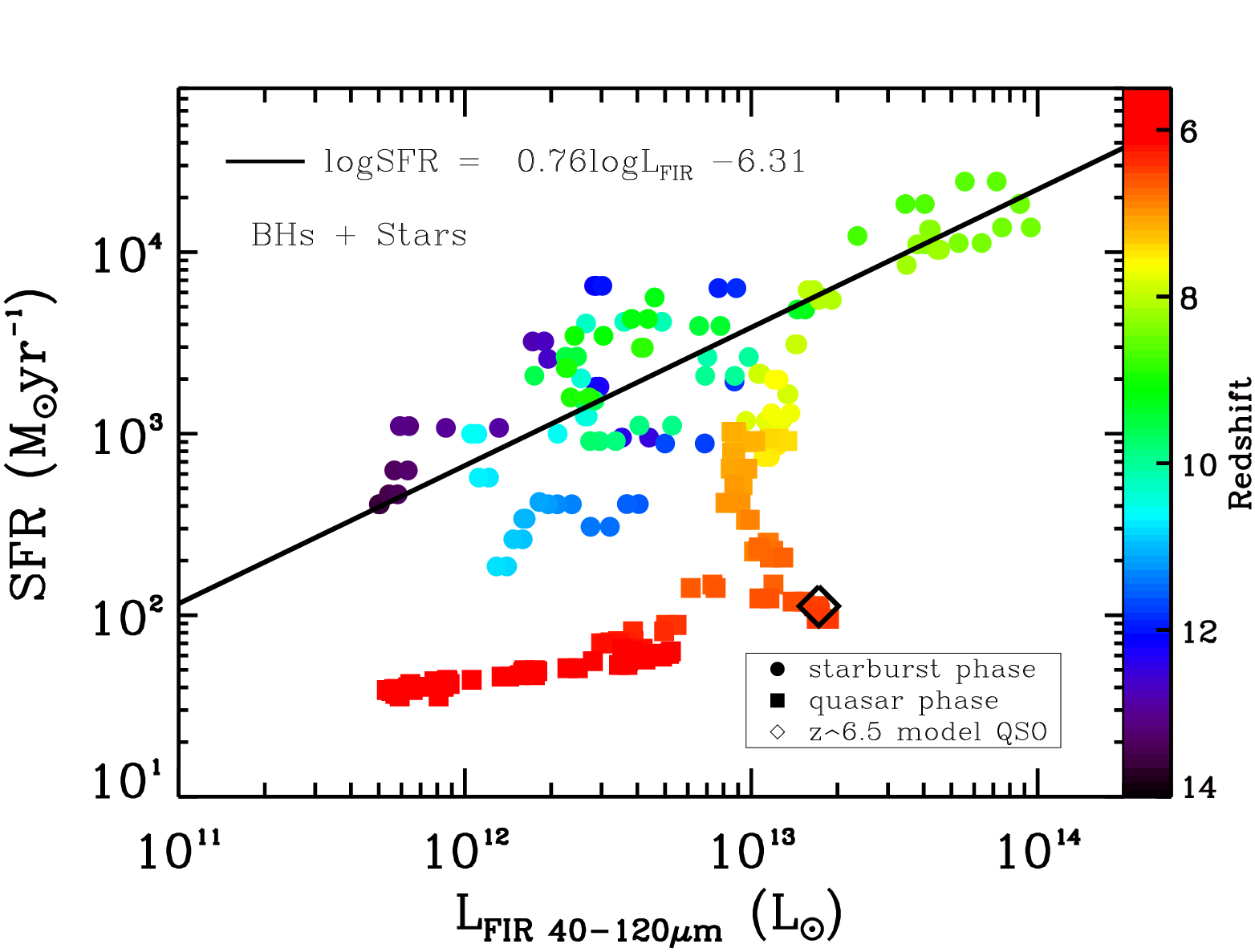} \\
\vspace{1cm}
\caption{Evolution of the SFR -- $\Lfir$ relation in our
simulations. Here we consider the relation in three cases where $\Lfir$ is
contributed by AGNs only (case 1, top panel), stars only (case 2, middle 
panel), and both AGNs and stars (case 3, bottom panel). The colored filled
symbols indicate the system at different redshifts, while the black open diamond
represents the model quasar at $z=6.5$.  The black curve in the middle panel
is the least-squares fit to all the data, while that in the bottom panel is the
fit to data points from $z \sim 14 - 7.5$ only (starburst phase). Case 1 has no
SFR -- $\Lfir$ correlation; case 2 has a tight, linear correlation similar to
that used in observations; and case 3 has a non-linear correlation.} 
\label{Fig_sfr_lir}
\end{center}
\end{figure}

AGN contamination in infrared observations of dusty, star-forming
quasar systems has been a long-standing problem. In our simulations, we
see that the contributions from AGNs and stars both vary according to
the activity of these two populations. As shown in
Figure~\ref{Fig_lir_stars_bhs}, during the starburst phase, not
surprisingly nearly all the infrared emission comes from dust heated
by stars. However, during the peak quasar phase, the contribution from
the AGN dominates the infrared light production in the system. This result
has significant implications for the interpretation of observational
data, such as estimates of the star formation rate.

In particular, owing to the lack of other indicators such as UV flux
or $\rm H{\alpha}$ emission, in observations of high-redshift objects
the far-infrared luminosity is commonly used to estimate the star
formation rate, assuming most or all of $\Lfir$ is contributed by
young stars.  For example, \zquasar\ has $\Lfir \sim 2\times 10^{13}\, \Lsun$, 
and the star formation rate is estimated to be $\sim 3\times 10^3\,
\Msun\, \yr^{-1}$ \citep{Bertoldi2003A, Carilli2004}. However, these
studies cannot rule out a substantial contribution (e.g. $\sim 50\%$) to
$\Lfir$ by the AGN. If $\Lfir$ is indeed heavily contaminated by AGN
activity, then the SFR -- $\Lfir$ relation would be invalid, and the actual 
star formation rate would be much lower.
 
To demonstrate this, we show in Figure~\ref{Fig_sfr_lir} the SFR -- $\Lfir$
relation in three cases: (1) only black holes are included as a 
heating source of the dust (top panel); (2) only stars are included (middle
panel); and (3) both black holes and stars are included (bottom panel). The
top panel shows that there is no correlation between the SFR and the $\Lfir$
which is produced by AGNs. However, if all the $\Lfir$ is 
produced solely from stars, then there is a tight, linear correlation. If the
$\Lfir$ is contributed by both AGN and stars, then the correlation would
change. The bottom panel in Figure~\ref{Fig_sfr_lir} shows that during the
starburst phase where stars dominate dust heating, there is
still a correlation between SFR and $\Lfir$ with modified slope and
normalization. However, during the peak quasar phase where the 
AGN dominates dust heating, there is no correlation at all.  To summarize,
Figure~\ref{Fig_sfr_lir} gives the following SFR -- $\Lfir$ relations:  

\begin{eqnarray}
\label{eq:SFR} 
\frac{\rm SFR}{\Msun\, \yr^{-1}} & = & 2.3\times 10^{-10}\left(\frac{L_{\rm
FIR}}{\Lsun}\right)^{1.01}\; (\rm {stars-only}) \, ,\\
\frac{\rm SFR}{\Msun\, \yr^{-1}} & = & 4.9\times 10^{-7}\left(\frac{L_{\rm
FIR}}{\Lsun}\right)^{0.76}\; (\rm {AGN + stars}) \, .
\end{eqnarray} 

Here, the FIR is in the range of $40 - 120\mu$~m (rest frame). This
correlation in the stars-only case is very close to that found by
\cite{Kennicutt1998B}, $\frac{\rm SFR}{\Msun\, \yr^{-1}} \simeq 1.7\times
10^{-10}\frac{\Lfir}{\Lsun}$ for starburst galaxies. Note the slight 
difference in the normalization might owe to different stellar IMFs used --
we use the top heavy Kroupa IMF \citep{Kroupa2002}, while \cite{Kennicutt1998B}
adopts a Salpeter IMF \citep{Salpeter1955}.  Also, the $\Lfir$ in that work
refers the infrared luminosity integrated over the range of $8 -
1000\mu$~m.

In our simulation, the model quasar at $z=6.5$ has $\Lfir \simeq
1.8\times10^{13}\, \Lsun$, which is close to that of \zquasar\ derived 
from radio observation \citep{Bertoldi2003A, Carilli2004}. If we assume a 
perfect SFR -- $\Lfir$ correlation as in the middle panel of
Figure~\ref{Fig_sfr_lir} (Equation~[\ref{eq:SFR}]), then the SFR would be $\sim 
4\times 10^3\, \Msun\, \yr^{-1}$, similar to the estimate of \cite{Bertoldi2003A,
  Carilli2004}. However, the SFR of the model quasar from our
simulations is about $\sim 112\, \Msun\, \yr^{-1}$, which is significantly
lower than that derived from an ideal SFR -- $\Lfir$ correlation. This
discrepancy is largely due to significant contribution to 
the $\Lfir$ by AGN activity. Several studies by \cite{Bertoldi2003A,
Carilli2004} and \cite{Wang2007} find that most $z \sim 6$ quasars
do not seem to show a $\Lfir$ -- $\LB$ correlation, but instead follow a
strong $\Lfir$ -- $\Lradio$ correlation as measured in local star-forming
galaxies \citep{Condon1992}. These authors therefore suggest that these $z
\sim 6$ quasars are strong starbursts, and that most of the $\Lfir$ may come
from stars. However, the $\Lfir$ -- $\Lradio$ correlation may not guarantee the
stellar origin of the $\Lfir$ as the physical basis for this correlation is
unknown. Our results suggest that in quasar systems, $\Lfir$ alone may no
longer be a reliable estimator for star formation rates; other diagnostics
should also be considered. We will study this topic with more detail in a
future paper by investigating the correlations in a multiple luminosity plane,
$\Lb$ -- $\Lfir$ -- $\Lx$ -- $\Lradio$, of starburst galaxies and quasars.

\subsection{Evolution of the Dust Properties}

\begin{figure}
\begin{center}
\includegraphics[width=3.5in]{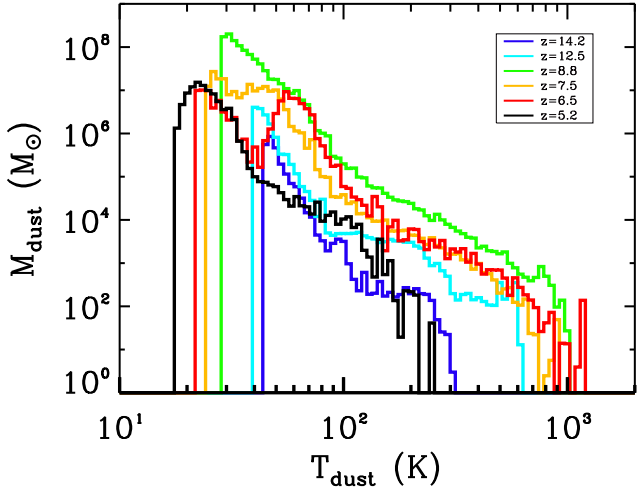}
\vspace{0.5cm}
\caption{Evolution of the dust mass and temperature as a function of
  redshift. The dust is heated by both stars and AGN. As the system evolves
  from starburst to quasar phases, the amount of hot dust increases. The dust
  reaches the highest temperature during the peak quasar phase. After that, the
  dust cools owing to the decline of AGN radiation.}   
\label{Fig_mdust_z}
\end{center}
\end{figure}

Figure~\ref{Fig_mdust_z} shows the evolution of the dust properties,
including the dust mass and temperature, as a function of redshift. As
the system evolves from starburst to quasar phases, the amount of hot
dust increases accordingly owing to the enhanced heating from the
central AGN, as well as replenishment of gas /dust from incoming new
galaxy progenitors during mergers. The amount of cold dust
($T \lesssim 100$ K) reaches a maximum at $z \sim 8.8$ when the last major
merger takes place, then decreases steadily as the dust is heated to
higher temperatures, or as the gas is consumed by star formation and black
hole accretion. Strong star formation is able to heat the dust nearly
to $T \sim 1000$~K. However, the hottest dust at $T \sim 1200$~K 
is associated only with the peak quasar phase at $z \sim 6.5$ when the hot
dust is directly heated by the AGN. After that, the amount of hot dust drops
dramatically as a result of the rapidly declining radiation from the
central AGN.

\begin{figure}
\begin{center}
\includegraphics[width=3.0in]{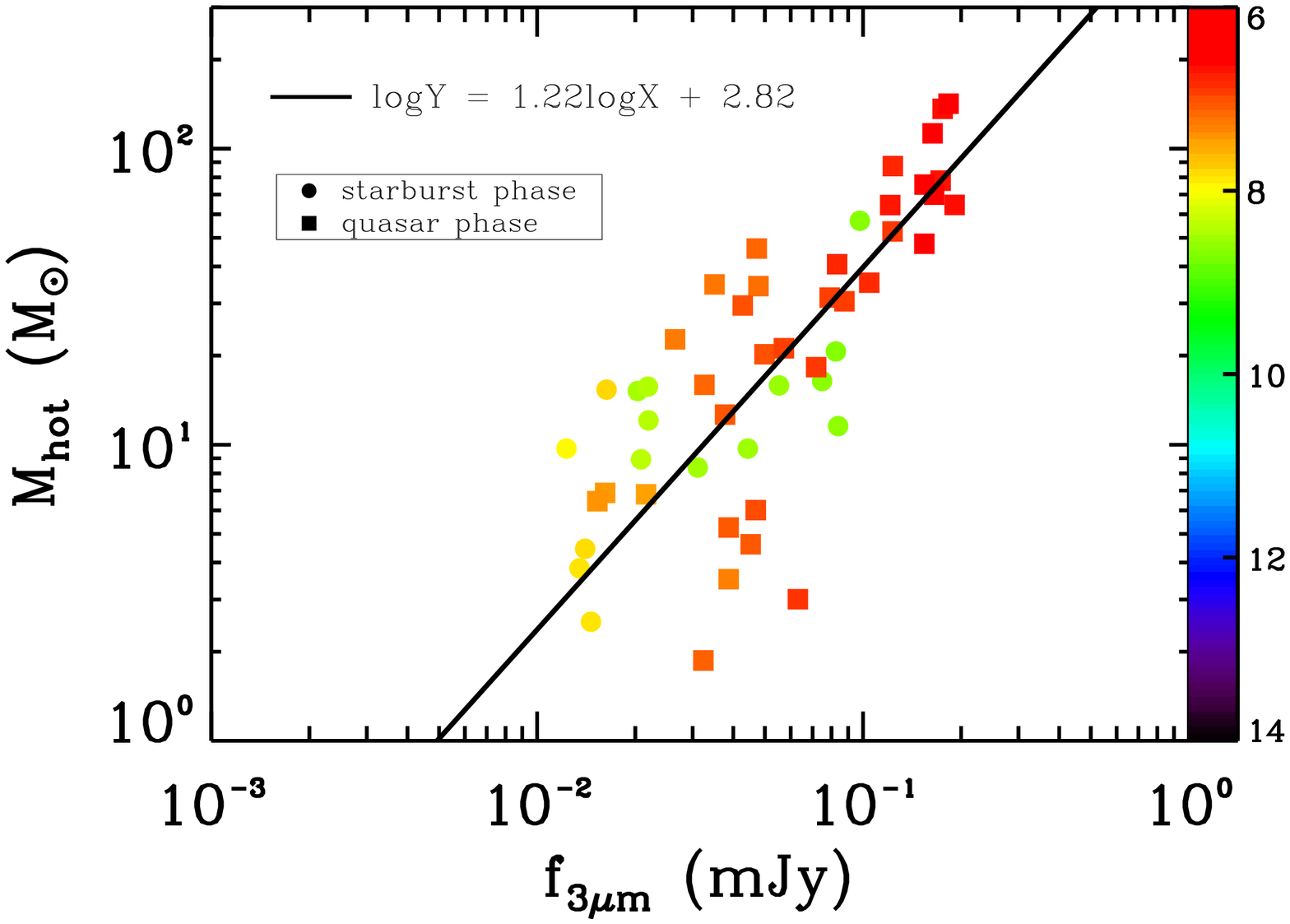}\\
\vspace{0.5cm}
\includegraphics[width=3.0in]{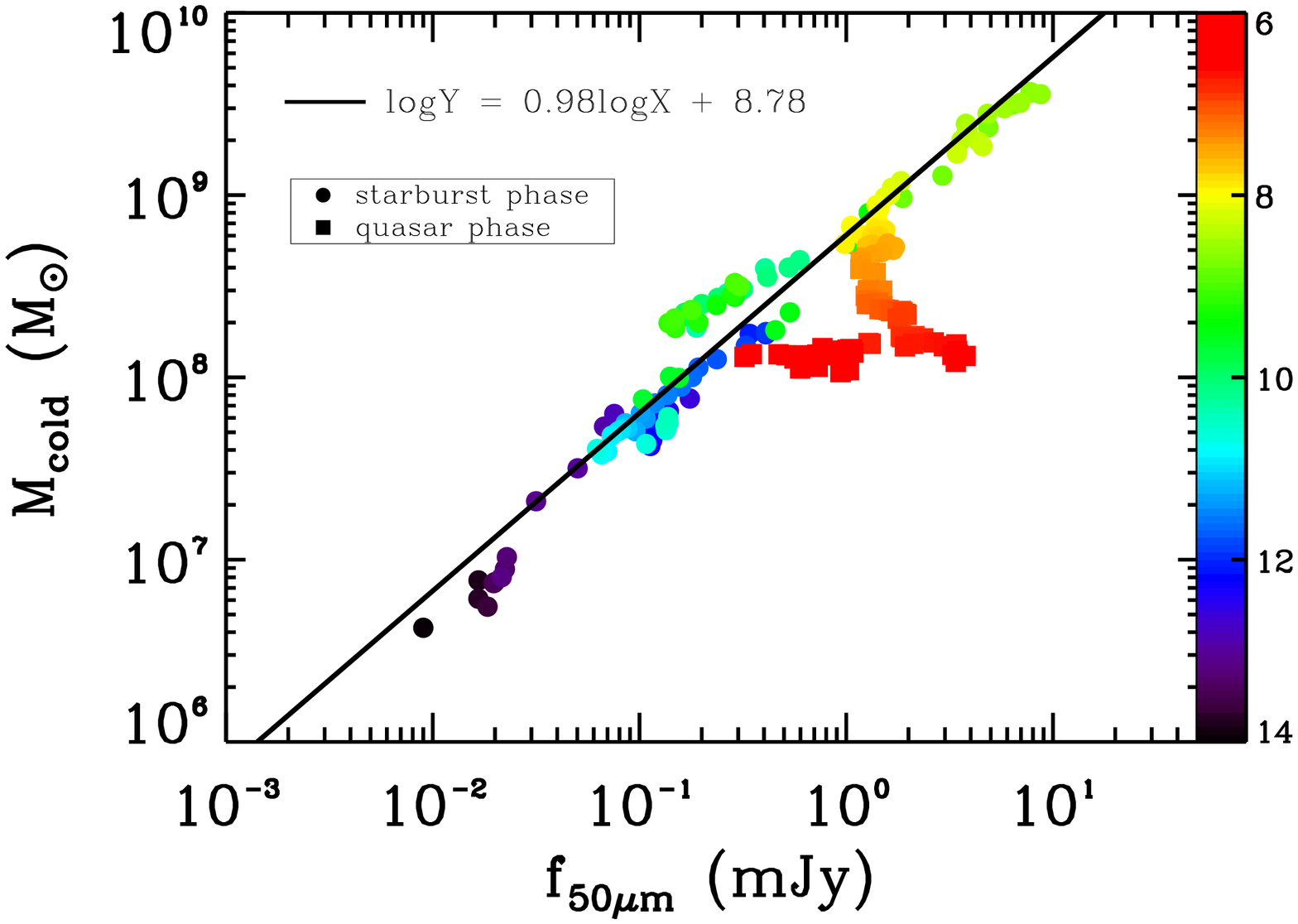}
\vspace{0.8cm}
\caption{Relations between rest-frame flux $f_{3 \mu\rm m}$ and hot dust
  mass $M_{\rm hot}$ (top panel), and $f_{50 \mu\rm m}$) and cold dust mass
  $M_{\rm cold}$ (bottom panel). The black curve is the least-squares fit to
  all the data in top panel, and data during starburst phase at redshift
  $z\sim 14 - 7.5$ in bottom panel.}  
\label{Fig_f3_f50_mdust}
\end{center}
\end{figure}

In observations, the dust mass is usually determined by assuming that the dust
radiates as a black- or grey-body \citep{Hughes1997, Jiang2006}. In the
emergent SEDs from our radiative transfer calculations, the cold dust 
bumps at different redshifts appear to peak around $\sim 50 \mu\rm m$
(rest-frame), while the hot dust bumps peak around $\sim 3 \mu\rm
m$. Figure~\ref{Fig_f3_f50_mdust} shows the relations between the dust mass
and fluxes at these two wavelengths. The top panel shows that $M_{\rm hot}$
($T \gtrsim 10^3$~K) increases with $f_{3\mu \rm m}$ flux, while $M_{\rm
cold}$ ($T \lesssim 10^2$~K) correlates with $f_{50\mu \rm m}$
flux linearly during the starburst phase:

\begin{eqnarray} 
\frac{M_{\rm hot}}{\Msun} & = & 4.2\times 10^{2}\left(\frac{f_{3\mu \rm m}}{\rm
  mJy}\right)^{1.22} \,, \\
\frac{M_{\rm cold}}{\Msun}& = & 6.0\times 10^{8}\left(\frac{f_{50\mu \rm
  m}}{\rm mJy}\right)^{0.98}\,. 
\end{eqnarray} 

However, during the quasar phase, there appears to be an excess in
$f_{50\mu \rm m}$, so the $M_{\rm cold}$ -- $f_{50\mu \rm m}$
correlation does not apply. These results suggest that rest-frame
$3\mu \rm m$ and $50\mu \rm m$ might serve as good diagnostics for
dust, the former for hot dust in quasar systems, and the latter for
cold dust in starburst galaxies.

\begin{figure}
\begin{center}
\includegraphics[width=3.0in]{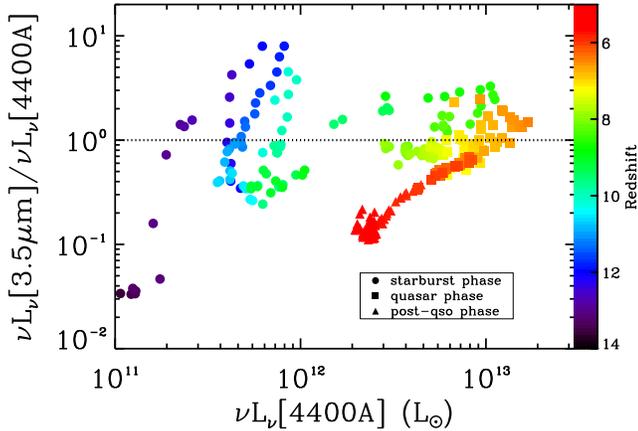} 
\vspace{1cm}
\caption{Evolution of the relation between rest-frame 3.5 $\mu$m
  luminosity and B-Band luminosity (4400 \AA) for the quasar system in our
  simulations. The filled circles represent starburst phases, squares
  represents quasar phases, and triangles represent post-quasar phases. Colors 
  indicates redshift of the object. } 
\label{Fig_NIR_B}
\end{center}
\end{figure}

It has been suggested by \cite{Jiang2006} that the NIR-to-optical flux
ratio may be used to probe dust properties. From a sample of
thirteen $z\sim 6$ quasars observed with {\em Spitzer}, these authors
find that two quasars have a remarkably low flux ratio (rest-frame
$3.5\mu \rm m$ to B-band) compared to other quasars at different
redshifts, and they are also weak in FIR. Furthermore, such a low flux
ratio was not seen in low-redshift quasar counterparts. These findings lead to
the suggestion that these two quasars may have different dust properties from
others \citep{Jiang2006}.  

From our model, we find that the dust properties are associated with the 
evolutionary stages of the host. Figure~\ref{Fig_NIR_B} shows the evolution of  
the ratio $\Lnir/\Lb$ from the quasar system at different stages of its
life. Early on, during the starburst phase, the ratio $\Lnir/\Lb$ is low. As
the system proceeds to its peak quasar activity, the heating from
the central AGN increases the hot dust emission around $3\, \mu$m,
boosting the $\Lnir/\Lb$ ratio. After that, in the post-quasar
phase, as the radiation from the AGN declines, the hot dust emission
drops rapidly, resulting in a lower $\Lnir/\Lb$.
According to our model, there might be two possible explanations for the two 
outliers in the \cite{Jiang2006} sample. One possibility is that they may be
young quasars that are still in the starburst phase but have not yet reached
peak quasar activity, so the light from star formation may be dominant or
comparable to that from the accreting SMBH still buried in dense gas. This may
explain the low NIR flux, as well as the B-band luminosity and the narrow
${\rm Ly}{\alpha}$ emission line, which are primarily produced by the
starburst. From Figure~\ref{Fig_LIR_color}, we note that there are ``valleys''
between major starbursts (or mergers) where the infrared luminosities are
low. This may explain the weak FIR in these two quasars. Another possibility
is that these two objects may be in the post-quasar phase, which would also
have low $\Lnir/\Lb$ and FIR (see also Figure~\ref{Fig_LIR_color} and
\ref{Fig_sfr_lir}).   

However, our model predicts that the same evolution should apply to every
luminous quasar formed in gas-rich mergers. It cannot explain statistically 
why such a low $\Lnir/\Lb$ ratio is only found in the $z \sim 6$ quasars, but
not at low redshifts. Perhaps these two quasars indeed have unusual dust
properties that cannot be explained by our current model. More observations of
dust in high-z objects are needed to resolve the issues surrounding
the IR-weak quasars in \cite{Jiang2006}, and to test our model.

\section{Discussion}

Our three-dimensional Monte Carlo radiative transfer code ART$^2$
makes it possible to calculate self-consistently the radiative transfer and
dust emission from galaxies and quasars which have large dynamic ranges and
irregular geometries. It works efficiently on hydrodynamics simulations of  
galaxies and mergers, and has flexible grid resolution that can be
adjusted to any SPH resolution in hydrodynamic simulations. 

To date, only a limited number of RT calculations have been performed on
galaxy mergers \citep{Jonsson2006, Chakrabarti2007A, Chakrabarti2007B}. In
particular, the parallel SUNRISE code developed by \cite{Jonsson2006} marked a  
milestone in this direction. It has an adaptive grid similar to ours
and works efficiently on the arbitrary geometry in galaxy mergers. However,
radiative equilibrium is not yet included in this code, so SUNRISE can not
currently calculate dust emission self-consistently. 

On the other hand, \cite{Chakrabarti2007A, Chakrabarti2007B} had a similar
approach to ours. They apply the code of \cite{Whitney2003A}, which employs
the same radiative equilibrium algorithm as in \cite{Bjorkman2001}, to
hydrodynamic simulations of galaxy mergers with black holes. However, the
spherical, logarithmically-spaced grid used by these authors is not optimal
for describing an inhomogeneous distribution of gas and dust with multiple
density centers in galaxies and mergers, as exemplified by
Figure~\ref{Fig_grid}. Moreover, our methodology is based on a treatment of
the multiphase ISM that includes extinction from dust in the diffuse phase,
which was ignored in the work of \cite{Chakrabarti2007A, Chakrabarti2007B}.
Furthermore, we adopt an AGN input spectrum that includes emission from dust
on scales that the simulations do not resolve, unlike \cite{Chakrabarti2007A,
Chakrabarti2007B} who employ a power-law AGN spectrum.  The tests
described above demonstrate that the near- and mid-IR emission is sensitive to 
the resolution of the RT calculation (Figure~\ref{Fig_res}),
extinction by dust in the diffuse phase of the ISM (Figure~\ref{Fig_phase}),
and the template spectrum of the AGN (Figure~\ref{Fig_inspec}).  The SEDs in
the work of \cite{Chakrabarti2007A, Chakrabarti2007B} show only cold dust
emission characteristic of a starburst during all evolutionary phases, even
when the quasar peaks (e.g., Figure~20 in \citealt{Chakrabarti2007B}), and,
in particular, do not at any time exhibit hot dust emission.  We
attribute the differences between our computed SEDs and those of
\cite{Chakrabarti2007A, Chakrabarti2007B} mainly to the different grid method
employed, the handling of extinction by the diffuse ISM, and our choice of an
AGN input spectrum that includes emission on scales that cannot be resolved in
our hydrodynamic simulations.

In our radiative transfer calculations, we do not explicitly include a dust
torus near the AGN. The resolution of our RT grid is constrained by
that of the hydrodynamic simulations, which is of $\sim 30$~pc. As a result,
we can not resolve the near vicinity of the AGN. In the Unification Scheme
for AGN \citep{Miller1983}, a parsec-scale dust torus, both optically and
geometrically thick, has been used to successfully explain the different
appearances of Type 1 and Type 2 AGNs. For example, Type 1 AGNs are typically
viewed face-on and show blue UV bump in their spectra, while Type 2 AGNs are
viewed edge-on without the blue UV bump. It is believed that the NIR and MIR
emission of an AGN comes from the hot dust in the torus structure, as
suggested by various observations, for examples, the MIR and NIR detection of a
Seyfert 2 galaxy NGC 1068 \citep{Jaffe2004, Wittkowski2004}, and the MIR
emission of a lensed quasar Q2237+0305 by \cite{Wyithe2002} in the CfA
Redshift Survey \citep{Huchra1985}. The radiative transfer modeling by 
\cite{Hoenig2006}, which concentrates on the inner parsecs of the AGN, has
successfully reproduced the SED of NGC 1068 in the NIR and MIR with a
three-dimensional, clumpy dust torus. Such a scale is, however, below the
resolution of our modeling, which follows the dynamical evolution of the
quasar system on a galactic scale. Moreover, it is not clear whether such a
well-defined torus structure would exist in luminous quasars at high
redshifts. Nevertheless, we have adopted a composite AGN spectrum
\citep{Hopkins2007A} as a sub-resolution recipe for the IR emission from the 
hottest dust near the AGN. Such an approach is supported by the fact 
that the composite spectrum represents the average near- to mid-IR emission in 
thousands of quasars \citep{Richards2006}, and that the torus does not
contribute much to the far-IR emission, which comes mainly from cold dust on
kiloparsec scales. Therefore, for spatial scales of interest, the
results should be reliable, and they are expected be similar to those from
calculations with a torus included. We defer such a calculation to the future
when sub-parsec resolution is feasible in hydrodynamic simulations of quasar
formation.       

We should point out that dust destruction is not explicitly included in our
modeling. It has been suggested (e.g., \citealt{McKee1989, Draine1993, 
Draine2003, Nozawa2006}) that dust destruction might be efficient in
non-radiative shocks by non-thermal and thermal sputtering owing to a high
shock velocity ($> 100$ km s$^{-1}$) and a high gas temperature ($> 10^6$
K). However, the sputtering timescale for this temperature is of $\sim 3\times
10^6$ yr for a gas density $n=1\, {\rm cm}^{-1}$, and even shorter if the gas
density is higher \citep{Burke1974, Reynolds1997}. It has also been suggested
that dust grains may be destroyed behind radiative shocks
\citep{Todini2001, Bianchi2007}. However, it is expected that the efficiency cannot 
be significant at high redshifts as magnetic fields are likely weak
\citep{Gnedin2000}. In our dust treatment in the RT analysis, the
dust is re-calculated based on the gas content in every snapshot
from the hydrodynamic simulations, which has a time interval of 2
Myr. This is equivalent to dust destruction within this
timescale. Moreover, we assume different dust-to-gas ratios for the
cold and hot gas, which accounts for dust survival in
these different phases of the interstellar medium.

One limitation of our ART$^2$ code is that it does not include the transient
heating of small dust grains (size $<200${\AA}), which may cause temperature
fluctuation and produce enhanced near-infrared emission \citep{Li2001,
Misselt2001}, in particular for the PAH feature. Although PAH is not
studied in this present work, the code can be improved in the future by
implementing a temperature distribution for those small grains. Moreover, 
one potential caveat of our RT work is that it is a post-processing
procedure. It would be ideal to couple the radiation with
hydrodynamics; i.e., having the radiative transfer done simultaneously
with the hydrodynamic simulations (as in e.g., \citealt{Yoshida2007})
as the dynamical evolution of the gas might affect the formation and
properties of the dust. However, this effect is currently uncertain because
such a technique is not yet available for dust emission in galaxy
simulations. Although ART$^2$ is not parallelized, it works fairly efficiently 
owing to the adaptive refinement tree. 

We have done extensive resolution studies which include both the number of
photon packets and the refinement level of the grid, and parameters for the
ISM model such as the cloud mass spectrum and size distribution, as well as
different dust models. We find convergence of the modeled SEDs with parameters
within observational ranges (see \S~2 and \S~3). Therefore, we conclude that
the RT calculations we present here with ART$^2$ are robust, and that our
model is more self-consistent, more realistic, and more versatile than
previous approaches.

\section{Summary}

We have implemented a three-dimensional Monte Carlo radiative transfer
code, ART$^2$ -- All-wavelength Radiative Transfer with Adaptive Refinement
Tree, and use it to calculate the dust emission and multi-wavelength
properties of quasars and galaxies. ART$^2$ includes the following essential
implementations: 

\begin{enumerate}

\item A radiative equilibrium algorithm developed by \cite{Bjorkman2001}. It
  conserves the total photon energy, and corrects the dust temperature
  without iteration. This algorithm calculates dust emission efficiently and
  self-consistently.  

\item An adaptive grid scheme in 3-D Cartesian coordinates similar to that of
  \cite{Jonsson2006}, which handles an
  arbitrary geometry and covers a large dynamic range over several orders of
  magnitude. This is indispensable for capturing the inhomogeneous and clumpy
  density distribution in galaxies and galaxy mergers. It easily achieves a
  $\sim$10 pc-scale grid resolution equivalent to a $\sim 10000^3$ uniform grid,
  which is prohibitive with current computation schemes.  

\item A two-phase ISM model in which the cold, dense clouds are embedded in
  a hot, diffuse medium in pressure equilibrium
  \citep{Springel2003A}. Moreover, the cold clouds follow a mass spectrum and a  
  size distribution similar to the \cite{Larson1981} scaling relations
  inferred for giant molecular clouds. This model ensures an appropriate
  sub-grid recipe for the ISM physics, which is important for studying dust
  properties in galaxies.  

\item A supernova-origin dust model in which the dust is produced by Type-II
  supernovae, and the size distribution of grains follows that derived by
  \cite{Todini2001}. This model may be especially  relevant for dust in
  high-redshift, young objects. In traditional dust models, the dust is
  produced by old, low mass stars such as AGB stars, which are typically over
  1 Gyr old. However, quasar systems at $\sim 6.4$ are only a few hundred
  Myr old, and there would be insufficient AGB stars to produce the abundant dust
  as observed in \zquasar. 

\item The input spectra include those from both stars, calculated
  using STARBURST99 \citep{Leitherer1999, Vazquez2005}, and black holes
  represented by a composite AGN template developed by
  \cite{Hopkins2007A}. The composite spectrum includes a broken power-law  
  as the intrinsic black hole spectrum, and infrared components that come from 
  the torus near the AGN. The torus structure is unresolved in our
  hydrodynamic simulations. This template is a sub-grid recipe to include the
  hot dust emission within pc-scales which is below the resolution   
  of our hydrodynamic simulations.   
 
\end{enumerate}

ART$^2$ works efficiently on hydrodynamic simulations of galaxies and
quasars performed with GADGET2 \citep{Springel2005D}. By applying
ART$^2$ to the quasar simulations of \cite{Li2007}, in which luminous
quasars at $z \gtrsim 6$ form rapidly through hierarchical mergers of
gas-rich galaxies, we are able to calculate the multi-wavelength SEDs (from
optical to millimeter) and dust properties of the model quasar at $z
\sim 6.5$ and its galaxy progenitors at even higher redshifts.

We find that a supernova-origin dust model may be able to explain the
dust properties as observed in the high-redshift quasars. Our
calculations reproduce the observed SED and properties such as the
dust mass and temperature of \zquasar, the most distant Sloan quasar.
The dust and infrared emission in quasar hosts are closely associated
with the formation and evolution of the system. As the system
transforms from starburst to quasar phases, the evolution of the SEDs
is characterized by a transition from a cold to warm ULIRG. During the
starburst phase at $z \gtrsim 7.5$, the SEDs of the quasar progenitors
exhibit cold dust bumps (peak at $\sim 50\, \mu$m rest frame) that are
characteristic of starburst galaxies. During the peak quasar phase ($z
\sim 7.5 - 6$), the SEDs show a prominent hot dust bump (peaks at $\sim
3\, \mu$m rest frame) as observed in luminous quasars at both low and high
redshifts (e.g, \citealt{Richards2006, Jiang2006}).  

Furthermore, we find that during the quasar phase, AGN activity
dominates the heating of dust and the resulting infrared
luminosities. This has significant implications for the interpretation
of observables from the quasar host. The hottest dust ($T \gtrsim
10^3$~K) is present only during the peak quasar activity, and
correlates strongly with the near-IR flux. The SFR -- $\Lfir$
correlation depends sensitively on the relative heating contributed from AGN
and stars. If the dust heating is dominated by stars, as in the
starburst phase, then there is a tight, linear SFR -- $\Lfir$
correlation similar to the one widely used to interpret observations
\citep{Kennicutt1998B}. However, if both AGN and stars contribute to
the dust emission in the far-IR, then the correlation becomes
non-linear with a modified normalization.  If the AGN dominates the
$\Lfir$, as in the peak quasar phase, we find no correlation at all.
Finally, we find correlations between dust masses and rest-frame
fluxes at $3\, \mu$m and $50\, \mu$m. The $f_{3\, \mu \rm m}$ flux may
serve as a good diagnostic for hot dust in quasars, while $f_{50\,
\mu \rm m}$ may be used to estimate the amount of cold dust in
starburst galaxies, as $f_{50\, \mu \rm m}$ correlates linearly with
the cold dust mass.

Our model demonstrates that massive star formation at higher
redshifts ($ z \gtrsim 10$) is necessary in order to produce the observed
dust properties in these quasars. This suggests that the quasar
hosts should have already built up a large stellar population by $z
\sim 6$. Our results support a merger-driven origin for luminous
quasars at high redshifts, and provide further evidence for the
hypothesis of starburst-to-quasar evolution.

\acknowledgments{Special thanks to Kenny Wood\footnotemark[2] and Barbara
  Whitney\footnotemark[3] for generously making their Monte Carlo radiative
  equilibrium codes publicly available;
  \footnotetext[2]{http://www-star.st-and.ac.uk/$\sim$kw25/research/montecarlo/montecarlo.html}
  \footnotetext[3]{http://gemelli.colorado.edu/$\sim$bwhitney/codes/codes.html}
  and to Patrik Jonsson for sharing publicly his fantastic
  radiative transfer code SUNRISE\footnotemark[4]
  \footnotetext[4]{http://www.ucolick.org/$\sim$patrik/sunrise/}.  
  We thank Gurtina Besla, Stephanie Bush, Sukanya Chakrabarti,
  Suvendra Dutta, Giovanni Fazio, Jiasheng Huang, Dusan Keres, Desika
  Narayanan, Erik Rosolowsky and Josh Younger for many stimulating discussions. 
  YL thanks Lee Armus, Frank Bertoldi, Bruce Draine, Patrik Jonsson, 
  Carol Lonsdale, Mike Nolta, Casey Papovich, George Rybicki, Nick
  Scoville, Rodger Thompson and Lin Yan for inspiration and
  discussions.  Finally, we thank the referee for a thoughtful report that has
  helped improve this manuscript. YL gratefully acknowledges a Keck
  Fellowship sponsored by the Keck Foundation, as well as an Institute for
  Theory and Computation Fellowship, under which support most of this work was
  done. The computations reported here were performed at the Center for 
  Parallel Astrophysical Computing at Harvard-Smithsonian Center for
  Astrophysics. This work was supported in part by NSF grant 03-07690 and NASA 
  ATP grant NAG5-13381.}   

\bibliography{ms}

\begin{thebibliography}{248}
\expandafter\ifx\csname natexlab\endcsname\relax\def\natexlab#1{#1}\fi

\bibitem[{{Abel} {et~al.}(2002){Abel}, {Bryan}, \& {Norman}}]{Abel2002}
{Abel}, T., {Bryan}, G.~L., \& {Norman}, M.~L. 2002, Science, 295, 93

\bibitem[{{Andre} {et~al.}(1996){Andre}, {Ward-Thompson}, \&
  {Motte}}]{Andre1996}
{Andre}, P., {Ward-Thompson}, D., \& {Motte}, F. 1996, \aap, 314, 625

\bibitem[{{Armus} {et~al.}(2004){Armus}, {Charmandaris}, {Spoon}, {Houck},
  {Soifer}, {Brandl}, {Appleton}, {Teplitz}, {Higdon}, {Weedman}, {Devost},
  {Morris}, {Uchida}, {van Cleve}, {Barry}, {Sloan}, {Grillmair}, {Burgdorf},
  {Fajardo-Acosta}, {Ingalls}, {Higdon}, {Hao}, {Bernard-Salas}, {Herter},
  {Troeltzsch}, {Unruh}, \& {Winghart}}]{Armus2004}
{Armus}, L., {Charmandaris}, V., {Spoon}, H.~W.~W., {Houck}, J.~R., {Soifer},
  B.~T., {Brandl}, B.~R., {Appleton}, P.~N., {Teplitz}, H.~I., {Higdon},
  S.~J.~U., {Weedman}, D.~W., {Devost}, D., {Morris}, P.~W., {Uchida}, K.~I.,
  {van Cleve}, J., {Barry}, D.~J., {Sloan}, G.~C., {Grillmair}, C.~J.,
  {Burgdorf}, M.~J., {Fajardo-Acosta}, S.~B., {Ingalls}, J.~G., {Higdon}, J.,
  {Hao}, L., {Bernard-Salas}, J., {Herter}, T., {Troeltzsch}, J., {Unruh}, B.,
  \& {Winghart}, M. 2004, \apjs, 154, 178

\bibitem[{{Baes} {et~al.}(2005){Baes}, {Stamatellos}, {Davies}, {Whitworth},
  {Sabatini}, {Roberts}, {Linder}, \& {Evans}}]{Baes2005}
{Baes}, M., {Stamatellos}, D., {Davies}, J.~I., {Whitworth}, A.~P., {Sabatini},
  S., {Roberts}, S., {Linder}, S.~M., \& {Evans}, R. 2005, New Astronomy, 10,
  523

\bibitem[{{Ballesteros-Paredes} \& {Mac Low}(2002)}]{Ballesteros-Paredes2002}
{Ballesteros-Paredes}, J., \& {Mac Low}, M.-M. 2002, \apj, 570, 734

\bibitem[{{Barnes} \& {Hernquist}(1992)}]{Barnes1992}
{Barnes}, J.~E., \& {Hernquist}, L. 1992, \araa, 30, 705

\bibitem[{{Barnes} \& {Hernquist}(1996)}]{Barnes1996}
---. 1996, \apj, 471, 115

\bibitem[{{Barnes} \& {Hernquist}(1991)}]{Barnes1991}
{Barnes}, J.~E., \& {Hernquist}, L.~E. 1991, \apjl, 370, L65

\bibitem[{{Barth} {et~al.}(2003){Barth}, {Martini}, {Nelson}, \&
  {Ho}}]{Barth2003}
{Barth}, A.~J., {Martini}, P., {Nelson}, C.~H., \& {Ho}, L.~C. 2003, \apjl,
  594, L95

\bibitem[{{Beelen} {et~al.}(2006){Beelen}, {Cox}, {Benford}, {Dowell},
  {Kov{\'a}cs}, {Bertoldi}, {Omont}, \& {Carilli}}]{Beelen2006}
{Beelen}, A., {Cox}, P., {Benford}, D.~J., {Dowell}, C.~D., {Kov{\'a}cs}, A.,
  {Bertoldi}, F., {Omont}, A., \& {Carilli}, C.~L. 2006, \apj, 642, 694

\bibitem[{{Bertoldi} {et~al.}(2003{\natexlab{a}}){Bertoldi}, {Carilli}, {Cox},
  {Fan}, {Strauss}, {Beelen}, {Omont}, \& {Zylka}}]{Bertoldi2003A}
{Bertoldi}, F., {Carilli}, C.~L., {Cox}, P., {Fan}, X., {Strauss}, M.~A.,
  {Beelen}, A., {Omont}, A., \& {Zylka}, R. 2003{\natexlab{a}}, \aap, 406, L55

\bibitem[{{Bertoldi} {et~al.}(2003{\natexlab{b}}){Bertoldi}, {Cox}, {Neri},
  {Carilli}, {Walter}, {Omont}, {Beelen}, {Henkel}, {Fan}, {Strauss}, \&
  {Menten}}]{Bertoldi2003B}
{Bertoldi}, F., {Cox}, P., {Neri}, R., {Carilli}, C.~L., {Walter}, F., {Omont},
  A., {Beelen}, A., {Henkel}, C., {Fan}, X., {Strauss}, M.~A., \& {Menten},
  K.~M. 2003{\natexlab{b}}, \aap, 409, L47

\bibitem[{{Bianchi} {et~al.}(2000){Bianchi}, {Ferrara}, {Davies}, \&
  {Alton}}]{Bianchi2000}
{Bianchi}, S., {Ferrara}, A., {Davies}, J.~I., \& {Alton}, P.~B. 2000, \mnras,
  311, 601

\bibitem[{{Bianchi} \& {Schneider}(2007)}]{Bianchi2007}
{Bianchi}, S., \& {Schneider}, R. 2007, \mnras, 378, 973

\bibitem[{{Bjorkman} \& {Wood}(2001)}]{Bjorkman2001}
{Bjorkman}, J.~E., \& {Wood}, K. 2001, \apj, 554, 615

\bibitem[{{Blitz} {et~al.}(2007){Blitz}, {Fukui}, {Kawamura}, {Leroy},
  {Mizuno}, \& {Rosolowsky}}]{Blitz2007}
{Blitz}, L., {Fukui}, Y., {Kawamura}, A., {Leroy}, A., {Mizuno}, N., \&
  {Rosolowsky}, E. 2007, in Protostars and Planets V, ed. B.~{Reipurth},
  D.~{Jewitt}, \& K.~{Keil}, 81--96

\bibitem[{{Blitz} \& {Rosolowsky}(2006)}]{Blitz2006}
{Blitz}, L., \& {Rosolowsky}, E. 2006, \apj, 650, 933

\bibitem[{{Bondi}(1952)}]{Bondi1952}
{Bondi}, H. 1952, \mnras, 112, 195

\bibitem[{{Bondi} \& {Hoyle}(1944)}]{BondiHoyle1944}
{Bondi}, H., \& {Hoyle}, F. 1944, \mnras, 104, 273

\bibitem[{{Brand} {et~al.}(2006){Brand}, {Dey}, {Weedman}, {Desai}, {Le
  Floc'h}, {Jannuzi}, {Soifer}, {Brown}, {Eisenhardt}, {Gorjian}, {Papovich},
  {Smith}, {Willner}, \& {Cool}}]{Brand2006}
{Brand}, K., {Dey}, A., {Weedman}, D., {Desai}, V., {Le Floc'h}, E., {Jannuzi},
  B.~T., {Soifer}, B.~T., {Brown}, M.~J.~I., {Eisenhardt}, P., {Gorjian}, V.,
  {Papovich}, C., {Smith}, H.~A., {Willner}, S.~P., \& {Cool}, R.~J. 2006,
  \apj, 644, 143

\bibitem[{{Brandt} {et~al.}(2002){Brandt}, {Schneider}, {Fan}, {Strauss},
  {Gunn}, {Richards}, {Anderson}, {Vanden Berk}, {Bahcall}, {Brinkmann},
  {Brunner}, {Chen}, {Hennessy}, {Lamb}, {Voges}, \& {York}}]{Brandt2002}
{Brandt}, W.~N., {Schneider}, D.~P., {Fan}, X., {Strauss}, M.~A., {Gunn},
  J.~E., {Richards}, G.~T., {Anderson}, S.~F., {Vanden Berk}, D.~E., {Bahcall},
  N.~A., {Brinkmann}, J., {Brunner}, R., {Chen}, B., {Hennessy}, G.~S., {Lamb},
  D.~Q., {Voges}, W., \& {York}, D.~G. 2002, \apjl, 569, L5

\bibitem[{{Bromm} \& {Larson}(2004)}]{Bromm2004}
{Bromm}, V., \& {Larson}, R.~B. 2004, \araa, 42, 79

\bibitem[{{Burke} \& {Silk}(1974)}]{Burke1974}
{Burke}, J.~R., \& {Silk}, J. 1974, \apj, 190, 1

\bibitem[{{Calzetti} {et~al.}(2000){Calzetti}, {Armus}, {Bohlin}, {Kinney},
  {Koornneef}, \& {Storchi-Bergmann}}]{Calzetti2000}
{Calzetti}, D., {Armus}, L., {Bohlin}, R.~C., {Kinney}, A.~L., {Koornneef}, J.,
  \& {Storchi-Bergmann}, T. 2000, \apj, 533, 682

\bibitem[{{Calzetti} {et~al.}(1994){Calzetti}, {Kinney}, \&
  {Storchi-Bergmann}}]{Calzetti1994}
{Calzetti}, D., {Kinney}, A.~L., \& {Storchi-Bergmann}, T. 1994, \apj, 429, 582

\bibitem[{{Carilli} {et~al.}(2001){Carilli}, {Bertoldi}, {Rupen}, {Fan},
  {Strauss}, {Menten}, {Kreysa}, {Schneider}, {Bertarini}, {Yun}, \&
  {Zylka}}]{Carilli2001}
{Carilli}, C.~L., {Bertoldi}, F., {Rupen}, M.~P., {Fan}, X., {Strauss}, M.~A.,
  {Menten}, K.~M., {Kreysa}, E., {Schneider}, D.~P., {Bertarini}, A., {Yun},
  M.~S., \& {Zylka}, R. 2001, \apj, 555, 625

\bibitem[{{Carilli} {et~al.}(2004){Carilli}, {Walter}, {Bertoldi}, {Menten},
  {Fan}, {Lewis}, {Strauss}, {Cox}, {Beelen}, {Omont}, \&
  {Mohan}}]{Carilli2004}
{Carilli}, C.~L., {Walter}, F., {Bertoldi}, F., {Menten}, K.~M., {Fan}, X.,
  {Lewis}, G.~F., {Strauss}, M.~A., {Cox}, P., {Beelen}, A., {Omont}, A., \&
  {Mohan}, N. 2004, \aj, 128, 997

\bibitem[{{Chakrabarti} {et~al.}(2007{\natexlab{a}}){Chakrabarti}, {Cox},
  {Hernquist}, {Hopkins}, {Robertson}, \& {Di Matteo}}]{Chakrabarti2007A}
{Chakrabarti}, S., {Cox}, T.~J., {Hernquist}, L., {Hopkins}, P.~F.,
  {Robertson}, B., \& {Di Matteo}, T. 2007{\natexlab{a}}, \apj, 658, 840

\bibitem[{{Chakrabarti} {et~al.}(2007{\natexlab{b}}){Chakrabarti}, {Cox},
  {Hopkins}, \& {Hernquist}}]{Chakrabarti2007B}
{Chakrabarti}, S., et al.
  2007{\natexlab{b}}, astro-ph/0610860

\bibitem[{{Charmandaris} {et~al.}(2004){Charmandaris}, {Uchida}, {Weedman},
  {Herter}, {Houck}, {Teplitz}, {Armus}, {Brandl}, {Higdon}, {Soifer},
  {Appleton}, {van Cleve}, \& {Higdon}}]{Charmandaris2004}
{Charmandaris}, V., {Uchida}, K.~I., {Weedman}, D., {Herter}, T., {Houck},
  J.~R., {Teplitz}, H.~I., {Armus}, L., {Brandl}, B.~R., {Higdon}, S.~J.~U.,
  {Soifer}, B.~T., {Appleton}, P.~N., {van Cleve}, J., \& {Higdon}, J.~L. 2004,
  \apjs, 154, 142

\bibitem[{{Code} \& {Whitney}(1995)}]{Code1995}
{Code}, A.~D., \& {Whitney}, B.~A. 1995, \apj, 441, 400

\bibitem[{{Colgan} {et~al.}(1994){Colgan}, {Haas}, {Erickson}, {Lord}, \&
  {Hollenbach}}]{Colgan1994}
{Colgan}, S.~W.~J., {Haas}, M.~R., {Erickson}, E.~F., {Lord}, S.~D., \&
  {Hollenbach}, D.~J. 1994, \apj, 427, 874

\bibitem[{{Condon}(1992)}]{Condon1992}
{Condon}, J.~J. 1992, \araa, 30, 575

\bibitem[{{Dame} {et~al.}(1986){Dame}, {Elmegreen}, {Cohen}, \&
  {Thaddeus}}]{Dame1986}
{Dame}, T.~M., {Elmegreen}, B.~G., {Cohen}, R.~S., \& {Thaddeus}, P. 1986,
  \apj, 305, 892

\bibitem[{{Di Matteo} {et~al.}(2007){Di Matteo}, {Colberg}, {Springel},
  {Hernquist}, \& {Sijacki}}]{DiMatteo2007}
{Di Matteo}, T., {Colberg}, J., {Springel}, V., {Hernquist}, L., \& {Sijacki},
  D. 2007, submitted to ApJ, astro-ph/0705.2269, 705

\bibitem[{{Di Matteo} {et~al.}(2005){Di Matteo}, {Springel}, \&
  {Hernquist}}]{DiMatteo2005}
{Di Matteo}, T., {Springel}, V., \& {Hernquist}, L. 2005, \nat, 433, 604

\bibitem[{{Dorschner} \& {Henning}(1995)}]{Dorschner1995}
{Dorschner}, J., \& {Henning}, T. 1995, \aapr, 6, 271

\bibitem[{{Draine}(2003)}]{Draine2003}
{Draine}, B.~T. 2003, \araa, 41, 241

\bibitem[{{Draine} \& {Lazarian}(1998)}]{Draine1998}
{Draine}, B.~T., \& {Lazarian}, A. 1998, \apj, 508, 157

\bibitem[{{Draine} \& {Li}(2001)}]{Draine2001}
{Draine}, B.~T., \& {Li}, A. 2001, \apj, 551, 807

\bibitem[{{Draine} \& {Li}(2007)}]{Draine2007}
---. 2007, \apj, 657, 810

\bibitem[{{Draine} \& {McKee}(1993)}]{Draine1993}
{Draine}, B.~T., \& {McKee}, C.~F. 1993, \araa, 31, 373

\bibitem[{{Dullemond} \& {Turolla}(2000)}]{Dullemond2000}
{Dullemond}, C.~P., \& {Turolla}, R. 2000, \aap, 360, 1187

\bibitem[{{Dunne} {et~al.}(2003){Dunne}, {Eales}, {Ivison}, {Morgan}, \&
  {Edmunds}}]{Dunne2003}
{Dunne}, L., {Eales}, S., {Ivison}, R., {Morgan}, H., \& {Edmunds}, M. 2003,
  \nat, 424, 285

\bibitem[{{Dunne} \& {Eales}(2001)}]{Dunne2001}
{Dunne}, L., \& {Eales}, S.~A. 2001, \mnras, 327, 697

\bibitem[{{Dwek}(2004)}]{Dwek2004}
{Dwek}, E. 2004, \apj, 607, 848

\bibitem[{{Dwek}(2005)}]{Dwek2005}
{Dwek}, E. 2005, in AIP Conf. Proc. 761: The Spectral Energy Distributions of
  Gas-Rich Galaxies: Confronting Models with Data, ed. C.~C. {Popescu} \& R.~J.
  {Tuffs}, 103--+

\bibitem[{{Dwek} {et~al.}(2007){Dwek}, {Galliano}, \& {Jones}}]{Dwek2007}
{Dwek}, E., {Galliano}, F., \& {Jones}, A.~P. 2007, \apj, 662, 927

\bibitem[{{Efstathiou} \& {Rowan-Robinson}(1990)}]{Efstathiou1990}
{Efstathiou}, A., \& {Rowan-Robinson}, M. 1990, \mnras, 245, 275

\bibitem[{{Efstathiou} \& {Rowan-Robinson}(1991)}]{Efstathiou1991}
---. 1991, \mnras, 252, 528

\bibitem[{{Elmegreen}(1989)}]{Elmegreen1989}
{Elmegreen}, B.~G. 1989, \apj, 338, 178

\bibitem[{{Elmegreen}(2002)}]{Elmegreen2002}
---. 2002, \apj, 564, 773

\bibitem[{{Elmegreen} \& {Falgarone}(1996)}]{Elmegreen1996}
{Elmegreen}, B.~G., \& {Falgarone}, E. 1996, \apj, 471, 816

\bibitem[{{Elvis} {et~al.}(2002){Elvis}, {Marengo}, \& {Karovska}}]{Elvis2002}
{Elvis}, M., {Marengo}, M., \& {Karovska}, M. 2002, \apjl, 567, L107

\bibitem[{{Elvis} {et~al.}(1994){Elvis}, {Wilkes}, {McDowell}, {Green},
  {Bechtold}, {Willner}, {Oey}, {Polomski}, \& {Cutri}}]{Elvis1994}
{Elvis}, M., {Wilkes}, B.~J., {McDowell}, J.~C., {Green}, R.~F., {Bechtold},
  J., {Willner}, S.~P., {Oey}, M.~S., {Polomski}, E., \& {Cutri}, R. 1994,
  \apjs, 95, 1

\bibitem[{{Fan}(2006)}]{Fan2006C}
{Fan}, X. 2006, Memorie della Societa Astronomica Italiana, 77, 635

\bibitem[{{Fan} {et~al.}(2006{\natexlab{a}}){Fan}, {Carilli}, \&
  {Keating}}]{Fan2006B}
{Fan}, X., {Carilli}, C.~L., \& {Keating}, B. 2006{\natexlab{a}}, \araa, 44,
  415

\bibitem[{{Fan} {et~al.}(2004){Fan}, {Hennawi}, {Richards}, {Strauss},
  {Schneider}, {Donley}, {Young}, {Annis}, {Lin}, {Lampeitl}, {Lupton}, {Gunn},
  {Knapp}, {Brandt}, {Anderson}, {Bahcall}, {Brinkmann}, {Brunner}, {Fukugita},
  {Szalay}, {Szokoly}, \& {York}}]{Fan2004}
{Fan}, X., {Hennawi}, J.~F., {Richards}, G.~T., {Strauss}, M.~A., {Schneider},
  D.~P., {Donley}, J.~L., {Young}, J.~E., {Annis}, J., {Lin}, H., {Lampeitl},
  H., {Lupton}, R.~H., {Gunn}, J.~E., {Knapp}, G.~R., {Brandt}, W.~N.,
  {Anderson}, S., {Bahcall}, N.~A., {Brinkmann}, J., {Brunner}, R.~J.,
  {Fukugita}, M., {Szalay}, A.~S., {Szokoly}, G.~P., \& {York}, D.~G. 2004,
  \aj, 128, 515

\bibitem[{{Fan} {et~al.}(2006{\natexlab{b}}){Fan}, {Strauss}, {Becker},
  {White}, {Gunn}, {Knapp}, {Richards}, {Schneider}, {Brinkmann}, \&
  {Fukugita}}]{Fan2006A}
{Fan}, X., {Strauss}, M.~A., {Becker}, R.~H., {White}, R.~L., {Gunn}, J.~E.,
  {Knapp}, G.~R., {Richards}, G.~T., {Schneider}, D.~P., {Brinkmann}, J., \&
  {Fukugita}, M. 2006{\natexlab{b}}, \aj, 132, 117

\bibitem[{{Fan} {et~al.}(2003){Fan}, {Strauss}, {Schneider}, {Becker}, {White},
  {Haiman}, {Gregg}, {Pentericci}, {Grebel}, {Narayanan}, {Loh}, {Richards},
  {Gunn}, {Lupton}, {Knapp}, {Ivezi{\'c}}, {Brandt}, {Collinge}, {Hao},
  {Harbeck}, {Prada}, {Schaye}, {Strateva}, {Zakamska}, {Anderson},
  {Brinkmann}, {Bahcall}, {Lamb}, {Okamura}, {Szalay}, \& {York}}]{Fan2003}
{Fan}, X., {Strauss}, M.~A., {Schneider}, D.~P., {Becker}, R.~H., {White},
  R.~L., {Haiman}, Z., {Gregg}, M., {Pentericci}, L., {Grebel}, E.~K.,
  {Narayanan}, V.~K., {Loh}, Y.-S., {Richards}, G.~T., {Gunn}, J.~E., {Lupton},
  R.~H., {Knapp}, G.~R., {Ivezi{\'c}}, {\v Z}., {Brandt}, W.~N., {Collinge},
  M., {Hao}, L., {Harbeck}, D., {Prada}, F., {Schaye}, J., {Strateva}, I.,
  {Zakamska}, N., {Anderson}, S., {Brinkmann}, J., {Bahcall}, N.~A., {Lamb},
  D.~Q., {Okamura}, S., {Szalay}, A., \& {York}, D.~G. 2003, \aj, 125, 1649

\bibitem[{{Finkbeiner}(1999)}]{Finkbeiner1999A}
{Finkbeiner}, D.~P. 1999, PhD thesis, AA(UNIVERSITY OF CALIFORNIA, BERKELEY)

\bibitem[{{Finkbeiner} {et~al.}(1999){Finkbeiner}, {Davis}, \&
  {Schlegel}}]{Finkbeiner1999B}
{Finkbeiner}, D.~P., {Davis}, M., \& {Schlegel}, D.~J. 1999, \apj, 524, 867

\bibitem[{{Finkbeiner} {et~al.}(2002){Finkbeiner}, {Schlegel}, {Frank}, \&
  {Heiles}}]{Finkbeiner2002}
{Finkbeiner}, D.~P., {Schlegel}, D.~J., {Frank}, C., \& {Heiles}, C. 2002,
  \apj, 566, 898

\bibitem[{{Folini} {et~al.}(2003){Folini}, {Walder}, {Psarros}, \&
  {Desboeufs}}]{Folini2003}
{Folini}, D., {Walder}, R., {Psarros}, M., \& {Desboeufs}, A. 2003, in ASP
  Conf. Ser. 288: Stellar Atmosphere Modeling, ed. I.~{Hubeny}, D.~{Mihalas},
  \& K.~{Werner}, 433--+

\bibitem[{{Freudling} {et~al.}(2003){Freudling}, {Corbin}, \&
  {Korista}}]{Freudling2003}
{Freudling}, W., {Corbin}, M.~R., \& {Korista}, K.~T. 2003, \apjl, 587, L67

\bibitem[{{Fuller} \& {Myers}(1992)}]{Fuller1992}
{Fuller}, G.~A., \& {Myers}, P.~C. 1992, \apj, 384, 523

\bibitem[{{Gao} {et~al.}(2005){Gao}, {White}, {Jenkins}, {Frenk}, \&
  {Springel}}]{Gao2005}
{Gao}, L., {White}, S.~D.~M., {Jenkins}, A., {Frenk}, C.~S., \& {Springel}, V.
  2005, \mnras, 363, 379

\bibitem[{{Gao} {et~al.}(2007){Gao}, {Yoshida}, {Abel}, {Frenk}, {Jenkins}, \&
  {Springel}}]{Gao2007}
{Gao}, L., {Yoshida}, N., {Abel}, T., {Frenk}, C.~S., {Jenkins}, A., \&
  {Springel}, V. 2007, \mnras, 378, 449

\bibitem[{{Gehrz}(1989)}]{Gehrz1989}
{Gehrz}, R. 1989, in IAU Symp. 135: Interstellar Dust, ed. L.~J. {Allamandola}
  \& A.~G.~G.~M. {Tielens}, 445--+

\bibitem[{{Gehrz} \& {Ney}(1987)}]{Gehrz1987}
{Gehrz}, R.~D., \& {Ney}, E.~P. 1987, Proceedings of the National Academy of
  Science, 84, 6961

\bibitem[{{Glikman} {et~al.}(2006){Glikman}, {Helfand}, \&
  {White}}]{Glikman2006}
{Glikman}, E., {Helfand}, D.~J., \& {White}, R.~L. 2006, \apj, 640, 579

\bibitem[{{Gnedin} {et~al.}(2000){Gnedin}, {Ferrara}, \&
  {Zweibel}}]{Gnedin2000}
{Gnedin}, N.~Y., {Ferrara}, A., \& {Zweibel}, E.~G. 2000, \apj, 539, 505

\bibitem[{{Gunn} \& {Peterson}(1965)}]{Gunn1965}
{Gunn}, J.~E., \& {Peterson}, B.~A. 1965, \apj, 142, 1633

\bibitem[{{Haas} {et~al.}(2003){Haas}, {Klaas}, {M{\"u}ller}, {Bertoldi},
  {Camenzind}, {Chini}, {Krause}, {Lemke}, {Meisenheimer}, {Richards}, \&
  {Wilkes}}]{Haas2003}
{Haas}, M., {Klaas}, U., {M{\"u}ller}, S.~A.~H., {Bertoldi}, F., {Camenzind},
  M., {Chini}, R., {Krause}, O., {Lemke}, D., {Meisenheimer}, K., {Richards},
  P.~J., \& {Wilkes}, B.~J. 2003, \aap, 402, 87

\bibitem[{{Hall} {et~al.}(2006){Hall}, {Gallagher}, {Richards}, {Alexander},
  {Anderson}, {Bauer}, {Brandt}, \& {Schneider}}]{Hall2006}
{Hall}, P.~B., {Gallagher}, S.~C., {Richards}, G.~T., {Alexander}, D.~M.,
  {Anderson}, S.~F., {Bauer}, F., {Brandt}, W.~N., \& {Schneider}, D.~P. 2006,
  \aj, 132, 1977

\bibitem[{{Hao} {et~al.}(2005){Hao}, {Spoon}, {Sloan}, {Marshall}, {Armus},
  {Tielens}, {Sargent}, {van Bemmel}, {Charmandaris}, {Weedman}, \&
  {Houck}}]{Hao2005}
{Hao}, L., {Spoon}, H.~W.~W., {Sloan}, G.~C., {Marshall}, J.~A., {Armus}, L.,
  {Tielens}, A.~G.~G.~M., {Sargent}, B., {van Bemmel}, I.~M., {Charmandaris},
  V., {Weedman}, D.~W., \& {Houck}, J.~R. 2005, \apjl, 625, L75

\bibitem[{{Hao} {et~al.}(2007){Hao}, {Weedman}, {Spoon}, {Marshall},
  {Levenson}, {Elitzur}, \& {Houck}}]{Hao2007}
{Hao}, L., {Weedman}, D.~W., {Spoon}, H.~W.~W., {Marshall}, J.~A., {Levenson},
  N.~A., {Elitzur}, M., \& {Houck}, J.~R. 2007, \apjl, 655, L77

\bibitem[{{Harries}(2000)}]{Harries2000}
{Harries}, T.~J. 2000, \mnras, 315, 722

\bibitem[{{Harries} {et~al.}(2004){Harries}, {Monnier}, {Symington}, \&
  {Kurosawa}}]{Harries2004}
{Harries}, T.~J., {Monnier}, J.~D., {Symington}, N.~H., \& {Kurosawa}, R. 2004,
  \mnras, 350, 565

\bibitem[{{Hernquist}(1989)}]{Hernquist1989B}
{Hernquist}, L. 1989, \nat, 340, 687

\bibitem[{{Hernquist}(1990)}]{Hernquist1990}
---. 1990, \apj, 356, 359

\bibitem[{{Hernquist} \& {Katz}(1989)}]{Hernquist1989A}
{Hernquist}, L., \& {Katz}, N. 1989, \apjs, 70, 419

\bibitem[{{Hernquist} \& {Mihos}(1995)}]{Hernquist1995}
{Hernquist}, L., \& {Mihos}, J.~C. 1995, \apj, 448, 41

\bibitem[{{Hines} {et~al.}(2006){Hines}, {Krause}, {Rieke}, {Fan}, {Blaylock},
  \& {Neugebauer}}]{Hines2006}
{Hines}, D.~C., {Krause}, O., {Rieke}, G.~H., {Fan}, X., {Blaylock}, M., \&
  {Neugebauer}, G. 2006, \apjl, 641, L85

\bibitem[{{Hirashita} {et~al.}(2005){Hirashita}, {Nozawa}, {Kozasa}, {Ishii},
  \& {Takeuchi}}]{Hirashita2005}
{Hirashita}, H., {Nozawa}, T., {Kozasa}, T., {Ishii}, T.~T., \& {Takeuchi},
  T.~T. 2005, \mnras, 357, 1077

\bibitem[{{H{\"o}nig} {et~al.}(2006){H{\"o}nig}, {Beckert}, {Ohnaka}, \&
  {Weigelt}}]{Hoenig2006}
{H{\"o}nig}, S.~F., {Beckert}, T., {Ohnaka}, K., \& {Weigelt}, G. 2006, \aap,
  452, 459

\bibitem[{{Hopkins} {et~al.}(2007{\natexlab{a}}){Hopkins}, {Cox}, {Keres}, \&
  {Hernquist}}]{Hopkins2007C}
{Hopkins}, P.~F., {Cox}, T.~J., {Keres}, D., \& {Hernquist}, L.
  2007{\natexlab{a}}, astro-ph/0706.1246, 706

\bibitem[{{Hopkins} {et~al.}(2005{\natexlab{a}}){Hopkins}, {Hernquist}, {Cox},
  {Di Matteo}, {Martini}, {Robertson}, \& {Springel}}]{Hopkins2005C}
{Hopkins}, P.~F., {Hernquist}, L., {Cox}, T.~J., {Di Matteo}, T., {Martini},
  P., {Robertson}, B., \& {Springel}, V. 2005{\natexlab{a}}, \apj, 630, 705

\bibitem[{{Hopkins} {et~al.}(2005{\natexlab{b}}){Hopkins}, {Hernquist}, {Cox},
  {Di Matteo}, {Robertson}, \& {Springel}}]{Hopkins2005B}
{Hopkins}, P.~F., {Hernquist}, L., {Cox}, T.~J., {Di Matteo}, T., {Robertson},
  B., \& {Springel}, V. 2005{\natexlab{b}}, \apj, 630, 716

\bibitem[{{Hopkins} {et~al.}(2006){Hopkins}, {Hernquist}, {Cox}, {Di Matteo},
  {Robertson}, \& {Springel}}]{Hopkins2006A}
---. 2006, \apjs, 163, 1

\bibitem[{{Hopkins} {et~al.}(2007{\natexlab{b}}){Hopkins}, {Hernquist}, {Cox},
  \& {Keres}}]{Hopkins2007B}
{Hopkins}, P.~F., {Hernquist}, L., {Cox}, T.~J., \& {Keres}, D.
  2007{\natexlab{b}}, astro-ph/0706.1243, 706

\bibitem[{{Hopkins} {et~al.}(2007{\natexlab{c}}){Hopkins}, {Hernquist}, {Cox},
  {Robertson}, \& {Krause}}]{Hopkins2007D}
{Hopkins}, P.~F., {Hernquist}, L., {Cox}, T.~J., {Robertson}, B., \& {Krause},
  E. 2007{\natexlab{c}}, \apj, 669, 67

\bibitem[{{Hopkins} {et~al.}(2005{\natexlab{c}}){Hopkins}, {Hernquist},
  {Martini}, {Cox}, {Robertson}, {Di Matteo}, \& {Springel}}]{Hopkins2005A}
{Hopkins}, P.~F., {Hernquist}, L., {Martini}, P., {Cox}, T.~J., {Robertson},
  B., {Di Matteo}, T., \& {Springel}, V. 2005{\natexlab{c}}, \apjl, 625, L71

\bibitem[{{Hopkins} {et~al.}(2007{\natexlab{d}}){Hopkins}, {Richards}, \&
  {Hernquist}}]{Hopkins2007A}
{Hopkins}, P.~F., {Richards}, G.~T., \& {Hernquist}, L. 2007{\natexlab{d}},
  \apj, 654, 731

\bibitem[{{Hopkins} {et~al.}(2004){Hopkins}, {Strauss}, {Hall}, {Richards},
  {Cooper}, {Schneider}, {Vanden Berk}, {Jester}, {Brinkmann}, \&
  {Szokoly}}]{Hopkins2004}
{Hopkins}, P.~F., {Strauss}, M.~A., {Hall}, P.~B., {Richards}, G.~T., {Cooper},
  A.~S., {Schneider}, D.~P., {Vanden Berk}, D.~E., {Jester}, S., {Brinkmann},
  J., \& {Szokoly}, G.~P. 2004, \aj, 128, 1112

\bibitem[{{Houck} {et~al.}(2004){Houck}, {Roellig}, {van Cleve}, {Forrest},
  {Herter}, {Lawrence}, {Matthews}, {Reitsema}, {Soifer}, {Watson}, {Weedman},
  {Huisjen}, {Troeltzsch}, {Barry}, {Bernard-Salas}, {Blacken}, {Brandl},
  {Charmandaris}, {Devost}, {Gull}, {Hall}, {Henderson}, {Higdon}, {Pirger},
  {Schoenwald}, {Sloan}, {Uchida}, {Appleton}, {Armus}, {Burgdorf},
  {Fajardo-Acosta}, {Grillmair}, {Ingalls}, {Morris}, \& {Teplitz}}]{Houck2004}
{Houck}, J.~R., {Roellig}, T.~L., {van Cleve}, J., {Forrest}, W.~J., {Herter},
  T., {Lawrence}, C.~R., {Matthews}, K., {Reitsema}, H.~J., {Soifer}, B.~T.,
  {Watson}, D.~M., {Weedman}, D., {Huisjen}, M., {Troeltzsch}, J., {Barry},
  D.~J., {Bernard-Salas}, J., {Blacken}, C.~E., {Brandl}, B.~R.,
  {Charmandaris}, V., {Devost}, D., {Gull}, G.~E., {Hall}, P., {Henderson},
  C.~P., {Higdon}, S.~J.~U., {Pirger}, B.~E., {Schoenwald}, J., {Sloan}, G.~C.,
  {Uchida}, K.~I., {Appleton}, P.~N., {Armus}, L., {Burgdorf}, M.~J.,
  {Fajardo-Acosta}, S.~B., {Grillmair}, C.~J., {Ingalls}, J.~G., {Morris},
  P.~W., \& {Teplitz}, H.~I. 2004, \apjs, 154, 18

\bibitem[{{Houck} {et~al.}(2005){Houck}, {Soifer}, {Weedman}, {Higdon},
  {Higdon}, {Herter}, {Brown}, {Dey}, {Jannuzi}, {Le Floc'h}, {Rieke}, {Armus},
  {Charmandaris}, {Brandl}, \& {Teplitz}}]{Houck2005}
{Houck}, J.~R., {Soifer}, B.~T., {Weedman}, D., {Higdon}, S.~J.~U., {Higdon},
  J.~L., {Herter}, T., {Brown}, M.~J.~I., {Dey}, A., {Jannuzi}, B.~T., {Le
  Floc'h}, E., {Rieke}, M., {Armus}, L., {Charmandaris}, V., {Brandl}, B.~R.,
  \& {Teplitz}, H.~I. 2005, \apjl, 622, L105

\bibitem[{{Huchra} {et~al.}(1985){Huchra}, {Gorenstein}, {Kent}, {Shapiro},
  {Smith}, {Horine}, \& {Perley}}]{Huchra1985}
{Huchra}, J., {Gorenstein}, M., {Kent}, S., {Shapiro}, I., {Smith}, G.,
  {Horine}, E., \& {Perley}, R. 1985, \aj, 90, 691

\bibitem[{{Hughes} {et~al.}(1997){Hughes}, {Dunlop}, \&
  {Rawlings}}]{Hughes1997}
{Hughes}, D.~H., {Dunlop}, J.~S., \& {Rawlings}, S. 1997, \mnras, 289, 766

\bibitem[{{Hummer} \& {Rybicki}(1971)}]{Hummer1971}
{Hummer}, D.~G., \& {Rybicki}, G.~B. 1971, \mnras, 152, 1

\bibitem[{{Ivezic} {et~al.}(1997){Ivezic}, {Groenewegen}, {Men'shchikov}, \&
  {Szczerba}}]{Ivezic1997}
{Ivezic}, Z., {Groenewegen}, M.~A.~T., {Men'shchikov}, A., \& {Szczerba}, R.
  1997, \mnras, 291, 121

\bibitem[{{Jaffe} {et~al.}(2004){Jaffe}, {Meisenheimer}, {R{\"o}ttgering},
  {Leinert}, {Richichi}, {Chesneau}, {Fraix-Burnet}, {Glazenborg-Kluttig},
  {Granato}, {Graser}, {Heijligers}, {K{\"o}hler}, {Malbet}, {Miley},
  {Paresce}, {Pel}, {Perrin}, {Przygodda}, {Schoeller}, {Sol}, {Waters},
  {Weigelt}, {Woillez}, \& {de Zeeuw}}]{Jaffe2004}
{Jaffe}, W., {Meisenheimer}, K., {R{\"o}ttgering}, H.~J.~A., {Leinert}, C.,
  {Richichi}, A., {Chesneau}, O., {Fraix-Burnet}, D., {Glazenborg-Kluttig}, A.,
  {Granato}, G.-L., {Graser}, U., {Heijligers}, B., {K{\"o}hler}, R., {Malbet},
  F., {Miley}, G.~K., {Paresce}, F., {Pel}, J.-W., {Perrin}, G., {Przygodda},
  F., {Schoeller}, M., {Sol}, H., {Waters}, L.~B.~F.~M., {Weigelt}, G.,
  {Woillez}, J., \& {de Zeeuw}, P.~T. 2004, \nat, 429, 47

\bibitem[{{Jiang} {et~al.}(2006){Jiang}, {Fan}, {Hines}, {Shi}, {Vestergaard},
  {Bertoldi}, {Brandt}, {Carilli}, {Cox}, {Le Floc'h}, {Pentericci},
  {Richards}, {Rieke}, {Schneider}, {Strauss}, {Walter}, \&
  {Brinkmann}}]{Jiang2006}
{Jiang}, L., {Fan}, X., {Hines}, D.~C., {Shi}, Y., {Vestergaard}, M.,
  {Bertoldi}, F., {Brandt}, W.~N., {Carilli}, C.~L., {Cox}, P., {Le Floc'h},
  E., {Pentericci}, L., {Richards}, G.~T., {Rieke}, G.~H., {Schneider}, D.~P.,
  {Strauss}, M.~A., {Walter}, F., \& {Brinkmann}, J. 2006, \aj, 132, 2127

\bibitem[{{Jonsson}(2006)}]{Jonsson2006}
{Jonsson}, P. 2006, \mnras, 372, 2

\bibitem[{{Kennicutt}(1998{\natexlab{a}})}]{Kennicutt1998}
{Kennicutt}, R.~C. 1998{\natexlab{a}}, \apj, 498, 541

\bibitem[{{Kennicutt}(1998{\natexlab{b}})}]{Kennicutt1998B}
{Kennicutt}, Jr., R.~C. 1998{\natexlab{b}}, \araa, 36, 189

\bibitem[{{Kim} {et~al.}(1994){Kim}, {Martin}, \& {Hendry}}]{Kim1994}
{Kim}, S.-H., {Martin}, P.~G., \& {Hendry}, P.~D. 1994, \apj, 422, 164

\bibitem[{{Klaas} {et~al.}(2001){Klaas}, {Haas}, {M{\"u}ller}, {Chini},
  {Schulz}, {Coulson}, {Hippelein}, {Wilke}, {Albrecht}, \&
  {Lemke}}]{Klaas2001}
{Klaas}, U., {Haas}, M., {M{\"u}ller}, S.~A.~H., {Chini}, R., {Schulz}, B.,
  {Coulson}, I., {Hippelein}, H., {Wilke}, K., {Albrecht}, M., \& {Lemke}, D.
  2001, \aap, 379, 823

\bibitem[{{Krause} {et~al.}(2004){Krause}, {Birkmann}, {Rieke}, {Lemke},
  {Klaas}, {Hines}, \& {Gordon}}]{Krause2004}
{Krause}, O., {Birkmann}, S.~M., {Rieke}, G.~H., {Lemke}, D., {Klaas}, U.,
  {Hines}, D.~C., \& {Gordon}, K.~D. 2004, \nat, 432, 596

\bibitem[{{Kroupa}(2002)}]{Kroupa2002}
{Kroupa}, P. 2002, Science, 295, 82

\bibitem[{{Kuraszkiewicz} {et~al.}(2003){Kuraszkiewicz}, {Wilkes}, {Hooper},
  {McLeod}, {Wood}, {Bjorkman}, {Delain}, {Hughes}, {Elvis}, {Impey},
  {Lonsdale}, {Malkan}, {McDowell}, \& {Whitney}}]{Kuraszkiewicz2003}
{Kuraszkiewicz}, J.~K., {Wilkes}, B.~J., {Hooper}, E.~J., {McLeod}, K.~K.,
  {Wood}, K., {Bjorkman}, J., {Delain}, K.~M., {Hughes}, D.~H., {Elvis}, M.~S.,
  {Impey}, C.~D., {Lonsdale}, C.~J., {Malkan}, M.~A., {McDowell}, J.~C., \&
  {Whitney}, B. 2003, \apj, 590, 128

\bibitem[{{Kurosawa} \& {Hillier}(2001)}]{Kurosawa2001}
{Kurosawa}, R., \& {Hillier}, D.~J. 2001, \aap, 379, 336

\bibitem[{{Laor} \& {Draine}(1993)}]{Laor1993}
{Laor}, A., \& {Draine}, B.~T. 1993, \apj, 402, 441

\bibitem[{{Larson}(1981)}]{Larson1981}
{Larson}, R.~B. 1981, \mnras, 194, 809

\bibitem[{{Lefevre} {et~al.}(1982){Lefevre}, {Bergeat}, \&
  {Daniel}}]{Lefevre1982}
{Lefevre}, J., {Bergeat}, J., \& {Daniel}, J.-Y. 1982, \aap, 114, 341

\bibitem[{{Lefevre} {et~al.}(1983){Lefevre}, {Daniel}, \&
  {Bergeat}}]{Lefevre1983}
{Lefevre}, J., {Daniel}, J.-Y., \& {Bergeat}, J. 1983, \aap, 121, 51

\bibitem[{{Leitherer} {et~al.}(1999){Leitherer}, {Schaerer}, {Goldader},
  {Delgado}, {Robert}, {Kune}, {de Mello}, {Devost}, \&
  {Heckman}}]{Leitherer1999}
{Leitherer}, C., {Schaerer}, D., {Goldader}, J.~D., {Delgado}, R.~M.~G.,
  {Robert}, C., {Kune}, D.~F., {de Mello}, D.~F., {Devost}, D., \& {Heckman},
  T.~M. 1999, \apjs, 123, 3

\bibitem[{{Leung}(1976)}]{Leung1976}
{Leung}, C.~M. 1976, \apj, 209, 75

\bibitem[{{Levenson} {et~al.}(2007){Levenson}, {Sirocky}, {Hao}, {Spoon},
  {Marshall}, {Elitzur}, \& {Houck}}]{Levenson2007}
{Levenson}, N.~A., {Sirocky}, M.~M., {Hao}, L., {Spoon}, H.~W.~W., {Marshall},
  J.~A., {Elitzur}, M., \& {Houck}, J.~R. 2007, \apjl, 654, L45

\bibitem[{{Li} \& {Draine}(2001)}]{Li2001}
{Li}, A., \& {Draine}, B.~T. 2001, \apjl, 550, L213

\bibitem[{{Li} \& {Draine}(2002)}]{Li2002}
---. 2002, \apj, 572, 232

\bibitem[{{Li} {et~al.}(2007){Li}, {Hernquist}, {Robertson}, {Cox}, {Hopkins},
  {Springel}, {Gao}, {Di Matteo}, {Zentner}, {Jenkins}, \& {Yoshida}}]{Li2007}
{Li}, Y., {Hernquist}, L., {Robertson}, B., {Cox}, T.~J., {Hopkins}, P.~F.,
  {Springel}, V., {Gao}, L., {Di Matteo}, T., {Zentner}, A.~R., {Jenkins}, A.,
  \& {Yoshida}, N. 2007, \apj, in press, astro-ph/0608190

\bibitem[{{Li} {et~al.}(2003){Li}, {Klessen}, \& {Mac Low}}]{Li2003}
{Li}, Y., {Klessen}, R.~S., \& {Mac Low}, M.-M. 2003, \apj, 592, 975

\bibitem[{{Li} {et~al.}(2004){Li}, {Mac Low}, \& {Klessen}}]{Li2004}
{Li}, Y., {Mac Low}, M.-M., \& {Klessen}, R.~S. 2004, \apjl, 614, L29

\bibitem[{{Li} {et~al.}(2005{\natexlab{a}}){Li}, {Mac Low}, \&
  {Klessen}}]{Li2005A}
---. 2005{\natexlab{a}}, \apjl, 620, L19

\bibitem[{{Li} {et~al.}(2005{\natexlab{b}}){Li}, {Mac Low}, \&
  {Klessen}}]{Li2005B}
---. 2005{\natexlab{b}}, \apj, 626, 823

\bibitem[{{Li} {et~al.}(2006){Li}, {Mac Low}, \& {Klessen}}]{Li2006}
---. 2006, \apj, 639, 879

\bibitem[{{Lopez} {et~al.}(1995){Lopez}, {Mekarnia}, \& {Lefevre}}]{Lopez1995}
{Lopez}, B., {Mekarnia}, D., \& {Lefevre}, J. 1995, \aap, 296, 752

\bibitem[{{Lucy}(1999)}]{Lucy1999}
{Lucy}, L.~B. 1999, \aap, 344, 282

\bibitem[{{Madau}(1995)}]{Madau1995}
{Madau}, P. 1995, \apj, 441, 18

\bibitem[{{Magorrian} {et~al.}(1998){Magorrian}, {Tremaine}, {Richstone},
  {Bender}, {Bower}, {Dressler}, {Faber}, {Gebhardt}, {Green}, {Grillmair},
  {Kormendy}, \& {Lauer}}]{Magorrian1998}
{Magorrian}, J., {Tremaine}, S., {Richstone}, D., {Bender}, R., {Bower}, G.,
  {Dressler}, A., {Faber}, S.~M., {Gebhardt}, K., {Green}, R., {Grillmair}, C.,
  {Kormendy}, J., \& {Lauer}, T. 1998, \aj, 115, 2285

\bibitem[{{Maiolino} {et~al.}(2001){Maiolino}, {Marconi}, {Salvati},
  {Risaliti}, {Severgnini}, {Oliva}, {La Franca}, \& {Vanzi}}]{Maiolino2001}
{Maiolino}, R., {Marconi}, A., {Salvati}, M., {Risaliti}, G., {Severgnini}, P.,
  {Oliva}, E., {La Franca}, F., \& {Vanzi}, L. 2001, \aap, 365, 28

\bibitem[{{Maiolino} {et~al.}(2006){Maiolino}, {Nagao}, {Marconi}, {Schneider},
  {Bianchi}, {Pedani}, {Pipino}, {Matteucci}, {Cox}, \&
  {Caselli}}]{Maiolino2006}
{Maiolino}, R., {Nagao}, T., {Marconi}, A., {Schneider}, R., {Bianchi}, S.,
  {Pedani}, M., {Pipino}, A., {Matteucci}, F., {Cox}, P., \& {Caselli}, P.
  2006, Memorie della Societa Astronomica Italiana, 77, 643

\bibitem[{{Maiolino} {et~al.}(2004){Maiolino}, {Schneider}, {Oliva}, {Bianchi},
  {Ferrara}, {Mannucci}, {Pedani}, \& {Roca Sogorb}}]{Maiolino2004}
{Maiolino}, R., {Schneider}, R., {Oliva}, E., {Bianchi}, S., {Ferrara}, A.,
  {Mannucci}, F., {Pedani}, M., \& {Roca Sogorb}, M. 2004, \nat, 431, 533

\bibitem[{{Marchenko}(2006)}]{Marchenko2006}
{Marchenko}, S.~V. 2006, in ASP Conf. Ser. 353: Stellar Evolution at Low
  Metallicity: Mass Loss, Explosions, Cosmology, ed. H.~J.~G.~L.~M. {Lamers},
  N.~{Langer}, T.~{Nugis}, \& K.~{Annuk}, 299--+

\bibitem[{{Marconi} \& {Hunt}(2003)}]{Marconi2003}
{Marconi}, A., \& {Hunt}, L.~K. 2003, \apjl, 589, L21

\bibitem[{{Marconi} {et~al.}(2004){Marconi}, {Risaliti}, {Gilli}, {Hunt},
  {Maiolino}, \& {Salvati}}]{Marconi2004}
{Marconi}, A., {Risaliti}, G., {Gilli}, R., {Hunt}, L.~K., {Maiolino}, R., \&
  {Salvati}, M. 2004, \mnras, 351, 169

\bibitem[{{Mathis}(1990)}]{Mathis1990}
{Mathis}, J.~S. 1990, \araa, 28, 37

\bibitem[{{Mathis} {et~al.}(1977){Mathis}, {Rumpl}, \&
  {Nordsieck}}]{Mathis1977}
{Mathis}, J.~S., {Rumpl}, W., \& {Nordsieck}, K.~H. 1977, \apj, 217, 425

\bibitem[{{McKee}(1989)}]{McKee1989}
{McKee}, C. 1989, in IAU Symposium, Vol. 135, Interstellar Dust, ed. L.~J.
  {Allamandola} \& A.~G.~G.~M. {Tielens}, 431--+

\bibitem[{{McKee} \& {Ostriker}(1977)}]{McKee1977}
{McKee}, C.~F., \& {Ostriker}, J.~P. 1977, \apj, 218, 148

\bibitem[{{Men'shchikov} \& {Henning}(1997)}]{Menshchikov1997}
{Men'shchikov}, A.~B., \& {Henning}, T. 1997, \aap, 318, 879

\bibitem[{{Mihos} \& {Hernquist}(1996)}]{Mihos1996}
{Mihos}, J.~C., \& {Hernquist}, L. 1996, \apj, 464, 641

\bibitem[{{Miller} \& {Antonucci}(1983)}]{Miller1983}
{Miller}, J.~S., \& {Antonucci}, R.~R.~J. 1983, \apjl, 271, L7

\bibitem[{{Misselt} {et~al.}(2001){Misselt}, {Gordon}, {Clayton}, \&
  {Wolff}}]{Misselt2001}
{Misselt}, K.~A., {Gordon}, K.~D., {Clayton}, G.~C., \& {Wolff}, M.~J. 2001,
  \apj, 551, 277

\bibitem[{{Morgan} {et~al.}(2003){Morgan}, {Dunne}, {Eales}, {Ivison}, \&
  {Edmunds}}]{Morgan2003B}
{Morgan}, H.~L., {Dunne}, L., {Eales}, S.~A., {Ivison}, R.~J., \& {Edmunds},
  M.~G. 2003, \apjl, 597, L33

\bibitem[{{Morgan} \& {Edmunds}(2003)}]{Morgan2003A}
{Morgan}, H.~L., \& {Edmunds}, M.~G. 2003, \mnras, 343, 427

\bibitem[{{Moseley} {et~al.}(1989){Moseley}, {Dwek}, {Glaccum}, {Graham}, \&
  {Loewenstein}}]{Moseley1989}
{Moseley}, S.~H., {Dwek}, E., {Glaccum}, W., {Graham}, J.~R., \& {Loewenstein},
  R.~F. 1989, \nat, 340, 697

\bibitem[{{Narayanan} {et~al.}(2006{\natexlab{a}}){Narayanan}, {Cox},
  {Robertson}, {Dav{\'e}}, {Di Matteo}, {Hernquist}, {Hopkins}, {Kulesa}, \&
  {Walker}}]{Narayanan2006b}
{Narayanan}, D., {Cox}, T.~J., {Robertson}, B., {Dav{\'e}}, R., {Di Matteo},
  T., {Hernquist}, L., {Hopkins}, P., {Kulesa}, C., \& {Walker}, C.~K.
  2006{\natexlab{a}}, \apjl, 642, L107

\bibitem[{{Narayanan} {et~al.}(2006{\natexlab{b}}){Narayanan}, {Kulesa},
  {Boss}, \& {Walker}}]{Narayanan2006c}
{Narayanan}, D., {Kulesa}, C., {Boss}, A., \& {Walker}, C.~K.
  2006{\natexlab{b}}, \apj, 647

\bibitem[{{Narayanan} {et~al.}(2007){Narayanan}, {Li}, {Cox}, {Hernquist}, ,
  {Hopkins}, {Chakrabarti}, {Dav{\'e}}, {Di Matteo}, {Kulesa}, {Robertson},
  {Springel}, \& {Walker}}]{Narayanan2007}
{Narayanan}, D., {Li}, Y., {Cox}, T.~J., {Hernquist}, L., , {Hopkins}, P.,
  {Chakrabarti}, S., {Dav{\'e}}, R., {Di Matteo}, T., {Kulesa}, C.,
  {Robertson}, B., {Springel}, V., \& {Walker}, C.~K. 2007, \apj, submitted

\bibitem[{{Nenkova} {et~al.}(2002){Nenkova}, {Ivezi{\'c}}, \&
  {Elitzur}}]{Nenkova2002}
{Nenkova}, M., {Ivezi{\'c}}, {\v Z}., \& {Elitzur}, M. 2002, \apjl, 570, L9

\bibitem[{{Norman} \& {Scoville}(1988)}]{Norman1988}
{Norman}, C., \& {Scoville}, N. 1988, \apj, 332, 124

\bibitem[{{Nozawa} {et~al.}(2006){Nozawa}, {Kozasa}, \& {Habe}}]{Nozawa2006}
{Nozawa}, T., {Kozasa}, T., \& {Habe}, A. 2006, \apj, 648, 435

\bibitem[{{Nozawa} {et~al.}(2007){Nozawa}, {Kozasa}, {Habe}, {Dwek}, {Umeda},
  {Tominaga}, {Maeda}, \& {Nomoto}}]{Nozawa2007}
{Nozawa}, T., {Kozasa}, T., {Habe}, A., {Dwek}, E., {Umeda}, H., {Tominaga},
  N., {Maeda}, K., \& {Nomoto}, K. 2007, \apj, 666, 955

\bibitem[{{Nozawa} {et~al.}(2003){Nozawa}, {Kozasa}, {Umeda}, {Maeda}, \&
  {Nomoto}}]{Nozawa2003}
{Nozawa}, T., {Kozasa}, T., {Umeda}, H., {Maeda}, K., \& {Nomoto}, K. 2003,
  \apj, 598, 785

\bibitem[{{Papovich} {et~al.}(2006){Papovich}, {Moustakas}, {Dickinson}, {Le
  Floc'h}, {Rieke}, {Daddi}, {Alexander}, {Bauer}, {Brandt}, {Dahlen}, {Egami},
  {Eisenhardt}, {Elbaz}, {Ferguson}, {Giavalisco}, {Lucas}, {Mobasher},
  {P{\'e}rez-Gonz{\'a}lez}, {Stutz}, {Rieke}, \& {Yan}}]{Papovich2006}
{Papovich}, C., {Moustakas}, L.~A., {Dickinson}, M., {Le Floc'h}, E., {Rieke},
  G.~H., {Daddi}, E., {Alexander}, D.~M., {Bauer}, F., {Brandt}, W.~N.,
  {Dahlen}, T., {Egami}, E., {Eisenhardt}, P., {Elbaz}, D., {Ferguson}, H.~C.,
  {Giavalisco}, M., {Lucas}, R.~A., {Mobasher}, B., {P{\'e}rez-Gonz{\'a}lez},
  P.~G., {Stutz}, A., {Rieke}, M.~J., \& {Yan}, H. 2006, \apj, 640, 92

\bibitem[{{Pascucci} {et~al.}(2004){Pascucci}, {Wolf}, {Steinacker},
  {Dullemond}, {Henning}, {Niccolini}, {Woitke}, \& {Lopez}}]{Pascucci2004}
{Pascucci}, I., {Wolf}, S., {Steinacker}, J., {Dullemond}, C.~P., {Henning},
  T., {Niccolini}, G., {Woitke}, P., \& {Lopez}, B. 2004, \aap, 417, 793

\bibitem[{{Pentericci} {et~al.}(2003){Pentericci}, {Rix}, {Prada}, {Fan},
  {Strauss}, {Schneider}, {Grebel}, {Harbeck}, {Brinkmann}, \&
  {Narayanan}}]{Pentericci2003}
{Pentericci}, L., {Rix}, H.-W., {Prada}, F., {Fan}, X., {Strauss}, M.~A.,
  {Schneider}, D.~P., {Grebel}, E.~K., {Harbeck}, D., {Brinkmann}, J., \&
  {Narayanan}, V.~K. 2003, \aap, 410, 75

\bibitem[{{Pinte} {et~al.}(2006){Pinte}, {M{\'e}nard}, {Duch{\^e}ne}, \&
  {Bastien}}]{Pinte2006}
{Pinte}, C., {M{\'e}nard}, F., {Duch{\^e}ne}, G., \& {Bastien}, P. 2006, \aap,
  459, 797

\bibitem[{{Polletta} {et~al.}(2000){Polletta}, {Courvoisier}, {Hooper}, \&
  {Wilkes}}]{Polletta2000}
{Polletta}, M., {Courvoisier}, T.~J.-L., {Hooper}, E.~J., \& {Wilkes}, B.~J.
  2000, \aap, 362, 75

\bibitem[{{Reynolds} {et~al.}(1997){Reynolds}, {Ward}, {Fabian}, \&
  {Celotti}}]{Reynolds1997}
{Reynolds}, C.~S., {Ward}, M.~J., {Fabian}, A.~C., \& {Celotti}, A. 1997,
  \mnras, 291, 403

\bibitem[{{Rice} {et~al.}(2003){Rice}, {Wood}, {Armitage}, {Whitney}, \&
  {Bjorkman}}]{Rice2003}
{Rice}, W.~K.~M., {Wood}, K., {Armitage}, P.~J., {Whitney}, B.~A., \&
  {Bjorkman}, J.~E. 2003, \mnras, 342, 79

\bibitem[{{Richards} {et~al.}(2006){Richards}, {Lacy}, {Storrie-Lombardi},
  {Hall}, {Gallagher}, {Hines}, {Fan}, {Papovich}, {Vanden Berk}, {Trammell},
  {Schneider}, {Vestergaard}, {York}, {Jester}, {Anderson}, {Budavari}, \&
  {Szalay}}]{Richards2006}
{Richards}, G.~T., {Lacy}, M., {Storrie-Lombardi}, L.~J., {Hall}, P.~B.,
  {Gallagher}, S.~C., {Hines}, D.~C., {Fan}, X., {Papovich}, C., {Vanden Berk},
  D.~E., {Trammell}, G.~B., {Schneider}, D.~P., {Vestergaard}, M., {York},
  D.~G., {Jester}, S., {Anderson}, S.~F., {Budavari}, T., \& {Szalay}, A.~S.
  2006, astro-ph/0601558

\bibitem[{{Rieke} \& {Lebofsky}(1981)}]{Rieke1981}
{Rieke}, G.~H., \& {Lebofsky}, M.~J. 1981, \apj, 250, 87

\bibitem[{{Robertson} {et~al.}(2006){Robertson}, {Hernquist}, {Cox}, {Di
  Matteo}, {Hopkins}, {Martini}, \& {Springel}}]{Robertson2006A}
{Robertson}, B., {Hernquist}, L., {Cox}, T.~J., {Di Matteo}, T., {Hopkins},
  P.~F., {Martini}, P., \& {Springel}, V. 2006, \apj, 641, 90

\bibitem[{{Robertson} {et~al.}(2007){Robertson}, {Li}, {Cox}, {Hernquist}, \&
  {Hopkins}}]{Robertson2007}
{Robertson}, B., {Li}, Y., {Cox}, T.~J., {Hernquist}, L., \& {Hopkins}, P.~F.
  2007, \apj, 667, 60

\bibitem[{{Robson} {et~al.}(2004){Robson}, {Priddey}, {Isaak}, \&
  {McMahon}}]{Robson2004}
{Robson}, I., {Priddey}, R.~S., {Isaak}, K.~G., \& {McMahon}, R.~G. 2004,
  \mnras, 351, L29

\bibitem[{{Roche} {et~al.}(1993){Roche}, {Aitken}, \& {Smith}}]{Roche1993}
{Roche}, P.~F., {Aitken}, D.~K., \& {Smith}, C.~H. 1993, \mnras, 261, 522

\bibitem[{{Rosolowsky}(2005)}]{Rosolowsky2005}
{Rosolowsky}, E. 2005, \pasp, 117, 1403

\bibitem[{{Rosolowsky}(2007)}]{Rosolowsky2007}
---. 2007, \apj, 654, 240

\bibitem[{{Rowan-Robinson}(1980)}]{Rowan-Robinson1980}
{Rowan-Robinson}, M. 1980, \apjs, 44, 403

\bibitem[{{Salpeter}(1955)}]{Salpeter1955}
{Salpeter}, E.~E. 1955, \apj, 121, 161

\bibitem[{{Sanders} \& {Mirabel}(1996)}]{Sanders1996}
{Sanders}, D.~B., \& {Mirabel}, I.~F. 1996, \araa, 34, 749

\bibitem[{{Sanders} {et~al.}(1985){Sanders}, {Scoville}, \&
  {Solomon}}]{Sanders1985}
{Sanders}, D.~B., {Scoville}, N.~Z., \& {Solomon}, P.~M. 1985, \apj, 289, 373

\bibitem[{{Sanders} {et~al.}(1988){Sanders}, {Soifer}, {Elias}, {Madore},
  {Matthews}, {Neugebauer}, \& {Scoville}}]{Sanders1988}
{Sanders}, D.~B., {Soifer}, B.~T., {Elias}, J.~H., {Madore}, B.~F., {Matthews},
  K., {Neugebauer}, G., \& {Scoville}, N.~Z. 1988, \apj, 325, 74

\bibitem[{{Savage} \& {Mathis}(1979)}]{Savage1979}
{Savage}, B.~D., \& {Mathis}, J.~S. 1979, \araa, 17, 73

\bibitem[{{Schlegel} {et~al.}(1998){Schlegel}, {Finkbeiner}, \&
  {Davis}}]{Schlegel1998}
{Schlegel}, D.~J., {Finkbeiner}, D.~P., \& {Davis}, M. 1998, \apj, 500, 525

\bibitem[{{Schmidt}(1959)}]{Schmidt1959}
{Schmidt}, M. 1959, \apj, 129, 243

\bibitem[{{Schneider} {et~al.}(2004){Schneider}, {Ferrara}, \&
  {Salvaterra}}]{Schneider2004}
{Schneider}, R., {Ferrara}, A., \& {Salvaterra}, R. 2004, \mnras, 351, 1379

\bibitem[{{Scoville}(2003)}]{Scoville2003}
{Scoville}, N. 2003, Journal of Korean Astronomical Society, 36, 167

\bibitem[{{Scoville} \& {Kwan}(1976)}]{Scoville1976}
{Scoville}, N.~Z., \& {Kwan}, J. 1976, \apj, 206, 718

\bibitem[{{Scoville} {et~al.}(1987){Scoville}, {Yun}, {Sanders}, {Clemens}, \&
  {Waller}}]{Scoville1987}
{Scoville}, N.~Z., {Yun}, M.~S., {Sanders}, D.~B., {Clemens}, D.~P., \&
  {Waller}, W.~H. 1987, \apjs, 63, 821

\bibitem[{{Shemmer} {et~al.}(2006){Shemmer}, {Brandt}, {Schneider}, {Fan},
  {Strauss}, {Diamond-Stanic}, {Richards}, {Anderson}, {Gunn}, \&
  {Brinkmann}}]{Shemmer2006}
{Shemmer}, O., {Brandt}, W.~N., {Schneider}, D.~P., {Fan}, X., {Strauss},
  M.~A., {Diamond-Stanic}, A.~M., {Richards}, G.~T., {Anderson}, S.~F., {Gunn},
  J.~E., \& {Brinkmann}, J. 2006, \apj, 644, 86

\bibitem[{{Shemmer} {et~al.}(2005){Shemmer}, {Brandt}, {Vignali}, {Schneider},
  {Fan}, {Richards}, \& {Strauss}}]{Shemmer2005}
{Shemmer}, O., {Brandt}, W.~N., {Vignali}, C., {Schneider}, D.~P., {Fan}, X.,
  {Richards}, G.~T., \& {Strauss}, M.~A. 2005, \apj, 630, 729

\bibitem[{{Siebenmorgen} {et~al.}(2005){Siebenmorgen}, {Haas}, {Kr{\"u}gel}, \&
  {Schulz}}]{Siebenmorgen2005}
{Siebenmorgen}, R., {Haas}, M., {Kr{\"u}gel}, E., \& {Schulz}, B. 2005, \aap,
  436, L5

\bibitem[{{Siebenmorgen} \& {Kr{\"u}gel}(2007)}]{Siebenmorgen2007}
{Siebenmorgen}, R., \& {Kr{\"u}gel}, E. 2007, \aap, 461, 445

\bibitem[{{Sijacki} {et~al.}(2007){Sijacki}, {Springel}, {Di Matteo}, \&
  {Hernquist}}]{Sijacki2007}
{Sijacki}, D., {Springel}, V., {Di Matteo}, T., \& {Hernquist}, L. 2007,
  submitted to MNRAS, astro-ph/0705.2238, 705

\bibitem[{{Solomon} {et~al.}(1987){Solomon}, {Rivolo}, {Barrett}, \&
  {Yahil}}]{Solomon1987}
{Solomon}, P.~M., {Rivolo}, A.~R., {Barrett}, J., \& {Yahil}, A. 1987, \apj,
  319, 730

\bibitem[{{Spergel} {et~al.}(2003){Spergel}, {Verde}, {Peiris}, {Komatsu},
  {Nolta}, {Bennett}, {Halpern}, {Hinshaw}, {Jarosik}, {Kogut}, {Limon},
  {Meyer}, {Page}, {Tucker}, {Weiland}, {Wollack}, \& {Wright}}]{Spergel2003}
{Spergel}, D.~N., {Verde}, L., {Peiris}, H.~V., {Komatsu}, E., {Nolta}, M.~R.,
  {Bennett}, C.~L., {Halpern}, M., {Hinshaw}, G., {Jarosik}, N., {Kogut}, A.,
  {Limon}, M., {Meyer}, S.~S., {Page}, L., {Tucker}, G.~S., {Weiland}, J.~L.,
  {Wollack}, E., \& {Wright}, E.~L. 2003, \apjs, 148, 175

\bibitem[{{Spoon} {et~al.}(2007){Spoon}, {Marshall}, {Houck}, {Elitzur}, {Hao},
  {Armus}, {Brandl}, \& {Charmandaris}}]{Spoon2007}
{Spoon}, H.~W.~W., {Marshall}, J.~A., {Houck}, J.~R., {Elitzur}, M., {Hao}, L.,
  {Armus}, L., {Brandl}, B.~R., \& {Charmandaris}, V. 2007, \apjl, 654, L49

\bibitem[{{Spoon} {et~al.}(2004){Spoon}, {Moorwood}, {Lutz}, {Tielens},
  {Siebenmorgen}, \& {Keane}}]{Spoon2004}
{Spoon}, H.~W.~W., {Moorwood}, A.~F.~M., {Lutz}, D., {Tielens}, A.~G.~G.~M.,
  {Siebenmorgen}, R., \& {Keane}, J.~V. 2004, \aap, 414, 873

\bibitem[{{Spoon} {et~al.}(2006){Spoon}, {Tielens}, {Armus}, {Sloan},
  {Sargent}, {Cami}, {Charmandaris}, {Houck}, \& {Soifer}}]{Spoon2006}
{Spoon}, H.~W.~W., {Tielens}, A.~G.~G.~M., {Armus}, L., {Sloan}, G.~C.,
  {Sargent}, B., {Cami}, J., {Charmandaris}, V., {Houck}, J.~R., \& {Soifer},
  B.~T. 2006, \apj, 638, 759

\bibitem[{{Springel}(2005)}]{Springel2005D}
{Springel}, V. 2005, \mnras, 364, 1105

\bibitem[{{Springel} {et~al.}(2005{\natexlab{a}}){Springel}, {Di Matteo}, \&
  {Hernquist}}]{Springel2005B}
{Springel}, V., {Di Matteo}, T., \& {Hernquist}, L. 2005{\natexlab{a}}, \mnras,
  361, 776

\bibitem[{{Springel} \& {Hernquist}(2002)}]{Springel2002}
{Springel}, V., \& {Hernquist}, L. 2002, \mnras, 333, 649

\bibitem[{{Springel} \& {Hernquist}(2003{\natexlab{a}})}]{Springel2003A}
---. 2003{\natexlab{a}}, \mnras, 339, 289

\bibitem[{{Springel} \& {Hernquist}(2003{\natexlab{b}})}]{Springel2003B}
---. 2003{\natexlab{b}}, \mnras, 339, 312

\bibitem[{{Springel} {et~al.}(2005{\natexlab{b}}){Springel}, {White},
  {Jenkins}, {Frenk}, {Yoshida}, {Gao}, {Navarro}, {Thacker}, {Croton},
  {Helly}, {Peacock}, {Cole}, {Thomas}, {Couchman}, {Evrard}, {Colberg}, \&
  {Pearce}}]{Springel2005A}
{Springel}, V., {White}, S.~D.~M., {Jenkins}, A., {Frenk}, C.~S., {Yoshida},
  N., {Gao}, L., {Navarro}, J., {Thacker}, R., {Croton}, D., {Helly}, J.,
  {Peacock}, J.~A., {Cole}, S., {Thomas}, P., {Couchman}, H., {Evrard}, A.,
  {Colberg}, J., \& {Pearce}, F. 2005{\natexlab{b}}, \nat, 435, 629

\bibitem[{{Spyromilio} {et~al.}(1993){Spyromilio}, {Stathakis}, \&
  {Meurer}}]{Spyromilio1993}
{Spyromilio}, J., {Stathakis}, R.~A., \& {Meurer}, G.~R. 1993, \mnras, 263, 530

\bibitem[{{Steffen} {et~al.}(2006){Steffen}, {Strateva}, {Brandt}, {Alexander},
  {Koekemoer}, {Lehmer}, {Schneider}, \& {Vignali}}]{Steffen2006}
{Steffen}, A.~T., {Strateva}, I., {Brandt}, W.~N., {Alexander}, D.~M.,
  {Koekemoer}, A.~M., {Lehmer}, B.~D., {Schneider}, D.~P., \& {Vignali}, C.
  2006, \aj, 131, 2826

\bibitem[{{Steinacker} {et~al.}(2006){Steinacker}, {Bacmann}, \&
  {Henning}}]{Steinacker2006}
{Steinacker}, J., {Bacmann}, A., \& {Henning}, T. 2006, \apj, 645, 920

\bibitem[{{Steinacker} {et~al.}(2003){Steinacker}, {Henning}, {Bacmann}, \&
  {Semenov}}]{Steinacker2003}
{Steinacker}, J., {Henning}, T., {Bacmann}, A., \& {Semenov}, D. 2003, \aap,
  401, 405

\bibitem[{{Strateva} {et~al.}(2005){Strateva}, {Brandt}, {Schneider}, {Vanden
  Berk}, \& {Vignali}}]{Strateva2005}
{Strateva}, I.~V., {Brandt}, W.~N., {Schneider}, D.~P., {Vanden Berk}, D.~G.,
  \& {Vignali}, C. 2005, \aj, 130, 387

\bibitem[{{Stratta} {et~al.}(2007){Stratta}, {Maiolino}, {Fiore}, \&
  {D'Elia}}]{Stratta2007}
{Stratta}, G., {Maiolino}, R., {Fiore}, F., \& {D'Elia}, V. 2007,
  astro-ph/0703349

\bibitem[{{Sturm} {et~al.}(2000){Sturm}, {Lutz}, {Tran}, {Feuchtgruber},
  {Genzel}, {Kunze}, {Moorwood}, \& {Thornley}}]{Sturm2000}
{Sturm}, E., {Lutz}, D., {Tran}, D., {Feuchtgruber}, H., {Genzel}, R., {Kunze},
  D., {Moorwood}, A.~F.~M., \& {Thornley}, M.~D. 2000, \aap, 358, 481

\bibitem[{{Sturm} {et~al.}(2005){Sturm}, {Schweitzer}, {Lutz}, {Contursi},
  {Genzel}, {Lehnert}, {Tacconi}, {Veilleux}, {Rupke}, {Kim}, {Sternberg},
  {Maoz}, {Lord}, {Mazzarella}, \& {Sanders}}]{Sturm2005}
{Sturm}, E., {Schweitzer}, M., {Lutz}, D., {Contursi}, A., {Genzel}, R.,
  {Lehnert}, M.~D., {Tacconi}, L.~J., {Veilleux}, S., {Rupke}, D.~S., {Kim},
  D.-C., {Sternberg}, A., {Maoz}, D., {Lord}, S., {Mazzarella}, J., \&
  {Sanders}, D.~B. 2005, \apjl, 629, L21

\bibitem[{{Sugerman} {et~al.}(2006){Sugerman}, {Ercolano}, {Barlow}, {Tielens},
  {Clayton}, {Zijlstra}, {Meixner}, {Speck}, {Gledhill}, {Panagia}, {Cohen},
  {Gordon}, {Meyer}, {Fabbri}, {Bowey}, {Welch}, {Regan}, \&
  {Kennicutt}}]{Sugerman2006}
{Sugerman}, B.~E.~K., {Ercolano}, B., {Barlow}, M.~J., {Tielens}, A.~G.~G.~M.,
  {Clayton}, G.~C., {Zijlstra}, A.~A., {Meixner}, M., {Speck}, A., {Gledhill},
  T.~M., {Panagia}, N., {Cohen}, M., {Gordon}, K.~D., {Meyer}, M., {Fabbri},
  J., {Bowey}, J.~E., {Welch}, D.~L., {Regan}, M.~W., \& {Kennicutt}, R.~C.
  2006, Science, 313, 196

\bibitem[{{Tan} \& {McKee}(2004)}]{Tan2004}
{Tan}, J.~C., \& {McKee}, C.~F. 2004, \apj, 603, 383

\bibitem[{{Telfer} {et~al.}(2002){Telfer}, {Zheng}, {Kriss}, \&
  {Davidsen}}]{Telfer2002}
{Telfer}, R.~C., {Zheng}, W., {Kriss}, G.~A., \& {Davidsen}, A.~F. 2002, \apj,
  565, 773

\bibitem[{{Todini} \& {Ferrara}(2001)}]{Todini2001}
{Todini}, P., \& {Ferrara}, A. 2001, \mnras, 325, 726

\bibitem[{{Vanden Berk} {et~al.}(2001){Vanden Berk}, {Richards}, \&
  {Bauer}}]{VandenBerk2001}
{Vanden Berk}, D.~E., {Richards}, G.~T., \& {Bauer}, A. 2001, \aj, 122, 549

\bibitem[{{V{\'a}zquez} \& {Leitherer}(2005)}]{Vazquez2005}
{V{\'a}zquez}, G.~A., \& {Leitherer}, C. 2005, \apj, 621, 695

\bibitem[{{Vignali} {et~al.}(2003){Vignali}, {Brandt}, {Schneider}, {Garmire},
  \& {Kaspi}}]{Vignali2003}
{Vignali}, C., {Brandt}, W.~N., {Schneider}, D.~P., {Garmire}, G.~P., \&
  {Kaspi}, S. 2003, \aj, 125, 418

\bibitem[{{Vignali} {et~al.}(2005){Vignali}, {Brandt}, {Schneider}, \&
  {Kaspi}}]{Vignali2005}
{Vignali}, C., {Brandt}, W.~N., {Schneider}, D.~P., \& {Kaspi}, S. 2005, \aj,
  129, 2519

\bibitem[{{Walter} {et~al.}(2003){Walter}, {Bertoldi}, {Carilli}, {Cox}, {Lo},
  {Neri}, {Fan}, {Omont}, {Strauss}, \& {Menten}}]{Walter2003}
{Walter}, F., {Bertoldi}, F., {Carilli}, C., {Cox}, P., {Lo}, K.~Y., {Neri},
  R., {Fan}, X., {Omont}, A., {Strauss}, M.~A., \& {Menten}, K.~M. 2003, \nat,
  424, 406

\bibitem[{{Walter} {et~al.}(2004){Walter}, {Carilli}, {Bertoldi}, {Menten},
  {Cox}, {Lo}, {Fan}, \& {Strauss}}]{Walter2004}
{Walter}, F., {Carilli}, C., {Bertoldi}, F., {Menten}, K., {Cox}, P., {Lo},
  K.~Y., {Fan}, X., \& {Strauss}, M.~A. 2004, \apjl, 615, L17

\bibitem[{{Wang} {et~al.}(2007){Wang}, {Carilli}, {Beelen}, {Bertoldi}, {Fan},
  {Walter}, {Menten}, {Omont}, {Cox}, {Strauss}, \& {Jiang}}]{Wang2007}
{Wang}, R., {Carilli}, C., {Beelen}, A., {Bertoldi}, F., {Fan}, X., {Walter},
  F., {Menten}, K.~M., {Omont}, A., {Cox}, P., {Strauss}, M.~A., \& {Jiang}, L.
  2007, \aj, in press, astro-ph/0704.2053, 704

\bibitem[{{Ward-Thompson} {et~al.}(1994){Ward-Thompson}, {Scott}, {Hills}, \&
  {Andre}}]{Ward-Thompson1994}
{Ward-Thompson}, D., {Scott}, P.~F., {Hills}, R.~E., \& {Andre}, P. 1994,
  \mnras, 268, 276

\bibitem[{{Weingartner} \& {Draine}(2001)}]{Weingartner2001}
{Weingartner}, J.~C., \& {Draine}, B.~T. 2001, \apj, 548, 296

\bibitem[{{Werner} {et~al.}(2004){Werner}, {Roellig}, {Low}, {Rieke}, {Rieke},
  {Hoffmann}, {Young}, {Houck}, {Brandl}, {Fazio}, {Hora}, {Gehrz}, {Helou},
  {Soifer}, {Stauffer}, {Keene}, {Eisenhardt}, {Gallagher}, {Gautier}, {Irace},
  {Lawrence}, {Simmons}, {Van Cleve}, {Jura}, {Wright}, \&
  {Cruikshank}}]{Werner2004}
{Werner}, M.~W., {Roellig}, T.~L., {Low}, F.~J., {Rieke}, G.~H., {Rieke}, M.,
  {Hoffmann}, W.~F., {Young}, E., {Houck}, J.~R., {Brandl}, B., {Fazio}, G.~G.,
  {Hora}, J.~L., {Gehrz}, R.~D., {Helou}, G., {Soifer}, B.~T., {Stauffer}, J.,
  {Keene}, J., {Eisenhardt}, P., {Gallagher}, D., {Gautier}, T.~N., {Irace},
  W., {Lawrence}, C.~R., {Simmons}, L., {Van Cleve}, J.~E., {Jura}, M.,
  {Wright}, E.~L., \& {Cruikshank}, D.~P. 2004, \apjs, 154, 1

\bibitem[{{White} {et~al.}(2005){White}, {Becker}, {Fan}, \&
  {Strauss}}]{White2005}
{White}, R.~L., {Becker}, R.~H., {Fan}, X., \& {Strauss}, M.~A. 2005, \aj, 129,
  2102

\bibitem[{{Whitney} \& {Hartmann}(1992)}]{Whitney1992}
{Whitney}, B.~A., \& {Hartmann}, L. 1992, \apj, 395, 529

\bibitem[{{Whitney} \& {Hartmann}(1993)}]{Whitney1993}
---. 1993, \apj, 402, 605

\bibitem[{{Whitney} {et~al.}(2004){Whitney}, {Indebetouw}, {Bjorkman}, \&
  {Wood}}]{Whitney2004}
{Whitney}, B.~A., {Indebetouw}, R., {Bjorkman}, J.~E., \& {Wood}, K. 2004,
  \apj, 617, 1177

\bibitem[{{Whitney} {et~al.}(2003{\natexlab{a}}){Whitney}, {Wood}, {Bjorkman},
  \& {Cohen}}]{Whitney2003B}
{Whitney}, B.~A., {Wood}, K., {Bjorkman}, J.~E., \& {Cohen}, M.
  2003{\natexlab{a}}, \apj, 598, 1079

\bibitem[{{Whitney} {et~al.}(2003{\natexlab{b}}){Whitney}, {Wood}, {Bjorkman},
  \& {Wolff}}]{Whitney2003A}
{Whitney}, B.~A., {Wood}, K., {Bjorkman}, J.~E., \& {Wolff}, M.~J.
  2003{\natexlab{b}}, \apj, 591, 1049

\bibitem[{{Whittet}(2003)}]{Whittet2003}
{Whittet}, D.~C.~B., ed. 2003, {Dust in the galactic environment}

\bibitem[{{Willott} {et~al.}(2007){Willott}, {Delorme}, {Omont}, {Bergeron},
  {Delfosse}, {Forveille}, {Albert}, {Reyle}, {Hill}, {Gully-Santiago},
  {Vinten}, {Crampton}, {Hutchings}, {Schade}, {Simard}, {Sawicki}, {Beelen},
  \& {Cox}}]{Willott2007}
{Willott}, C.~J., {Delorme}, P., {Omont}, A., {Bergeron}, J., {Delfosse}, X.,
  {Forveille}, T., {Albert}, L., {Reyle}, C., {Hill}, G.~J., {Gully-Santiago},
  M., {Vinten}, P., {Crampton}, D., {Hutchings}, J.~B., {Schade}, D., {Simard},
  L., {Sawicki}, M., {Beelen}, A., \& {Cox}, P. 2007, ArXiv e-prints, 706

\bibitem[{{Willott} {et~al.}(2003){Willott}, {McLure}, \&
  {Jarvis}}]{Willott2003}
{Willott}, C.~J., {McLure}, R.~J., \& {Jarvis}, M.~J. 2003, \apjl, 587, L15

\bibitem[{{Willott} {et~al.}(2005){Willott}, {Percival}, {McLure}, {Crampton},
  {Hutchings}, {Jarvis}, {Sawicki}, \& {Simard}}]{Willott2005}
{Willott}, C.~J., {Percival}, W.~J., {McLure}, R.~J., {Crampton}, D.,
  {Hutchings}, J.~B., {Jarvis}, M.~J., {Sawicki}, M., \& {Simard}, L. 2005,
  \apj, 626, 657

\bibitem[{{Witt}(1977)}]{Witt1977}
{Witt}, A.~N. 1977, \apjs, 35, 1

\bibitem[{{Witt} {et~al.}(1992){Witt}, {Thronson}, \& {Capuano}}]{Witt1992}
{Witt}, A.~N., {Thronson}, Jr., H.~A., \& {Capuano}, Jr., J.~M. 1992, \apj,
  393, 611

\bibitem[{{Wittkowski} {et~al.}(2004){Wittkowski}, {Kervella}, {Arsenault},
  {Paresce}, {Beckert}, \& {Weigelt}}]{Wittkowski2004}
{Wittkowski}, M., {Kervella}, P., {Arsenault}, R., {Paresce}, F., {Beckert},
  T., \& {Weigelt}, G. 2004, \aap, 418, L39

\bibitem[{{Wolf} {et~al.}(1999){Wolf}, {Henning}, \& {Stecklum}}]{Wolf1999}
{Wolf}, S., {Henning}, T., \& {Stecklum}, B. 1999, \aap, 349, 839

\bibitem[{{Wolfire} \& {Cassinelli}(1986)}]{Wolfire1986}
{Wolfire}, M.~G., \& {Cassinelli}, J.~P. 1986, \apj, 310, 207

\bibitem[{{Wood} {et~al.}(1996{\natexlab{a}}){Wood}, {Bjorkman}, {Whitney}, \&
  {Code}}]{Wood1996B}
{Wood}, K., {Bjorkman}, J.~E., {Whitney}, B., \& {Code}, A. 1996{\natexlab{a}},
  \apj, 461, 847

\bibitem[{{Wood} {et~al.}(1996{\natexlab{b}}){Wood}, {Bjorkman}, {Whitney}, \&
  {Code}}]{Wood1996A}
{Wood}, K., {Bjorkman}, J.~E., {Whitney}, B.~A., \& {Code}, A.~D.
  1996{\natexlab{b}}, \apj, 461, 828

\bibitem[{{Wood} \& {Jones}(1997)}]{Wood1997}
{Wood}, K., \& {Jones}, T.~J. 1997, \aj, 114, 1405

\bibitem[{{Wood} {et~al.}(1998){Wood}, {Kenyon}, {Whitney}, \&
  {Turnbull}}]{Wood1998}
{Wood}, K., {Kenyon}, S.~J., {Whitney}, B., \& {Turnbull}, M. 1998, \apj, 497,
  404

\bibitem[{{Wood} {et~al.}(2002){Wood}, {Lada}, {Bjorkman}, {Kenyon}, {Whitney},
  \& {Wolff}}]{Wood2002}
{Wood}, K., {Lada}, C.~J., {Bjorkman}, J.~E., {Kenyon}, S.~J., {Whitney}, B.,
  \& {Wolff}, M.~J. 2002, \apj, 567, 1183

\bibitem[{{Wood} \& {Loeb}(2000)}]{Wood2000}
{Wood}, K., \& {Loeb}, A. 2000, \apj, 545, 86

\bibitem[{{Wyithe} {et~al.}(2002){Wyithe}, {Agol}, \& {Fluke}}]{Wyithe2002}
{Wyithe}, J.~S.~B., {Agol}, E., \& {Fluke}, C.~J. 2002, \mnras, 331, 1041

\bibitem[{{York} {et~al.}(2000){York}, {Adelman}, \& {Anderson}}]{York2000}
{York}, D.~G., {Adelman}, J., \& {Anderson}, Jr., J.~E. et al. 2000, \aj, 120, 1579

\bibitem[{{Yorke}(1980)}]{Yorke1980}
{Yorke}, H.~W. 1980, \aap, 86, 286

\bibitem[{{Yoshida} {et~al.}(2003){Yoshida}, {Abel}, {Hernquist}, \&
  {Sugiyama}}]{Yoshida2003}
{Yoshida}, N., {Abel}, T., {Hernquist}, L., \& {Sugiyama}, N. 2003, \apj, 592,
  645

\bibitem[{{Yoshida} {et~al.}(2007){Yoshida}, {Oh}, {Kitayama}, \&
  {Hernquist}}]{Yoshida2007}
{Yoshida}, N., {Oh}, S.~P., {Kitayama}, T., \& {Hernquist}, L. 2007, \apj, in
  press, astro-ph/0610819

\bibitem[{{Yoshida} {et~al.}(2006){Yoshida}, {Omukai}, {Hernquist}, \&
  {Abel}}]{Yoshida2006}
{Yoshida}, N., {Omukai}, K., {Hernquist}, L., \& {Abel}, T. 2006, \apj, 652, 6

\end{thebibliography}

\end{document}